\pdfoutput=1
\documentclass[cernpreprint, atlasdraft=false, texlive=2023, UKenglish, texmf, orcidlogo]{atlasdoc}
 
\usepackage[subfigure]{atlaspackage}
\usepackage{atlasbiblatex}
 
\usepackage{atlasphysics}
 
\graphicspath{{logos/}{figures/}}
 
\usepackage{ANA-TOPQ-2020-06-PAPER-defs}
\usepackage{atlasjetetmiss}
\usepackage{atlasbsm}
\usepackage{multicol}
\usepackage{multirow}
\usepackage{float}
\usepackage{tabularx}
\usepackage{listings}
\usepackage{adjustbox}

\AtlasTitle{Evidence for the charge asymmetry in $pp \rightarrow t\bar{t}$ production at $\sqrt{s}= 13$~TeV with the ATLAS detector }
 
\AtlasAbstract{
Inclusive and differential measurements of the top--antitop ($t\bar{t}$) charge asymmetry $A_\text{C}^{t\bar{t}}$ and the leptonic asymmetry $A_\text{C}^{\ell\bar{\ell}}$ are presented in proton--proton collisions at $\sqrt{s} = 13$~TeV recorded by the ATLAS experiment at the CERN Large Hadron Collider. The measurement uses the complete Run~2 dataset, corresponding to an integrated luminosity of 139~fb$^{-1}$, combines data in the single-lepton and dilepton channels, and employs reconstruction techniques adapted to both the resolved and boosted topologies.
A Bayesian unfolding procedure is performed to correct for detector resolution and acceptance effects.
The combined inclusive $t\bar{t}$ charge asymmetry is measured to be $A_\text{C}^{t\bar{t}} = 0.0068 \pm 0.0015$, which differs from zero by 4.7~standard deviations.
Differential measurements are performed as a function of the invariant mass, transverse momentum and longitudinal boost of the $t\bar{t}$ system.
Both the inclusive and differential measurements are found to be compatible with the Standard Model predictions, at next-to-next-to-leading order in quantum chromodynamics perturbation theory with next-to-leading-order electroweak corrections.
The measurements are interpreted in the framework of the Standard Model effective field theory, placing competitive bounds on several Wilson coefficients. }
 
\author{The ATLAS Collaboration}

\AtlasJournal{JHEP}
\AtlasJournalRef{JHEP 08 (2023) 077}
\AtlasDOI{10.1007/JHEP08(2023)077}
\PreprintIdNumber{CERN-EP-2022-166}

\hypersetup{pdftitle={ATLAS document},pdfauthor={The ATLAS Collaboration}}
 
\begin{document}
 
\maketitle
 
\tableofcontents
 
\FloatBarrier
\section{Introduction}

 
The large mass of the top quark, which is close to the electroweak symmetry breaking scale, indicates that this particle could play a special role 
in beyond the Standard Model (BSM) theories. Due to the large top-pair production (\ttbar) cross-section for $\sqrt{s} = 13$~\TeV\ proton--proton ($pp$) collisions \cite{TOPQ-2018-17,CMS-TOP-20-001}, the Large Hadron Collider (LHC) experiments collected an unprecedented number of events in which a top-pair is produced. The top quark has a very short lifetime ($\tau_{t} \approx 0.5 \times 10^{-24}\;$s) and decays before hadronisation ($\tau^{\mathrm{had}} \sim 10^{-23}\;$s) \cite{ParticleDataGroup:2020ssz}; therefore, several of its properties may be measured precisely from studies of the top quark's decay products. These measurements test predictions of quantum chromodynamics (QCD), which provides the largest contribution to \ttbar production, and probe potential contributions from couplings between the top quark and BSM particles \cite{Antunano:2007da, Rosner:1996eb, Ferrario:2008wm}.

Production of top-quark pairs is symmetric at leading order (LO) under charge conjugation, and production via $gg \rightarrow t\bar{t}$ (the dominant mechanism at the LHC) is symmetric to all orders. An asymmetry between the $t$ and \tbar originates from interference of the higher-order amplitudes in the $\qqbar \rightarrow \ttbar$ and $qg \rightarrow t\bar{t}q$ production processes, with the contribution of \qqbar annihilation dominating~\cite{Kuhn:1998kw,Kuhn:2011ri}. As a consequence of these asymmetries, the top quark (top antiquark) is preferentially produced in the direction of the incoming quark (antiquark).
 
At a \ppbar collider the preferential direction of the incoming quark (antiquark) almost always coincides with that of the proton (antiproton). The forward--backward asymmetry \Afb is sizeable, ${\cal O}(10\%)$, and can be measured directly~\cite{Aguilar-Saavedra:2014kpa,Czakon:2014xsa,Aaltonen:2008hc,Abazov:2007ab}. The CDF and D0 collaborations initially reported measurements of \Afb that were in some tension with the next-to-leading-order (NLO) Standard Model (SM) prediction. However, the Tevatron legacy results are in good agreement with higher-order calculations~\cite{Aaltonen:2008hc,Abazov:2007ab,Aaltonen:2011kc,Abazov:2015fna,Aaltonen:2012it,Abazov:2014cca}.
 
The LHC is a $pp$ collider. Since the colliding beams are symmetric, it is not possible to measure \Afb as there is no preferential direction for either the top quark or the top antiquark. However, due to the difference in the proton parton distribution functions, on average the valence quarks carry a larger fraction of the proton momentum than the sea antiquarks.  This results in more forward-rapidity top quarks and more central-rapidity top antiquarks.  A central--forward charge asymmetry for $\ttbar$ production, referred to as the charge asymmetry (\Actt)~\cite{Aguilar-Saavedra:2014kpa, Jung:2011zv, Diener:2009ee} is defined as:
 
\begin{equation}
\Actt = \frac{N(\dY>0)-N(\dY<0)}{N(\dY>0)+N(\dY<0)},
\label{intro:eq:ac}
\end{equation}
 
where $\dY = |y_t|-|y_{\tbar}|$ is the difference between the absolute value of the top-quark rapidity\footnote{\AtlasCoordFootnote} $|y_t|$ and the absolute value of the top-antiquark rapidity $|y_{\tbar}|$. At the LHC, since the top-pair production contributions from \qqbar and $qg$ initial states are small, the dominant charge symmetric $gg \rightarrow \ttbar$ process dilutes the measurable asymmetry and the inclusive charge asymmetry is a subtle ${\cal{O}}(1\%)$ effect.
 
A SM calculation at next-to-next-to-leading-order (NNLO) accuracy in the strong coupling and with NLO electroweak (EW) corrections predicts the inclusive charge asymmetry defined in Eq.~(\ref{intro:eq:ac}) to be 0.0064$^{+0.0005}_{-0.0006}$~\cite{Czakon:2017lgo}\footnote{We thank A. Mitov and R. Poncelet for preparing numerical results for the inclusive and differential $A_\text{C}^{t\bar{t}}$ measurements using our binning. The detailed results are publicly available on \href{https://www.precision.hep.phy.cam.ac.uk/results/ttbar-nnloqcd-nloew/}{https://www.precision.hep.phy.cam.ac.uk/results/ttbar-nnloqcd-nloew/}.}, where the uncertainty includes renormalisation and factorisation scale variations and PDF uncertainties.
The `unexpanded' result of Ref.~\cite{Czakon:2017lgo} is used, in which both the numerator and denominator are evaluated through the corresponding order in perturbation theory and then the ratio is taken.
At the NLO, the contribution from EW corrections is about 13\% for the inclusive asymmetry and almost 20\% when the invariant mass of the $t\bar t$ system is above 1~\TeV~\cite{Bern:reuther2012sx} in the differential case.

In events with two final-state leptons, an additional leptonic charge asymmetry is defined using the pseudorapidity of the charged leptons:

\begin{equation}
\Acll = \frac{N(\Delta |\eta_{\ell \bar \ell}|>0)-N(\Delta |\eta_{\ell \bar \ell}|<0)}{N(\Delta |\eta_{\ell \bar \ell}|>0)+N(\Delta |\eta_{\ell \bar \ell}|<0)}\text{,}
\label{intro:eq:acll}
\end{equation}
 
with $\Delta |\eta_{\ell \bar \ell}| = |\eta_{\bar \ell}|-|\eta_{\ell}|$, and where $N(\Delta |\eta_{\ell \bar \ell}|>0)$ and $N(\Delta |\eta_{\ell \bar \ell}|<0)$ represent the number of events with positive and negative $\Delta |\eta_{\ell \bar \ell}|$, respectively.
The leptonic asymmetry is slightly diluted relative to the underlying top-quark asymmetry, but has the advantage that reconstruction of the top-quark pair is not required; only the directions and charges of the leptons are needed.
 
A SM calculation at NLO in QCD, including NLO EW corrections, predicts the inclusive \Acll asymmetry defined in Eq.~(\ref{intro:eq:acll}) to be 0.0040$^{+0.0002}_{-0.0001}$~\cite{Bern:reuther2012sx}\footnote{We thank W. Bernreuther and Z.-G. Si for providing the Standard Model prediction for the leptonic asymmetry for $pp$ collisions at the LHC with $\sqrt{s} = 13$~TeV, using our reduced phase-space.}. The uncertainty includes renormalisation and factorisation scale variations and PDF uncertainties.
 
Several BSM processes, such as anomalous vector or axial-vector couplings (e.g.\ axigluons), heavy \Zprime bosons, or processes which interfere with the SM, can alter \Actt~\cite{Jung:2011zv,Antunano:2007da,Djouadi:2009nb,Ferrario:2008wm,Jung:2009jz,Frampton:2009rk,Ferrario:2009bz,Grinstein:2011yv,AguilarSaavedra:2011hz,Ligeti:2011vt,AguilarSaavedra:2011vw,Rosner:1996eb}. Some BSM models also predict charge asymmetries which vary as a function of the invariant mass \mtt, the transverse momentum \pttt and the longitudinal boost of the $t\bar t$ system along the $z$-axis, \betatt. In particular, BSM effects are expected to be enhanced in specific kinematic regions, for example, when \betatt or \mtt is large~\cite{AguilarSaavedra:2011cp}.

The measurements performed so far by both the ATLAS and CMS collaborations at $\sqrt{s} = $~7, 8 and 13~\TeV, in different decay channels and topologies, have demonstrated good agreement with SM predictions~\cite{TOPQ-2012-17,TOPQ-2014-16,CMS-TOP-11-030,CMS-TOP-12-033,TOPQ-2013-09,TOPQ-2015-08,CMS-TOP-12-010,CMS-TOP-15-009,TOPQ-2015-10,CMS-TOP-13-013,CMS-TOP-17-014}. However, even for the combined ATLAS and CMS inclusive and differential measurements in the lepton+jet channel at $\sqrt{s}=7$ and $\sqrt{s}=8~\TeV$~\cite{TOPQ-2016-16-updated}, uncertainties in the inclusive regions and in kinematic regions such as those with high \mtt are statistically dominated and do not have sufficient sensitivity to exclude many BSM signals, nor to claim evidence for charge asymmetry at the LHC.
 
This paper reports the measurement of \Actt and \Acll with the full Run~2 data at $\sqrt{s}=13~\TeV$ collected in 2015--2018 by the ATLAS experiment at the LHC. The \Actt measurement is performed in \ttbar events with a single isolated lepton in the final state (single-lepton), in both the resolved and boosted topologies of the top-quark decays, and in \ttbar events with two isolated leptons in the final state (dilepton) in the resolved topology only. The \Actt measurements in all topologies are combined inclusively, as well as differentially as a function of \mtt, \betatt and \pttt. The inclusive and differential \Acll measurements (as a function of \mll, \ptll and \betall) are made in the dilepton final state in the resolved topology only. In the resolved topology, each of the top-quark decay products is reconstructed as an individual object, while in the boosted topology the products of the hadronically decaying top (anti)quark are collimated into one object, a \largeR jet.  A Bayesian unfolding procedure~\cite{Choudalakis} is applied to correct for acceptance and detector effects, resulting in parton-level \Ac measurements for comparison with fixed-order theory predictions.
 
The combined results of the \Actt{} inclusive measurement and the differential measurement versus \mtt are interpreted in the framework of the SM effective field theory (SMEFT)~\cite{BUCHMULLER1986621}. Bounds are placed on the Wilson coefficients of dimension-six operators that affect top-quark pair production at the LHC, including 14 four-fermion operators and an operator that modifies the top--gluon interaction~\cite{Grzadkowski:2010es, AguilarSaavedra:2018nen}. The bounds are compared with the bounds obtained from the energy asymmetry measurement~\cite{TOPQ-2019-28-updated} in $t\bar{t}$+jet production.
 
The paper is organised as follows. The ATLAS detector is described in Section~\ref{sec:detector}, while the object definitions and the signal and background modelling  are detailed in Sections \ref{sec:objects} and~\ref{sec:samples}, respectively. The event selection is described in Section~\ref{sec:selection}.
The unfolding procedure is presented in Section~\ref{sec:unfolding}. Section~\ref{sec:systematics} discusses the systematic uncertainties. The results, including the EFT interpretation of \Actt results, are shown in Section~\ref{sec:results}. The conclusions are given in Section~\ref{sec:conclusion}.


\FloatBarrier
\section{ATLAS detector}
\label{sec:detector}

The ATLAS detector~\cite{PERF-2007-01} at the LHC covers nearly the entire solid angle around the collision point.
It consists of an inner tracking detector surrounded by a thin superconducting solenoid, electromagnetic and hadron calorimeters,
and a muon spectrometer incorporating three large superconducting air-core toroidal magnets.
 
The inner-detector system (ID) is immersed in a \SI{2}{\tesla} axial magnetic field
and provides charged-particle tracking in the pseudorapidity range \(|\eta| < 2.5\).
The high-granularity silicon pixel detector covers the vertex region and typically provides four measurements per track,
the first hit normally being in the insertable B-layer (IBL) installed before Run~2~\cite{ATLAS-TDR-19,PIX-2018-001}.
It is followed by the silicon microstrip tracker (SCT), which usually provides eight measurements per track.
These silicon detectors are complemented by the transition radiation tracker (TRT),
which enables radially extended track reconstruction up to \(|\eta| = 2.0\).
The TRT also provides electron identification information
based on the fraction of hits (typically 30 in total) above a higher energy-deposit threshold corresponding to transition radiation.
 
The calorimeter system covers the pseudorapidity range \(|\eta| < 4.9\).
Within the region \(|\eta|< 3.2\), electromagnetic calorimetry is provided by barrel and
endcap high-granularity lead/liquid-argon (LAr) calorimeters,
with an additional thin LAr presampler covering \(|\eta| < 1.8\)
to correct for energy loss in material upstream of the calorimeters.
Hadron calorimetry is provided by the steel/scintillator-tile calorimeter,
segmented into three barrel structures within \(|\eta| < 1.7\), and two copper/LAr hadron endcap calorimeters.
The solid angle coverage is completed with forward copper/LAr and tungsten/LAr calorimeter modules
optimised for electromagnetic and hadronic energy measurements respectively.
 
The muon spectrometer (MS) comprises separate trigger and
high-precision tracking chambers measuring the deflection of muons in a magnetic field generated by the superconducting air-core toroidal magnets.
The field integral of the toroids ranges between \num{2.0} and \SI{6.0}{\tesla\metre}
across most of the detector.
A set of precision chambers covers the region \(|\eta| < 2.7\) with three layers of monitored drift tubes,
complemented by cathode-strip chambers in the forward region, where the background is highest.
The muon trigger system covers the range \(|\eta| < 2.4\) with resistive-plate chambers in the barrel, and thin-gap chambers in the endcap regions.
 
Interesting events are selected by the first-level trigger system implemented in custom hardware,
followed by selections made by algorithms implemented in software in the high-level trigger~\cite{TRIG-2016-01}.
The first-level trigger accepts events from the \SI{40}{\MHz} bunch crossings at a rate below \SI{100}{\kHz},
which the high-level trigger further reduces in order to record events to disk at about \SI{1}{\kHz}.
 
An extensive software suite~\cite{ATL-SOFT-PUB-2021-001} is used in data simulation, in the reconstruction and analysis of real
and simulated data, in detector operations, and in the trigger and data acquisition systems of the experiment.


\FloatBarrier
\section{Object definition and reconstruction}
\label{sec:objects}

This section defines the final-state objects at detector and parton level. The objects are used to characterise the event topologies and to define phase-space regions for the asymmetry measurements.
 
\subsection{Detector-level objects}
 
The primary vertex (PV) of an event is that which has the highest scalar sum of the squared transverse momenta, \pTsq{}, over all associated ID tracks~\cite{ATL-PHYS-PUB-2015-026} with $\pt>0.5~\GeV$. At least two tracks are required.

Electron candidates are reconstructed from clusters of energy deposits in the EM calorimeter that are matched to tracks in the ID~\cite{EGAM-2018-01}.
Candidates are required to have a transverse energy \ET\ greater than 28\,\GeV\ and  $|\eta_{\text{cluster}}| < 2.47$.
If $\eta_{\text{cluster}}$ is within the transition region between the barrel and the endcap of the LAr calorimeter ($1.37 < |\eta_{\text{cluster}}| < 1.52$) the electron candidate is removed.
A multivariate algorithm is used to select signal electrons, which have to satisfy a `tight' likelihood-based quality criterion including shower shape and track selection requirements~\cite{PERF-2017-01}.  The transverse impact parameter $d_0$ of each electron candidate divided by its estimated uncertainty is required to be $|d_0|/\sigma(d_0) < 5$, while the longitudinal impact parameter $z_0$ must satisfy $|z_0\sin{\theta}| < 0.5$\,mm.
The electron candidates have to pass \pt- and $\eta$-dependent isolation requirements based on their tracks and clusters.

Muon candidates are reconstructed from ID tracks combined with track segments or full tracks in the MS~\cite{MUON-2018-03}.
Candidates are required to fulfil the `medium' identification quality criteria.
Only muon candidates within $|\eta| < 2.5$, with $\pt>28~\GeV$, and satisfying impact parameter criteria of $|d_0|/\sigma(d_0) < 3$ and $|z_0\sin{\theta}| < 0.5$\,mm, are selected.
For muon candidates, the track isolation is defined similarly to electron candidates.

Jets are reconstructed from calibrated topological clusters of calorimeter-cell energies~\cite{PERF-2014-07} by using the \akt algorithm~\cite{Cacciari:2008gp,Fastjet} with a radius parameter $R = 0.4$ (\smallR jets). The jet energy is corrected for pile-up effects using a jet-area method~\cite{Cacciari:2008gn} and further corrected using a calibration based on both Monte Carlo (MC) simulations and data~\cite{PERF-2016-04}. Only jets with $\pt>25~\GeV$ and $|\eta| < 2.5$ are selected. Additionally, for jets with $\pt < 60~\GeV$ and $|\eta| < 2.4$, a `jet vertex tagger' (JVT)~\cite{PERF-2014-03} is used to discriminate between jets originating from the PV and those produced in pile-up collisions. The selected JVT working point provides an average efficiency of 92\% for hard-scatter jets and a rejection factor of 99\% for pile-up jets.

\SmallR jets containing $b$-hadrons are identified (`$b$-tagged') using a multivariate algorithm, MV2c10~\cite{PERF-2016-05}, which combines information about track impact parameters and secondary vertices reconstructed within the jet. In the simulated \ttbar events, the operating point for this algorithm provides an efficiency of 77\% to tag $b$-quark jets. The corresponding rejection factors for jets originating from a $c$-quark, light quark, or $\tau$-lepton are 5, 115, and 20, respectively. To account for possible mismodelling causing differences between the predicted and observed selection efficiencies for the different quark-flavour jets and for jets originating from hadronically decaying $\tau$-leptons, per-jet scale factors are obtained from \ttbar{} events in data~\cite{PERF-2016-05}.

\LargeR jets are reconstructed with the \akt algorithm from the individually calibrated topological cell clusters, using a radius parameter $R=1.0$, and then calibrated using simulation~\cite{PERF-2012-01}. They are subsequently \textit{trimmed}~\cite{Krohn:2009th} to remove the effects of pile-up and underlying events. Trimming is a technique in which the original constituents of the jets are reclustered using the \kt algorithm \cite{Ellis:1993tq} with a distance parameter \Rsub, in order to produce a collection of sub-jets. Sub-jets with a fraction of the \largeR jet \pt less than a calibrated threshold \fcut are removed. The trimming parameters used in this analysis are $\Rsub = 0.2$ and $\fcut = 5$\% based on previous studies~\cite{PERF-2015-03}. The \largeR jet moments (e.g.\ mass, $\tau_{32}$~\cite{Thaler:2010tr}) are calculated using only the constituents of the selected sub-jets.

The missing transverse momentum~\cite{PERF-2016-07}, with magnitude \MET, is calculated from a vectorial sum of all reconstructed objects.
These include calibrated electrons, muons, photons, hadronically decaying $\tau$-leptons, and reconstructed \smallR jets.
They are combined with soft hadronic activity measured as reconstructed charged-particle tracks associated with the PV but not with other objects.
 
In order to avoid double counting of the same energy clusters or tracks as different object types, an overlap removal procedure is applied. Firstly, electron candidates sharing a track with any muon candidates are removed. Secondly, if the distance between a \smallR jet and an electron candidate is $\DeltaR < 0.2$, the jet is removed.  If multiple \smallR jets are found with this requirement, only the closest \smallR jet is removed.  If the distance between a \smallR jet and an electron candidate is $0.2 < \DeltaR < 0.4$, then the electron candidate is removed. If the distance between a \smallR jet and any muon candidate is $\DeltaR < 0.4$, the muon candidate is removed if the \smallR jet has more than two associated tracks, otherwise the \smallR jet is removed. Finally, if the distance between a \largeR jet and an electron candidate is $\DeltaR < 1.0$, the \largeR jet is removed.
 
\subsection{Parton-level objects}
Unfolding requires a definition of parton-level $t$ and \tbar{} quarks. For this measurement, a parton-based definition using generator-level information is used.
The top-(anti)quark four-momentum is obtained from MC events, taking the last $t$ (\tbar{}) in the decay chain, just after final-state radiation (FSR) but before it decays.
The \Actt{} measurements are unfolded to parton level in the full phase-space; the unfolding procedure corrects for the effects of reconstruction and acceptance, parton shower and hadronisation.
 
In the measurement of the leptonic asymmetry, the leptons are defined as prompt particles from $W$-boson decays in the MC event record, where the $W$ bosons originate from top-quark decays. The lepton momenta are taken prior to QED radiation.
The unfolded \Acll is obtained in the reduced phase-space defined by the requirement $|\Delta |\eta_{\ell \bar \ell}||$~<~2.5.


\FloatBarrier
\section{Signal and background modelling}
\label{sec:samples}

All signal and background processes are modelled using MC simulations, with the exception of non-prompt-lepton and fake-lepton backgrounds, which are estimated using data-driven methods (see Section~\ref{subsec:Fakes}). All simulated samples used \EVTGEN[1.6.0]~\cite{EvtGen} to model the decays of heavy hadrons, with the exception of some background samples generated with \Sherpa~\cite{Bothmann:2019yzt}. Most of the MC samples were processed using a full simulation~\cite{SOFT-2010-01} of the detector response with the \GEANT toolkit~\cite{AGOSTINELLI2003250}.
Samples used to estimate systematic uncertainties in the modelling were obtained either by reweighting the default full-simulation samples or by using fast simulation software (\textsc{ATLFAST-II}~\cite{SOFT-2010-01}) to parameterise the showers in the calorimeter. To model additional $pp$ interactions from the same or neighbouring bunch crossings, the hard-scattering events were overlaid with a set of minimum-bias interactions generated using \Pythia[8]~\cite{Sjostrand:2014zea} and the \MSTW[lo]~\cite{Martin:2009iq} parton distribution function (PDF) set with the A3~\cite{ATL-PHYS-PUB-2016-017} tuned parameter settings. Finally, the simulated MC events are reconstructed using the same software as the data. More details of the MC samples for the signal and for each background are provided in the following.

\subsection{\ttbar signal}
\label{subsec:samples:ttsig}
All \ttbar samples were generated with a top-quark mass of $\mtop = 172.5\,\GeV$ and are normalised to the inclusive production cross-section of $\sigma(\ttbar) = \mathrm{832\pm51~pb}$.  This cross-section is calculated at NNLO in QCD including the resummation of next-to-next-to-leading-logarithm (NNLL) soft-gluon terms using \TOPpp[2.0]~\cite{Beneke:2011mq,Cacciari:2011hy,Baernreuther:2012ws,Czakon:2012zr,Czakon:2012pz,Czakon:2013goa,Czakon:2011xx}. The uncertainties in the cross-section due to the PDF and \alphas are calculated using the PDF4LHC prescription~\cite{Botje:2011sn} with the \MSTW[nnlo] 68\% CL~\cite{Martin:2009iq,Martin:2009bu}, \CT[10nnlo]~\cite{Lai:2010vv,Gao:2013xoa} and \NNPDF[2.3]~5f~FFN~\cite{Ball:2012cx} PDF sets, and are added in quadrature to the effect of the scale uncertainty.
 
The nominal \ttbar\ events were generated with the \POWHEGBOX[v2]~\cite{Frixione:2007nw,Nason:2004rx,Frixione:2007vw,Alioli:2010xd,Re:2010bp,Frederix:2012dh, Alioli:2009je} generator, which provides matrix elements at NLO in the strong coupling constant \alphas, with the \NNPDF[3.0nlo]~\cite{Ball:2014uwa} PDF and the \hdamp\ parameter set to $1.5$\,\mtop~\cite{ATL-PHYS-PUB-2016-020}. The \hdamp\ parameter controls the \pt\ of the first additional emission beyond the LO Feynman diagram in the parton shower and therefore regulates the high-\pt\ emission against which the \ttbar\ system recoils. The functional form of the renormalisation and factorisation scales (\muR and \muF) was set to the nominal scale of $\sqrt{m_{\textrm{top}}^2 + p_{\text{T,top}}^2}$. The events were interfaced with \Pythia[8.230]~\cite{Sjostrand:2014zea} for the parton shower (PS) and hadronisation, using the A14 set of tuned parameters~\cite{ATL-PHYS-PUB-2014-021}  and the \NNPDF[2.3lo]~\cite{Ball:2014uwa} PDF set.
 
To study the \ttbar\ modelling uncertainties, alternative MC samples are considered.  Particular samples were simulated using a parameterised detector shower response in \textsc{ATLFAST-II}, and compared with a version of the nominal sample which was produced in the same way.

The impact of choosing a different PS and hadronisation model is evaluated by comparing the nominal \ttbar\ signal sample with an alternative sample from the nominal matrix-element (ME) generator interfaced with \Herwig[7.04]~\cite{Bahr:2008pv,Bellm:2015jjp} instead of \Pythia[8.230], and using the H7UE set of tuned parameters~\cite{Bellm:2015jjp,Bellm:2017bvx} and the \MMHT[lo] PDF set~\cite{Harland-Lang:2014zoa}.

The \PROTOS generator~\cite{AguilarSaavedra:2008gt} with the \CTEQ[6L1] PDF set~\cite{Pumplin:2002vw} was used to generate \ttbar samples predicting different asymmetry values due to the inclusion of a new heavy axigluon. The generated samples contain only parton-level information, which is later used to reweight the nominal \ttbar \POWPY[8] sample.

\subsection{Single top}
Single-top $tW$ associated production was modelled using the \POWHEGBOX[v2] generator, which provides MEs at NLO in \alphas, using the five-flavour scheme with the \NNPDF[3.0nlo] PDF set. The functional form of  \muR and \muF was set to the nominal scale of $\sqrt{m_{\textrm{top}}^2 + p_{\text{T,top}}^2}$. The diagram removal (DR) scheme~\cite{Frixione:2008yi} was employed to treat the interference with \ttbar\ production~\cite{ATL-PHYS-PUB-2016-020}. Dedicated samples using a diagram subtraction (DS) scheme~\cite{Frixione:2008yi} are considered when evaluating the uncertainty due to the treatment of the overlap with \ttbar\ production.
 
Single-top $t$-channel ($s$-channel) production was modelled using the \POWHEGBOX[v2] generator, which provides MEs at NLO in \alphas, using the four-flavour (five-flavour) scheme with the \NNPDF[3.0nlo] PDF set. The functional form of  \muR and \muF is not determined by the heavy top quark, but was set to $\sqrt{m_{b}^2 + p_{{\textrm T},b}^2}$~\cite{Frederix:2012dh}, following the arguments presented in Ref.~\cite{PhysRevLett.102.182003}. For these processes, the events were interfaced with \Pythia[8.230] using the A14 parameter tune and the \NNPDF[2.3lo] PDF set.

Similarly to the \ttbar\ samples, the impact of choosing a different PS and hadronisation model is evaluated by comparing the nominal sample with an alternative sample from the nominal ME generator interfaced instead with \Herwig[7.04], using the H7UE set of tuned parameters and the \MMHT[lo] PDF set.

\subsection{\Wboson{} and \Zboson{} bosons with additional jets}
QCD \Vjets production ($V$ = $W$ or $Z$ boson) was simulated with the \Sherpa[2.2.1]~\cite{Bothmann:2019yzt} PS MC generator.
In this set-up, NLO MEs for up to two jets, and LO  MEs for up to four jets, were calculated with the Comix~\cite{Gleisberg:2008fv} and \OPENLOOPS~\cite{Cascioli:2011va,Denner:2016kdg} libraries.
The nominal \Sherpa PS~\cite{Schumann:2007mg}, based on Catani--Seymour dipoles and the cluster hadronisation model~\cite{Winter:2003tt}, was used. This employs a dedicated set of tuned parameters developed by the \Sherpa authors and based on the \NNPDF[3.0nnlo] PDF set~\cite{Ball:2014uwa}.
The \Vjets samples are normalised to a NNLO prediction~\cite{Anastasiou:2003ds}.
\subsubsection{\Zboson{} calibration in the dilepton channel}
Scaling factors are applied to the \Zjets background in the dilepton channel, to allow for the significant theoretical uncertainties in modelling the heavy-flavour contribution to this process. The modelling uncertainties mainly affect the same-flavour channels, for which the \Zll{} contribution to the background is larger.
A control region is created by applying a 10\,\GeV{} mass window around the \Zboson-boson mass to the invariant mass of the opposite-sign lepton pair; this region is named `OSZ'.
The \MET requirements (see Section~\ref{subsec:selection_dilep}) are removed, in order to increase the number of \Zjets events and their purity in the control region. Using this control region, a scale factor, $\mu_{Z}$, is defined as the ratio of the number of data events after the subtraction of other processes (based on MC predictions)
to the predicted number of \Zjets events.
 
The scale factors are derived separately for the 1$b$-tag-exclusive and 2$b$-tag-inclusive regions (with exactly one or at least two $b$-tagged jets, respectively) because of their different heavy-flavour compositions.  The validity of the derived scale factors is checked as a function of the kinematics of the \ttbar system and as a function of the loosened \MET requirement in the OSZ control region. Each scale factor is assigned a 30\% normalisation uncertainty, which fully covers the statistical and systematic errors, and all sources of kinematic or selection dependence. The measured scale factors are $1.21\pm 0.36$ and $1.39 \pm 0.42$ for the 1$b$-tag-exclusive and 2$b$-tag-inclusive regions, respectively.  These factors are applied to the  \Zjets background events predicted by the MC simulation.
 
The contribution of the $Z\rightarrow \tau\tau$ process in the $e\mu$ channel is very small, and is estimated using MC simulation only. No scale factors are applied, but a 30\% normalisation uncertainty is assigned for consistency with the same-flavour channels.

\subsection{Non-prompt and fake leptons background}
\label{subsec:Fakes}
\subsubsection{Single-lepton channel}
Non-prompt- and fake-lepton events (fake events), can enter the selected data samples if a non-prompt or fake lepton is reconstructed. Several production mechanisms or mistakes in event reconstruction can produce such leptons.  These include semileptonic decays of heavy-flavour hadrons, long-lived weakly decaying states (e.g.\ \pipm, \kaon mesons), \pizero mesons misreconstructed as electrons, electrons from photon conversions, or prompt photons. To estimate the total contribution of fake events a data-driven `matrix-method'~\cite{ATLAS-CONF-2014-058} is used.  Two categories of events are selected, satisfying `loose' (identification only) and `tight' (identification and isolation) lepton selection requirements, respectively. The real-lepton (fake-lepton) efficiency, $\epsilon_{\text{real}}$ ($\epsilon_{\text{fake}}$), is defined as the ratio of the number of events with a real (non-prompt/fake) lepton satisfying the tight selection to the number of events with a real (non-prompt/fake) lepton satisfying the loose selection. The real-lepton efficiency is measured in data by applying a tag-and-probe method to $\Zboson{}$ decays with two leptons and jets in the final state, while the fake-lepton efficiency is measured in control regions enriched in fake/non-prompt leptons.
The contribution of fake events is estimated from weighted data events passing loose lepton selection requirements, where the weight depends on the real- and fake-lepton efficiencies.
 
\subsubsection{Dilepton channel}
To allow for possible mismodelling of the non-prompt- and fake-lepton backgrounds in the dilepton channel, a scaling factor is extracted using a data-driven approach and is applied to the MC prediction. The MC contribution is evaluated from the generator's event record, using the history of the stable particles and summing over all MC samples.
 
A control region enriched in fake leptons is created by applying all the standard selection requirements, but requiring the two leptons to have same-sign charges; this region is denoted by `SS'.
The fake-lepton scale factor, $\mu_{\text{fake}}$, is derived by comparing the number of simulated events in the SS region with those observed in data,
after subtracting the MC-predicted number of
events with two same-sign prompt leptons.
 
The scale factors are measured separately in $ee+e\mu$ events (where the fake electrons dominate) and in $\mu\mu$ events, and are applied to each event containing fake leptons in the corresponding opposite-sign signal region.
In addition, the effect of charge misidentification is taken into consideration in the $ee+e\mu$ fake-lepton scale factors.  A correction for charge misidentification is first derived by comparing events in the OSZ region with events in a same-sign version of the $Z$-mass window, denoted by `SSZ', and between the MC prediction and data.  Unlike the OSZ region, the SSZ region consists primarily of events (dilepton \ttbar, single top and \Zjets) for which the charge of one of the leptons has been misassigned.  The \Zjets events have already been scaled by $\mu_{Z}$, to reduce the effect of heavy-flavour mismodelling on the estimate of charge misidentification.
 
The final $ee+e\mu$ fake-lepton scale factors, accounting also for charge misidentification, are measured to be $0.96 \pm 0.29$ and $0.94 \pm 0.28$ in the 1$b$-tag-exclusive and 2$b$-tag-inclusive regions, respectively. A normalisation uncertainty of 30\% accounts for the statistical uncertainty, charge misidentification and range of fake-lepton estimates across the \ttbar and dilepton kinematic variables.
The equivalent $\mu\mu$ fake-lepton scale factors are measured to be $2.3 \pm 1.2$ and $4.2 \pm 2.1$ in the 1$b$-tag-exclusive and 2$b$-tag-inclusive regions, respectively.  These include a $50\%$ normalisation uncertainty, due to the low number of events in the muonic SS region and high contamination from \ttV\ events.  Fake leptons from $\mu\mu$ events contribute less than $1\%$ to the total number of events in the signal region.

\subsection{Other backgrounds}
Diboson ($WW$, $ZZ$, $WZ$) samples were simulated with the \Sherpa[2.2.1] and~2.2.2~\cite{Bothmann:2019yzt} PS MC generator.
\Sherpa[2.2.2] was used for two- and three-lepton samples.
Additional hard-parton emissions~\cite{Gleisberg:2008fv} were matched to a PS based on Catani--Seymour dipoles, using a dedicated set of tuned parton-shower parameters developed by the \Sherpa authors, and the \NNPDF[3.0nnlo] PDF set.
Matrix element and PS matching~\cite{Hoeche:2011fd} was employed for different jet multiplicities, which were then merged into an inclusive sample using an improved CKKW matching
procedure~\cite{Catani:2001cc,Hoeche:2009rj}.  The procedure was extended to NLO using the MEPS@NLO prescription~\cite{Hoeche:2012yf}.
These simulations are at NLO for up to one additional parton and at LO for up to three additional parton emissions using factorised on-shell decays.
The virtual QCD corrections for matrix elements at NLO were provided by the \OPENLOOPS library.
The calculation was performed in the $G_\mu$ scheme, ensuring an optimal description of pure EW interactions at the EW scale.
 
The production of \ttV\ and \ttH\ events was modelled using the \MGNLO[2.3.3]~\cite{Alwall:2014hca} and \POWHEGBOX[v2] generators, respectively. The generators provide matrix elements at NLO in \alphas, with the \NNPDF[3.0nlo] PDF set. An exception is the simulation of \ttH\ events corresponding to data collected in 2018, modelled using the  \MGNLO[2.6.0] generator. For \ttV\ and \ttH\ production, the events were interfaced with \Pythia[8.210]~\cite{Sjostrand:2014zea} and \Pythia[8.230], respectively.  Each used the A14 tune and the \NNPDF[2.3lo] PDF set.
 
The rare SM processes leading to $tWZ$ and $tZ$ production are also considered in the dilepton channel. For both processes, the samples were simulated with the \MGNLO[2.3.3] generator at NLO with the \NNPDF[3.0nlo] PDF set. The events were interfaced with \Pythia[8.212]~\cite{Sjostrand:2014zea} using the A14 tune and the \NNPDF[2.3lo] PDF set.


\FloatBarrier
\section{Event selection and reconstruction}
\label{sec:selection}

\subsection{Common event selection}
The analysis uses data collected by the ATLAS detector between 2015 and 2018 from $pp$ collisions at a centre-of-mass energy of $\sqrt{s}=13\,\TeV$, corresponding to an integrated luminosity of 139\,fb$^{-1}$.
Only events recorded under stable beam conditions with all detector subsystems operational~\cite{DAPR-2018-01}, with a primary vertex and passing a single-electron or single-muon trigger are considered.
Multiple triggers are used to increase the selection efficiency. The lowest-threshold triggers utilise isolation requirements to reduce the trigger rate.  These have \pt thresholds of 20\,\GeV\ for muons and 24\,\GeV\ for electrons in 2015 data, and  26\,\GeV\  for both lepton types in 2016, 2017 and 2018 data~\cite{TRIG-2018-05,TRIG-2018-01}. They are complemented by other triggers with higher \pt{} thresholds and with no isolation requirements.
 
\subsection{Event selection and reconstruction in the single-lepton topology}
A common event selection is used for the resolved and boosted single-lepton topologies, requiring exactly one lepton candidate matched to the trigger lepton with a minimum \pt of $28\,\GeV$ and at least one of the \smallR jets to be $b$-tagged. Events containing additional leptons with $\pt > 25\,\GeV$ are rejected.
To reduce the impact of fake-events background, requirements on \MET{} and \MTW{} are applied.\footnote{$\MTW = \sqrt{2p_{\text{T}}^{\ell}\MET(1-\cos{\Delta\phi})}$ where $\Delta\phi$ is the angle between the lepton and \METvec{} in the transverse plane and $p_{\text{T}}^{\ell}$ is the transverse momentum of the lepton.} In the electron channel, both \MET{} and \MTW{} are required to be larger than 30\,\GeV{} because of the higher level of background with fake leptons (see Section~\ref{subsec:Fakes}), while in the muon channel a triangular requirement, $\MET+\MTW > 60\,\GeV$, is applied.
 
The selected events are divided into four regions, based on event topology (resolved, boosted) and on $b$-jet multiplicity ($1b$-tag-exclusive, $2b$-tag-inclusive). The events with an electron and those with a muon are summed.

\subsubsection{Resolved topology}
 
The single-lepton resolved topology requires at least four \smallR{} jets with $\pt > 25~\GeV$. Events which also pass the boosted selection (see Section~\ref{subsec:selection_boosted}) are removed.
 
The challenge in reconstructing an event in the single-lepton resolved topology is to correctly assign individual selected jets to the corresponding partons from the decaying top quarks. For this purpose a boosted decision tree (BDT), implemented within the TMVA package~\cite{Hocker:2007ht} is designed. The BDT combines kinematic variables and $b$-tagging information with weight information from the Kinematic Likelihood Fitter (KLFitter)~\cite{Erdmann:2013rxa}, into a single discriminant. Each permutation of jet-to-parton assignments is evaluated and the permutation with the highest BDT score is used for the \ttbar{} kinematic reconstruction.
 
Since the number of possible permutations increases rapidly with the jet multiplicity, only permutations of up to five jets are considered. If more than five jets are present in an event, the two jets with the highest $b$-tagging scores are considered, together with the remaining three highest-\pt jets. The \ttbar signal sample (see Section~\ref{subsec:samples:ttsig}) is used for the BDT training, with each jet-to-parton permutation for an event labelled as either `signal' or `background'.  Only permutations with four jets correctly assigned within $\Delta R = 0.3$ of the corresponding partons are labelled as signal; all other permutations are labelled as combinatorial background.
No attempt is made to individually match the partons from the hadronically decaying $W$ boson, as the twofold ambiguity does not affect the reconstruction.
The BDT aims to discriminate the signal from the combinatorial background and is trained separately for the $1b$-tag-exclusive and $2b$-tag-inclusive regions, but inclusively in the lepton flavours (electron, muon). A large number of background permutations per event have a  low likelihood calculated by the KLFitter, and these are easily distinguishable from the correct permutation. Including these permutations in the BDT training would reduce the overall performance. Therefore, the BDT is trained using only the background permutations for which the KLFitter calculates the highest likelihood.
 
Thirteen variables are used as input to the BDT:
\begin{itemize}
\item the reconstructed mass of the hadronically decaying top quark,
\item the logarithm of the KLFitter likelihood,
\item the reconstructed mass of the hadronically decaying $W$ boson,
\item a label depending on whether the $b$-jet candidate from the semileptonically decaying top quark is $b$-tagged or not,
\item a label depending on whether the $b$-jet candidate from the hadronically decaying top quark is $b$-tagged or not,
\item a label depending on whether a light-jet candidate from the $W$-boson decay is $b$-tagged or not,
\item the reconstructed mass of the semileptonically decaying top quark,
\item the $\Delta R$ between the $b$-jet candidate from the semileptonically decaying top quark and the lepton,
\item the $\Delta R$ between the two light-jet candidates from the $W$-boson decay,
\item the \pt of the system of the lepton and the $b$-jet candidate from the semileptonically decaying top quark,
\item the number of jets in the event,
\item the pseudorapidity of the hadronically decaying top quark,
\item the $\Delta R$ between the two $b$-jet candidates from the \ttbar decay.
\end{itemize}
 
The $b$-jet candidates in the variables above are those assigned by the best jet permutation, so a $b$-jet candidate is not necessarily $b$-tagged.
The reconstructed mass of the semileptonically decaying top quark requires the reconstruction of the neutrino four-momentum. Experimentally, only the magnitude of the neutrino three-momentum in the transverse direction, represented by \MET{}, and its angle in $\phi$ are accessible. Assuming no mass term for the neutrino, one parameter of the four-momentum must be determined. It is calculated using a $W$-boson mass constraint by exploiting the formula: $m^2_W = (p^{\ell} +p^{\nu})^2 = m_{\ell}^2 + 2(E^{\ell}E^{\nu} - \vec{p}^{\ell} \cdot \vec{p}^{\nu})$, where $p^{\ell}$ and  $p^{\nu}$ are the four momenta of the charged lepton and the neutrino, respectively. This leads to a quadratic equation for the longitudinal component of the neutrino three-momentum, $p^{\nu}_z$. If real solutions are found, the solution which produces a reconstructed top-quark mass closest to the expected top-quark mass is chosen. If imaginary solutions are found, the transverse components of the neutrino three-momentum are
varied until a real solution for $p^{\nu}_z$ is found.
 
\begin{figure}[htp!]
\centering
\vspace{-0.2cm}
\subfigure[][]{\includegraphics[width=0.48\textwidth]{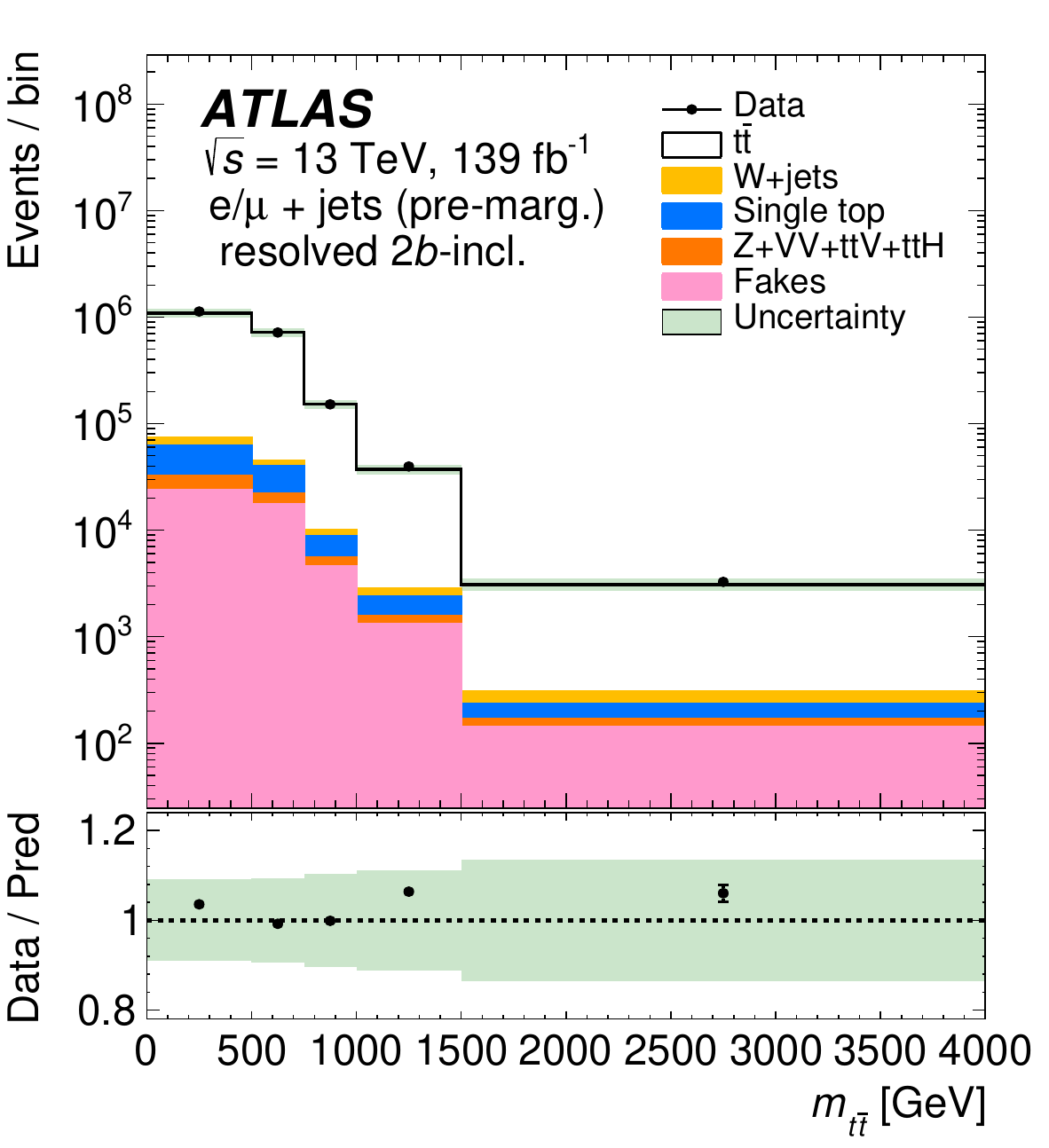}}
\subfigure[][]{\includegraphics[width=0.48\textwidth]{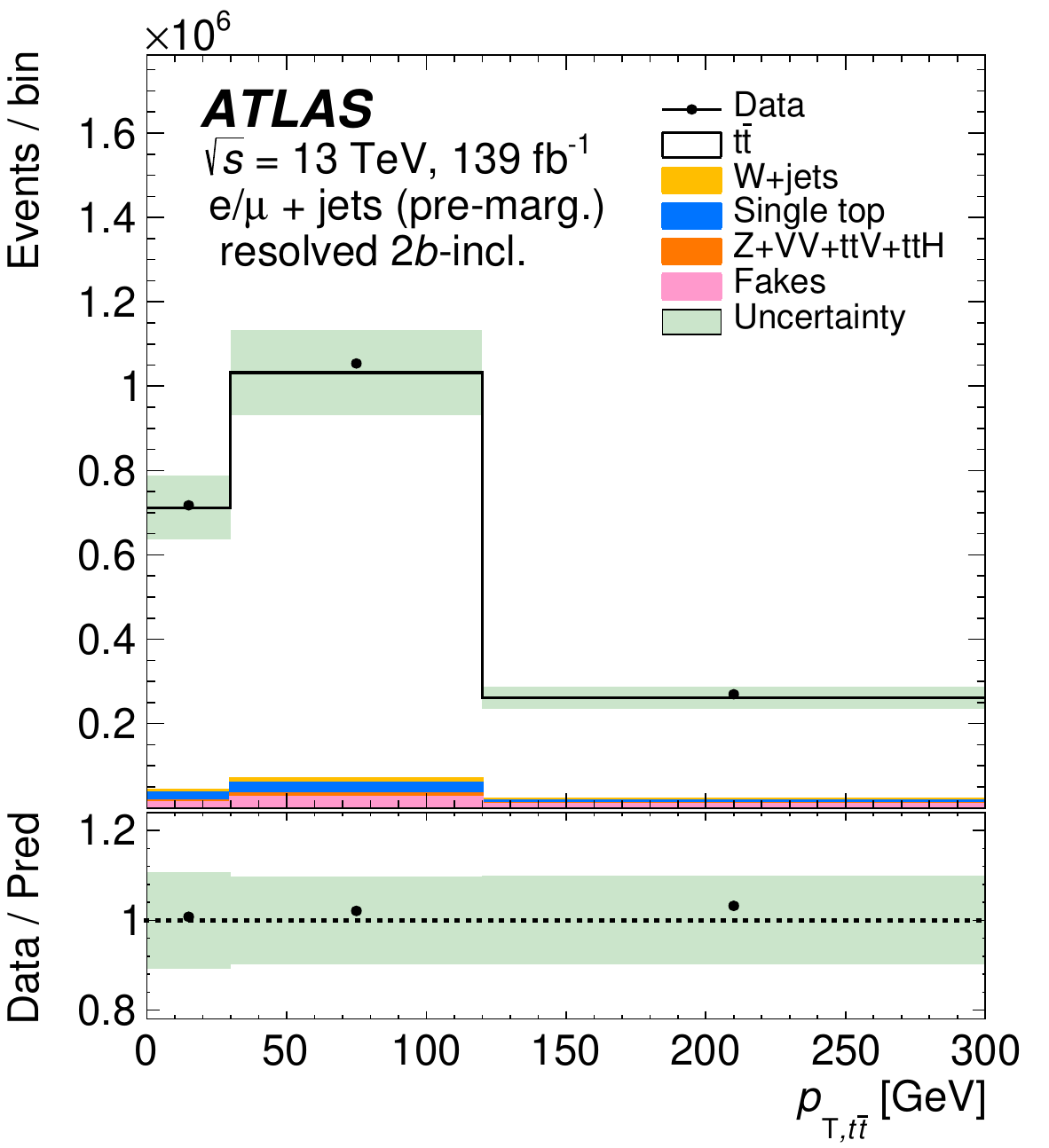}}
\vspace{-0.1cm}
\caption{
Comparison between the data and the prediction for the reconstructed (a) \mtt and (b) \pttt in the single-lepton resolved 2$b$-inclusive region. Statistical and systematic uncertainties are included in the presented total uncertainty.}
\label{fig:prefit_ljets_resolved_mtt_pttt}
\end{figure}

For the final selection, the best permutation in each event is required to have a BDT discriminant score $> 0.3$ in order to reject \ttbar combinatorial backgrounds and to suppress non-\ttbar background processes.  For \ttbar signal events where a jet-to-parton assignment is possible for all partons from the \ttbar decay, the correct assignment is found for 75\% of the events. This represents 50\% of correctly assigned jets with respect to all \ttbar signal events passing event selection. Non-\ttbar background processes naturally populate lower regions of the BDT score, because the jet permutations are unable to satisfy the expected decay kinematics of a \ttbar pair.
Imposing a threshold on the BDT score increases the ratio of signal to non-\ttbar background by a factor of ${\sim}2$.
Table~\ref{tab:ljets_event_yields} shows the event yields after the event selection for the single-lepton resolved topology. Distributions of \mtt and \pttt for the single-lepton resolved 2$b$-inclusive region are shown in Figure~\ref{fig:prefit_ljets_resolved_mtt_pttt}. The total uncertainty before marginalisation (see Section~\ref{sec:unfolding}) is presented. The uncertainty includes the experimental uncertainties, and the signal and background modelling uncertainties discussed in Section~\ref{sec:systematics}.
 
\begin{table}[H]
\caption{Event yields split by topology (resolved, boosted) and $b$-tag multiplicity (1$b$-excl., 2$b$-incl.) in the single-lepton channel. Statistical and systematic uncertainties are included in the presented total uncertainty.}
\label{tab:ljets_event_yields}
\begin{center}
\begin{tabular}{cr@{\,$\pm$\,}rr@{\,$\pm$\,}rr@{\,$\pm$\,}rr@{\,$\pm$\,}r}
\hline
\multirow{2}{*}{Process:} & \multicolumn{4}{c}{Single-lepton resolved} & \multicolumn{4}{c}{Single-lepton boosted} \\
& \multicolumn{2}{c}{$1b$-excl.} & \multicolumn{2}{c}{$2b$-incl.} & \multicolumn{2}{c}{$1b$-excl.} & \multicolumn{2}{c}{$2b$-incl.}\\
\hline
\ttbar                          & 1\,540\,000       & 140\,000 & 1\,870\,000      & 170\,000        & 50\,000 & 12\,000 & 74\,000 & 18\,000 \\
Single top                      & 90\,000         & 11\,000  & 51\,000        & 8\,000          & 3\,600  & 1\,100  & 3\,000  & 1\,100   \\
\Wjets                          & 180\,000        & 100\,000 & 20\,000        & 9\,000          & 8\,900  & 2\,600  & 1\,600  & 500   \\
$Z+VV+t\bar tX$                 & 48\,000         & 25\,000  & 14\,000        & 7\,000          & 2\,400  & 1\,200  & 1\,400  & 700    \\
Fake                            & 90\,000         & 50\,000  & 47\,000        & 24\,000         & 3\,000  & 1\,500  & 2\,300  & 1\,200         \\
\hline
Total Prediction                & 1\,940\,000       & 190\,000 & 2\,010\,000      & 180\,000        & 68\,000 & 14\,000 & 83\,000 & 18\,000 \\
\hline
Data & \multicolumn{2}{c}{1\,964\,127}    & \multicolumn{2}{c}{2\,041\,063} & \multicolumn{2}{c}{54\,750}      & \multicolumn{2}{c}{66\,571}     \\
\hline
\end{tabular}
\end{center}
\end{table}
 
\subsubsection{Boosted topology}
\label{subsec:selection_boosted}
In the single-lepton boosted topology, the reconstruction aims to identify one high-\pt{} and collimated hadronic top-quark decay and at least one \smallR{} jet with $\pt > 25\,\GeV$ close to the selected lepton, with $\DeltaR(\text{jet}_{R=0.4}\text{, }\ell)<1.5$. If multiple small-$R$ jets satisfy this condition, the one with highest \pt{} is considered for the subsequent semileptonically decaying top-quark reconstruction. In addition, at least one \largeR{} top-tagged jet with $\pt > 350\,\GeV$ and $|\eta|<2$ is required as the hadronically decaying top-quark candidate. The \largeR jet is evaluated by a top-tagging algorithm utilising jet mass and \tauthrtwo{} substructure~\cite{Thaler:2010tr} variables, where an operating point with an efficiency of 80\% is chosen. The top tagger is optimised using the same approach as described in Ref.~\cite{ATLAS-CONF-2017-064}. Since the two top quarks are expected to be back-to-back in the \ttbar{} rest frame, additional requirements related to the \largeR{} jet, the isolated lepton and the \smallR{} jet close to the lepton are applied: $\Delta\phi(\text{jet}_{R=1.0}\text{, }\ell) > 2.3$ and $\DeltaR(\text{jet}_{R=1.0}\text{, jet}_{R=0.4}) > 1.5$. Finally, a requirement on the invariant mass of the reconstructed \ttbar{} system of $\mtt > 500\,\GeV$ is applied. This criterion is imposed to remove poorly reconstructed events passing selection in boosted regions, removing approximately 0.1\% of events. Consequently, the $\mtt < 500\,\GeV$ bin in the \Ac{} vs \mtt{} measurement is removed in the boosted regions.
 
The four-momentum of the leading-\pt{} \largeR{} jet satisfying the selection criteria is taken as the four-momentum estimate of the hadronically decaying top quark. The semileptonically decaying top quark's four-momentum is constructed from the isolated lepton, the selected \smallR{} jet and the neutrino four-momentum. The neutrino four-momentum is calculated in the same way as in the single-lepton resolved topology.
Table~\ref{tab:ljets_event_yields} shows the event yields after the event selection for the single-lepton boosted topology. The MC prediction's overestimation of the yield by about 20\% in the single-lepton boosted regions is confirmed by differential cross-sections measurements~\cite{TOPQ-2019-23-updated,TOPQ-2018-11} at large values of top \pt (>300\,GeV).

\subsection{Event selection and reconstruction in the dilepton topology}
\label{subsec:selection_dilep}
The dilepton topology requires at least  two \smallR{} jets with $\pt > 25\,\GeV$ and two opposite-charge leptons. Minimum \pt requirements of $28$ and $25\,\GeV$ are applied for the leading and sub-leading leptons, respectively, and at least one of the leptons must be matched to the trigger lepton. At least one of the \smallR jets is required to be $b$-tagged. For the same-flavour channel, the reconstructed invariant mass of the dilepton system is selected to be outside a \Zboson-boson mass window ($|\mll-m_{Z}|>10~\GeV$) while a missing transverse momentum requirement of $\MET > 60$~(30)\,\GeV\ for the $1b$-tag-exclusive ($2b$-tag-inclusive) region is applied to further reduce the \Zjets contribution. In the $1b$-tag-exclusive region, the reconstructed invariant mass of the same-flavour dilepton system is required
to be greater than 15\,\GeV{} to suppress backgrounds from low-mass resonances. The \ttbar{} system is reconstructed using the Neutrino Weighting (NW) algorithm~\cite{D0:2015dxa}. The NW event weight is required to be greater than zero, in order to remove events with poorly reconstructed kinematics; these amount to approximately 15\% of signal events. Only the events used for the \Actt measurements are considered in the \Acll measurements.
The selected events are divided into four regions: by lepton flavour -- events with exactly one electron and one muon ($e\mu$ channel), or two same-flavour leptons ($ee+\mu\mu$ channel) -- and by $b$-jet multiplicity ($1b$-tag-exclusive, $2b$-tag-inclusive).
Table~\ref{tab:dilep_event_yields} shows the event yields after the event selection for the dilepton channel.
 
\begin{table}[H]
\caption{Event yields split by lepton flavour ($e\mu$, $ee+\mu\mu$) and $b$-tag multiplicity ($1b$-excl., $2b$-incl.) in the dilepton channel. Statistical and systematic uncertainties are included in the presented total uncertainty.}
\label{tab:dilep_event_yields}
\begin{center}
\begin{tabular}{cr@{\,$\pm$\,}rr@{\,$\pm$\,}rr@{\,$\pm$\,}rr@{\,$\pm$\,}r}
\hline
\multirow{2}{*}{Process:} & \multicolumn{4}{c}{$e\mu$} & \multicolumn{4}{c}{$ee+\mu\mu$} \\
& \multicolumn{2}{c}{$1b$-excl.} & \multicolumn{2}{c}{$2b$-incl.} & \multicolumn{2}{c}{$1b$-excl.} & \multicolumn{2}{c}{$2b$-incl.}\\
\hline
\ttbar                                & 225\,000 & 13\,000 & 262\,000 & 17\,000 & 105\,000 & 6\,000 & 192\,000 & 13\,000 \\
Single top ($tW$)                     &  13\,800 &  1\,200 &   5\,600 &     900 &   6\,100 &    600 &   4\,100 &     700 \\
Diboson                               &      650 &      70 &       45 &       6 &      296 &     34 &       55 &       6 \\
\Zjets                                &      610 &     200 &       77 &      25 &   3\,500 & 1\,400 &   3\,800 &  1\,200 \\
Rare SM ($t\bar tX$, $tWZ$, etc.)     &      690 &      90 &   1\,040 &     140 &      390 &     50 &      850 &     110 \\
Fake                                  &   2\,500 &     800 &   2\,100 &     700 &   1\,200 &    400 &   1\,900 &     800 \\
\hline
Total Prediction                      & 243\,000 & 13\,000 & 271\,000 & 17\,000 & 117\,000 & 7\,000 & 203\,000 & 13\,000 \\
\hline
Data                   & \multicolumn{2}{c}{   244\,258}                  & \multicolumn{2}{c}{   273\,856} & \multicolumn{2}{c}{   116\,096}                  & \multicolumn{2}{c}{   202\,967} \\
\hline
\end{tabular}
\end{center}
\end{table}


\FloatBarrier
\section{Unfolding}
\label{sec:unfolding}

The \dY distributions, used to extract \Ac, are smeared by acceptance and detector resolution effects. The true \dY, defined in MC simulation using the $t$-quark and $\bar{t}$-antiquark after final-state radiation but before decay, is estimated from data using an unfolding procedure. The fully Bayesian unfolding (FBU)~\cite{Choudalakis} method is used to unfold the observed data.
FBU is an application of Bayesian inference to the unfolding problem.  Given the observed data $\Data\in \mathcal{R}^{N_r}$ with $N_r$ bins, the task is to determine the true distribution $\Truth{}\in\mathcal{R}^{N_t}$ with $N_t$ bins. Assuming a response matrix $\TrasfMatrix\in\mathcal{R}^{N_r} \times \mathcal{R}^{N_t}$, which models the detector response to the true distribution, and an estimate of the background distribution $\Bckg{}\in\mathcal{R}^{N_r}$, the posterior probability of the true distribution follows the probability density:
 
\begin{equation*}
\conditionalProb{\Truth{}}{\Data{}}
\propto{}
\conditionalLhood{\Data{}}{\Truth{}}
\cdot{}
\pi{}\left(\Truth{}\right),
\end{equation*}
 
where $\conditionalProb{\Truth{}}{\Data{}}$ is the posterior probability of the true distribution $\Truth{}$ under the condition of \Data{}, $\conditionalLhood{\Data{}}{\Truth{}}$ is the likelihood function of $\Data{}$ for a given $\Truth{}$, and  $\pi{}\left(\Truth{}\right)$ is the prior probability density for the true distribution $\Truth{}$. The likelihood $\conditionalLhood{\Data}{\Truth{}}$ is defined as a product of Poissonian probabilities for each bin of the reconstructed spectrum:
\begin{equation*}
\conditionalLhood{\Data}{\Truth} = \prod\limits_{i}^{N_r} \frac{(r_i + b_i)^{d_i}}{d_i!} \mathrm{e}^{-(r_i + b_i)},
\end{equation*}
 
where $r_i$ and $b_i$ are the expected signal and background yields and $d_i$ is the observed data yield in bin $i$. The expected signal yield $r_i$ is given by the true distribution \Truth{} and the response matrix \TrasfMatrix{} according to the following equation:
\begin{equation*}
r_i = \sum\limits_j^{N_t} m_{ij} t_j,
\end{equation*}
 
where $m_{ij}$ are the elements of the response matrix and $t_j$ is the true prediction in bin $j$.
 
For this measurement, in all bins a uniform prior probability density is chosen for $\pi{}\left(\Truth{}\right)$, such that equal probabilities are assigned to all $\Truth{}$ spectra within a wide range. The response matrix is estimated from the simulated sample of \ttbar events, and the unfolded asymmetry \Ac is computed from $\conditionalProb{\Truth{}}{\Data{},\TrasfMatrix{}}$ as:

\begin{equation*}
\conditionalProb{\Ac{}}{\Data{}} = \int{ \delta(\Ac{} - \Ac{}(\Truth{})) \conditionalProb{\Truth{}}{\Data{},\TrasfMatrix{}} \dif \Truth{}} .
\end{equation*}

The treatment of systematic uncertainties is naturally included in the Bayesian inference approach by extending the likelihood \conditionalLhood{\Data{}}{\Truth{}} to include nuisance parameters. The marginal likelihood is defined as:

\begin{equation*}
\conditionalLhood{\Data{}}{\Truth{}}=
\int
\conditionalLhood{\Data{}}{\Truth{},\thetavec{}}
\cdot{}\mathcal{N}(\thetavec{})~\dif\thetavec{},
\end{equation*}
 
where \thetavec{} are the nuisance parameters, and $\mathcal{N}(\thetavec{})$ are their prior probability densities.  These are assumed to be Gaussian distributions with $\mu=0$ and $\sigma=1$, with the exception of background normalisation distributions, which are represented by truncated Gaussian distributions to prevent unphysical negative yields. One nuisance parameter is associated with each of the uncertainty sources.
 
In FBU, the marginalisation approach provides a framework in which to perform the unfolding and the background estimations in multiple data regions simultaneously. Given the data distribution $\Data{}_i$ measured in $N_\mathrm{reg}$ independent channels, the likelihood can be extended to combine the channels using a product of likelihoods as:
 
\begin{equation}
\label{eq:unf:comb}
\conditionalLhood{\{\Data{}_1\cdots{}\Data{}_{N_\mathrm{reg}}\}}{\Truth{}}=
\int
\prod_{i=1}^{N_\mathrm{reg}}\conditionalLhood{\Data{}_i}{\Truth{};\thetavec{}}
\cdot{} \mathcal{N}(\thetavec{})
~\dif\thetavec{},
\end{equation}
 
where the nuisance parameters are common to all analysed regions.
The likelihood is sampled around its minimum using an extended Markov-chain Monte Carlo method~\cite{JMLR:v15:hoffman14a} in order to estimate the posterior probability of all the parameters of interest.
The post-marginalisation uncertainty takes into account possible constraints and post-marginalisation correlations among individual nuisance parameters.


\FloatBarrier
\section{Systematic uncertainties}
\label{sec:systematics}

The inclusive and differential measurements are affected by several sources of systematic uncertainty, including experimental uncertainties, signal and background modelling uncertainties,  uncertainties due to limited MC sample sizes and uncertainties due to unfolding bias. The individual systematic uncertainty sources are described in this section. The effect of the signal modelling uncertainties is estimated by varying the response matrix. The effects of all other systematic uncertainties are obtained by taking the difference between the relevant systematically varied reconstructed distribution and nominal reconstructed distribution. Two-sided systematic uncertainties are symmetrised by taking half of the difference between the upward and downward variations.

A Bootstrap~\cite{Bohm:169869} method is applied in order to estimate the effect of limited MC sample sizes on the systematic uncertainties.
Only systematic variations which are found to be significant compared to their statistical precision are kept. When systematic effects are found not to be significant, their respective uncertainty is set to zero. Such a method cannot fully remove the effects of the limited MC sample sizes, which are particularly important for the generator modelling uncertainties.

\subsection{Experimental uncertainties}
 
The uncertainty in the combined 2015--2018 integrated luminosity is 1.7\% \cite{ATLAS-CONF-2019-021}, obtained using the LUCID-2 detector \cite{LUCID2} for the primary luminosity measurements.  This systematic uncertainty affects all processes modelled using MC simulation.
 
The uncertainty in the reweighting of the MC pile-up distribution to match the distribution in data is evaluated by varying the pile-up correction factors used to perform the reweighting within their uncertainties.
 
Uncertainties associated with the lepton selection arise from the trigger, reconstruction, identification and isolation efficiencies, and the lepton momentum scale and resolution~\cite{EGAM-2018-01,MUON-2018-03}. The uncertainties are obtained from data by using either a tag-and-probe method on events with \Zboson-boson, \Wboson-boson, or \jpsi decays~\cite{PERF-2017-01,MUON-2018-03}, or studies of reconstructed distributions of $\Zboson\rightarrow\ellell$, $\jpsi\rightarrow\ellell$ and $\Wboson\rightarrow\enu$ using methods similar to those in Refs.~\cite{MUON-2018-03,PERF-2017-03}.
 
Uncertainties associated with the \smallR jet selection arise from the jet energy scale (JES), the JVT requirement and the jet energy resolution (JER). The uncertainty in the JES is assessed in data~\cite{PERF-2016-04}, using MC-based corrections and in situ techniques. It is broken down into a set of 29 decorrelated nuisance parameters, with contributions from pile-up, jet-flavour composition, single-particle response, and punch-through. Each parameter has different jet \pt and $\eta$ dependencies~\cite{JETM-2018-05}.
The uncertainty in the JVT efficiency comes from the estimation of the residual contamination from pile-up jets after pile-up suppression, and a systematic uncertainty assessed by using different MC generators for simulation of $\Zboson\rightarrow\mu\mu$ and \ttbar events~\cite{PERF-2014-03}.  The uncertainty in the JER is determined by an eigenvector decomposition strategy similar to that for the JES systematic uncertainties. Eight nuisance parameters take into account various effects evaluated by comparing simulation with data. The overall effect of the JER uncertainty is parameterised in jet \pt and $\eta$~\cite{JETM-2018-05}.
 
The efficiency of the flavour-tagging algorithm is measured for each jet flavour using control samples in data and simulation.  The uncertainties are related to the $b$-jet tagging calibration for $b$-jets, $c$-jets and light-jets, and comprise nine, four and five eigenvector variations to the tagging efficiencies, respectively~\cite{PERF-2016-05}.
The uncertainties due to the high-\pt{} extrapolation of the $b$-jet and $c$-jet efficiencies are accounted for by two additional nuisance parameters.
 
The uncertainties of all the \MET components associated with reconstructed objects are propagated in a fully correlated way. Additional uncertainties in the scale and resolution of the soft-track components are considered~\cite{PERF-2016-07}.
 
The \largeR jets in the single-lepton channel have uncertainties associated with their momentum and scale resolution.  The scale of the detector response for all relevant jet momenta (\pt, jet mass $m_{\text{jet}}$, \tauthrtwo) is derived by comparing the calorimeter response with the tracker response for a matched reference track-jet~\cite{PERF-2015-03}.
To account for the resolution of the detector response, the jet \pt is smeared by an additional 2\% uncertainty. The jet mass resolution uncertainty is estimated by smearing the jet mass response by an additional 20\% (where the nominal response is parameterised in jet \pt and $m_{\text{jet}}/\pt$)~\cite{ATLAS-CONF-2017-063}. A set of 14 nuisance parameters is used to estimate uncertainties due to these effects in the single-lepton boosted channel.
 
\subsection{Signal modelling}
\label{subsec:SigModel}
 
During the unfolding procedure, the \ttbar{} signal normalisation is a free parameter common to all bins of the true \dY distribution, and its posterior probability is estimated.
Therefore, the overall normalisation (affecting all \dY bins simultaneously) of each signal modelling uncertainty that compares two specific generator configurations is removed. Only the shape difference affecting the \dY bins separately is considered in the following nuisance parameters (NP):
 
\begin{itemize}
\item The \ttbar{} PS+hadronisation NP takes into account the choice of PS, underlying event, and hadronisation models. The uncertainty is evaluated by comparing the \ttbar{} \POWPY[8] (nominal) sample with the \POWHER[7] sample.
 
\item The \ttbar \muF and \ttbar \muR NPs correspond to uncertainties arising from missing higher-order QCD corrections in the ME computation. They are estimated by varying one scale in the ME by factors of 0.5 and 2.0 in the nominal \ttbar\ signal sample, while the other scale is kept at its nominal value, and vice versa.
 
\item The \ttbar Var3c NP represents the uncertainty due to variation of the strong coupling in the initial-state PS. The uncertainty is estimated by varying the Var3c parameter from the A14 tune~\cite{ATL-PHYS-PUB-2014-021} in the nominal \ttbar\ signal sample.
 
\item The \ttbar \hdamp NP characterises the uncertainty associated with the choice of \hdamp parameter value, which determines the
kinematics of the leading emission~\cite{ATL-PHYS-PUB-2020-023}. In this case, the nominal \ttbar\ sample is compared with a dedicated alternative sample in which the value of the \hdamp\ parameter is increased to 3.0\,\mtop{}.
 
\item The \ttbar FSR NP corresponds to the uncertainty arising from FSR. It is obtained using PS weights in the nominal \ttbar{} sample, which vary the renormalisation scale for QCD emission in the FSR by a factor of 2.0 or 0.5 for reduced or enhanced radiation, respectively.

\end{itemize}
The uncertainty due to the choice of parton distribution functions is obtained using the PDF4LHC15 prescription~\cite{Rojo:2016ymp}, which utilises a set of 30 separate NPs.
 
\subsection{Background modelling}
In the single-lepton channel, the \Wjets process is the dominant background in the $1b$-tag-exclusive region. It is a charge asymmetric process, to which a total cross-section uncertainty of 5\%~\cite{STDM-2015-03} is assigned. Variations of \Wjets~\cite{ATL-PHYS-PUB-2017-006} production which alter both its predicted shape and acceptance are used to estimate a modelling uncertainty. By reweighting the nominal \Wjets prediction using dedicated MC generator weights, variations are considered for the renormalisation and factorisation scales, matrix-element-to-parton-shower matching CKKW scale~\cite{Catani:2001cc,Krauss_2002}, and the scale used for the resummation of soft gluon emission.
In addition, to account for potential mismodelling of the various flavour contributions of \Wjets, three flavour composition variations are considered. The $W{+}b/c$/light-jets events are filtered into separate samples and their individual normalisations are varied by $\pm \SI{30}{\%}$~\cite{ATL-PHYS-PUB-2017-006}.
 
Single-top production is the dominant background in the dilepton channel and is non-negligible especially in the $2b$-tag-inclusive single-lepton regions. Since the main contribution comes from the $t\Wboson$ channel, an uncertainty of 5.3\%~\cite{Aliev:2010zk,Kant:2014oha} is assigned to the predicted cross-section. In addition, the MC samples used for \ttbar{} and single-top $t\Wboson$ production contain an overlap in the final state, which is removed using the  DR scheme. To estimate the uncertainty, the difference between the predictions of the nominal $t\Wboson$ sample and the alternative sample using the DS scheme is considered. Furthermore, uncertainties due to PS and hadronisation, ISR and FSR are taken into account for the $tW$ channel as well as for the $s$- and $t$-channels. The PS and hadronisation uncertainty is evaluated by comparing the nominal sample with the sample produced with \POWHEGBOX and interfaced with \Herwig[7.04].
The uncertainties due to ISR and FSR are estimated using varied weights in the ME and in the PS\@. To simulate higher (lower) ISR,  \muR and \muF are varied by a factor of 0.5 (2.0) in the ME\@. The impact of increased or decreased FSR is evaluated using PS weights, which vary the renormalisation scale for QCD emission in the FSR by a factor of 0.5 or 2.0, respectively.
In the dilepton channel, $tW$ production contributes to the single-top background, while events from $s$- and $t$-channel production contribute to the fake-lepton background.
Therefore, in the single-lepton channel, the systematic variations associated with $tW$ production are separated from the systematic variations associated with $s$- and $t$-channel production.
 
In the dilepton channel, \Zjets\ production is the second most dominant background process in the same-flavour regions. By reweighting the nominal \Zjets prediction using dedicated MC generator weights, variations are considered for the renormalisation and factorisation scales. In addition, a 30\% normalisation uncertainty is assigned.
In the single-lepton channel, \Zjets\ production is considered together with other physics backgrounds, described below.
 
Diboson production is a non-negligible background in the dilepton channel.
An uncertainty of 6\% is assigned to the predicted diboson cross-section~\cite{ATL-PHYS-PUB-2017-005}.
In the single-lepton channel, diboson production is considered together with other physics backgrounds, described below.
 
In the matrix method, the real- and fake-lepton efficiencies are parameterised. To estimate the shape uncertainty, an alternative parameterisation of real- and fake-lepton efficiencies is compared with the nominal parameterisation in each region. In addition, a 50\% normalisation uncertainty is considered.
 
Other physics backgrounds making small contributions in the single-lepton channel include \Zjets, diboson, \ttV, and \ttH production. They are treated as a single background process in the unfolding procedure, and a cross-section normalisation uncertainty of 50\% is applied.
 
Rare SM processes such as \ttV\ and \ttH\ are assigned a cross-section uncertainty of 13\%~\cite{Frixione:2015zaa}.
 
\subsection{Method uncertainties}
 
Given the finite number of events in the \ttbar{} signal sample, the response matrix is estimated with limited statistical precision. To estimate the resulting uncertainty in \Ac, an Asimov dataset~\cite{Cowan:2010js} unfolding is repeated 500 times with smeared response matrices (according to the MC statistics) to obtain a distribution of pseudo-experiment results for \Ac. The width of this distribution is considered as the uncertainty, which is then summed in quadrature with the total uncertainty obtained from the unfolding. The Gaussian smearing is applied to each element of the response matrix.
 
To evaluate any source of bias from the unfolding procedure, the response of the unfolding is determined from eight pseudo-datasets generated with \PROTOS.
Each pseudo-dataset is composed of the nominal \ttbar signal, reweighted to simulate a specific injected asymmetry. The injected \Ac values range between $-0.05$ and 0.06, depending on the differential variable and bin. By unfolding the eight reweighted pseudo-datasets with the nominal response matrix, including all systematic uncertainties, the uncertainty associated with the unfolding response is calculated as $\Acmeas - (\Acmeas -b)/a$, where $a$ and $b$ are the slope and offset of a linear fit of the generator-level (intrinsic) \Ac to the unfolded \Ac of the eight reweighted pseudo-datasets, and \Acmeas is the measured asymmetry value in data. The unfolding bias uncertainty has a negligible impact on the total uncertainty.


\FloatBarrier
\section{Results}
\label{sec:results}

\FloatBarrier
\subsection{\Actt{} measurement}
 
The measured value of \Actt{} is inferred from the posterior distribution of \dY{} according to Eq.~(\ref{intro:eq:ac}), where the \dY{} posterior is obtained by marginalising the extended likelihood in Eq.~(\ref{eq:unf:comb}). The value of \Actt is measured separately for the single-lepton and dilepton channels, combining the four regions of each channel in the unfolding procedure. In order to obtain the final result, all the single-lepton and dilepton regions are combined in the unfolding procedure.
 
\begin{figure}[htp!]
\centering
\vspace{-0.2cm}
\subfigure[][]{\includegraphics[width=0.48\textwidth]{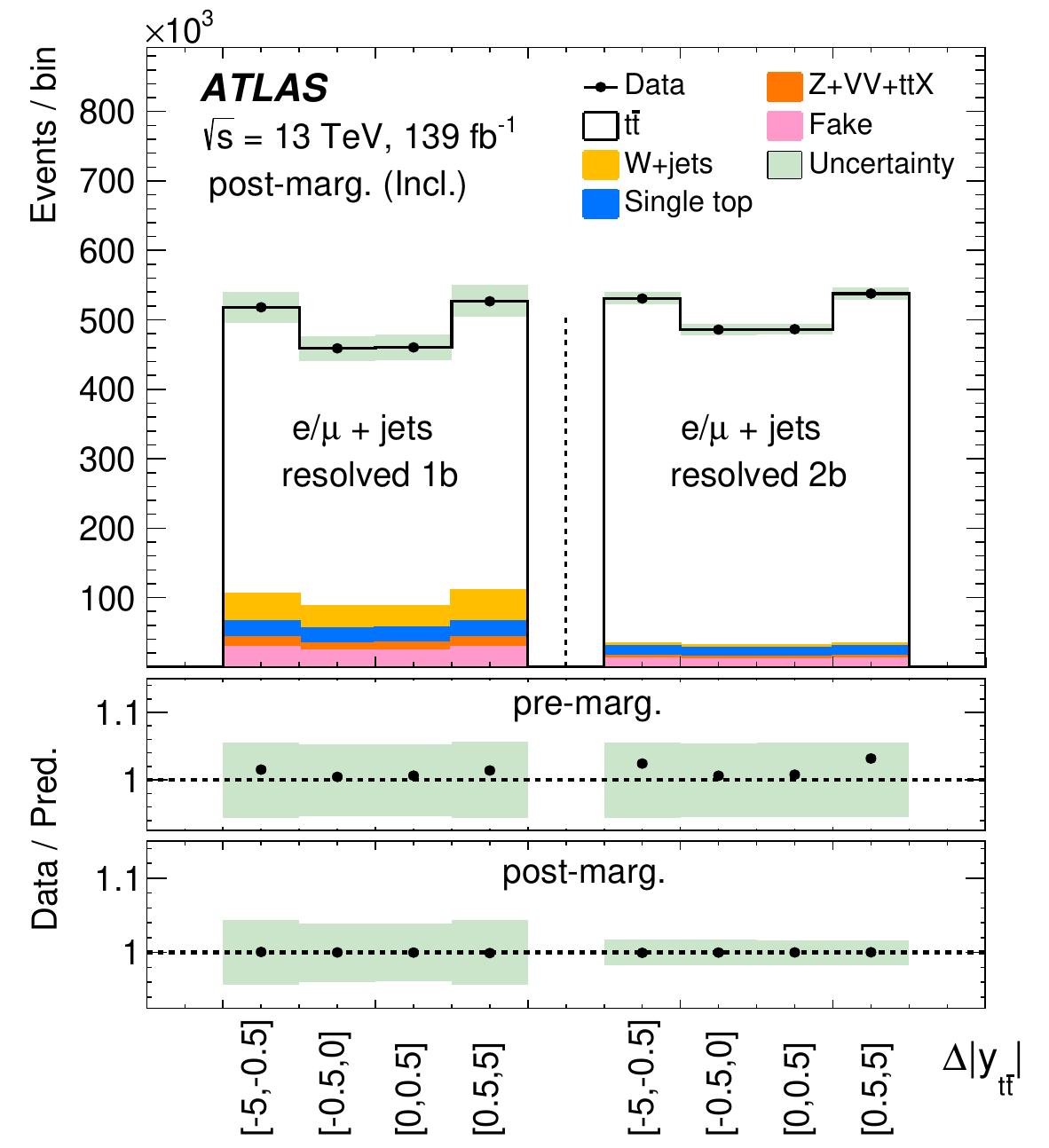}}
\subfigure[][]{\includegraphics[width=0.48\textwidth]{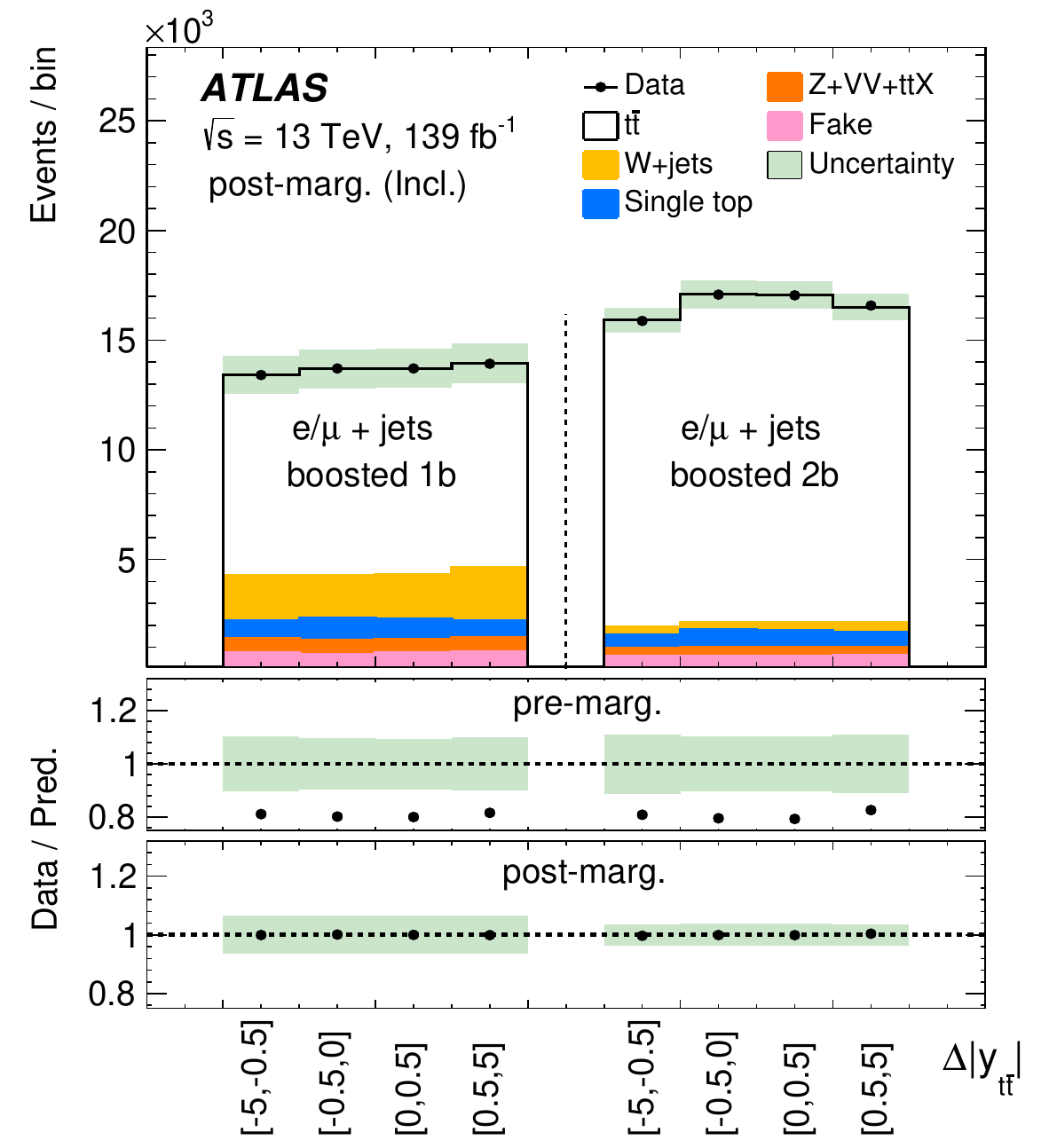}}\\
\subfigure[][]{\includegraphics[width=0.48\textwidth]{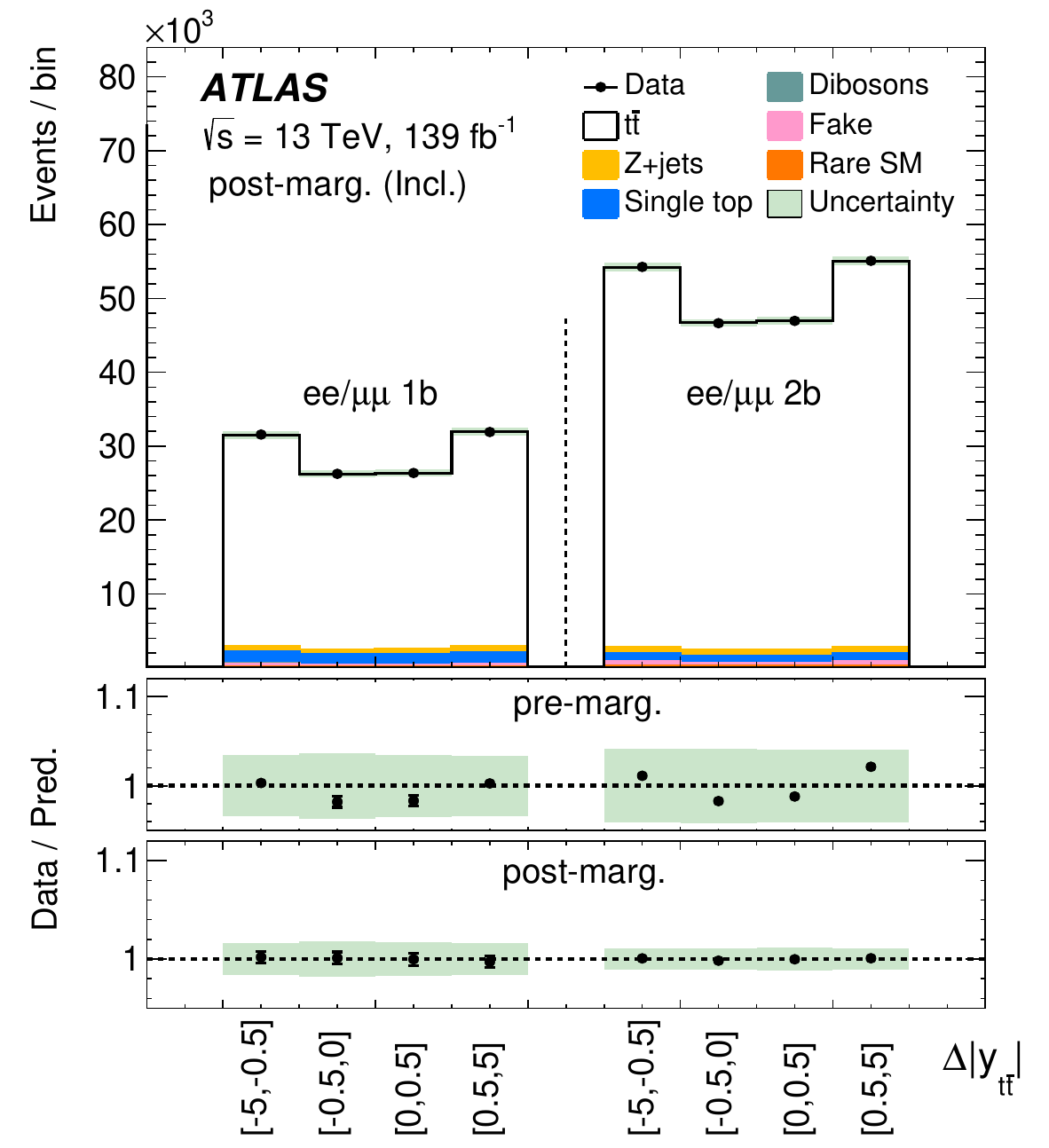}}
\subfigure[][]{\includegraphics[width=0.48\textwidth]{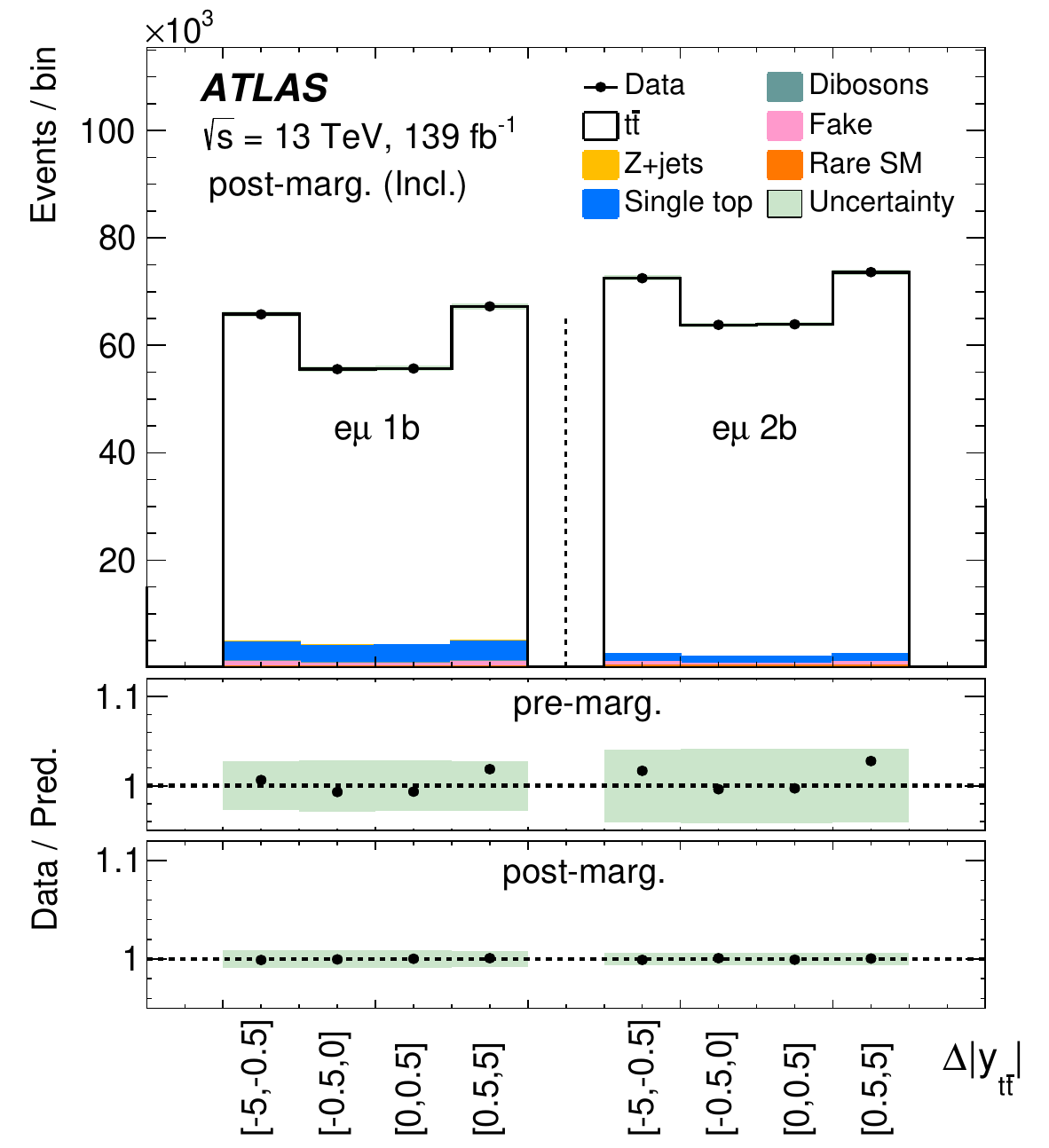}}
\vspace{-0.1cm}
\caption{
Comparison between the data and the prediction for the reconstructed \dYtt in the inclusive 
\Actt measurement in the (a) single-lepton resolved, (b) single-lepton boosted, (c) dilepton $ee+\mu\mu$ and (d) dilepton $e\mu$ region. The ratio of data to prediction  is shown before and after marginalisation within FBU. Shaded bands correspond to the total uncertainty of the prediction. The signal modelling uncertainties do not include normalisation effects because the overall acceptance difference cancels out in the observable. The normalisation difference of the order of 20\% between the data and MC prediction, present in the single-lepton boosted channel before marginalisation, is confirmed by differential cross-sections measurements. }
\label{fig:prefit_postfit1}
\end{figure}

To compute \Actt, at least two \dY bins are necessary.
The \dY spectrum is divided into four-bin histograms for all channels/regions and for all differential measurements.
While using fewer than four bins results in a smaller \Actt statistical uncertainty, using a larger number of bins allows a more accurate mapping of migrations that change the sign of \dY{} and allows an unbiased measurement of \Actt.
The \dY{} bin edges are hence chosen to minimise a possible bias in the measurement.
As an example, Figure~\ref{fig:prefit_postfit1} shows the four regions for the two channels in the inclusive case.
The optimisation of the \mtt, \pttt and \betatt differential bins is based on the number of events they contain.
The same \dY, \mtt, \pttt and \betatt bin sizes are used in the single-lepton and dilepton channels as well as in the combination.
 
As mentioned in Section~\ref{subsec:SigModel}, the \ttbar signal modelling uncertainties take into account only the shape differences affecting \dYtt{} bins. However, when combining $N_{\mathrm{reg}}$ regions, the different \ttbar signal models may have different acceptances for signal events to enter a particular region. To allow independent \ttbar signal normalisation in each of the $N_{\mathrm{reg}}$ combined regions, $N_{\mathrm{reg}} -1$ signal normalisation parameters\footnote{Since the overall \ttbar{} signal normalisation is a free parameter because of the unfolding, only $N_{\mathrm{reg}}-1$ additional degrees of freedom are necessary. The MC overestimation in the single-lepton boosted regions is absorbed by these normalisation parameters.} $\vect{K}_{\text{norm}}$ are inserted into the likelihood with a uniform prior $\pi{}(\vect{K}_{\text{norm}})$. The posterior probability density thus becomes:
 
\begin{equation}
\label{eq:results:like}
\begin{split}
\conditionalProb{\Truth{}}{\{\Data{}_1\cdots{}\Data{}_{N_{\textrm{reg}}}\}}=&
\int
\prod_{i=1}^{N_{\textrm{reg}}}\conditionalLhood{\Data{}_i}{\Reco{}_i(\Truth{}, \vect{K}_{\text{norm}};\thetavec{}_{\textrm r}),\Bckg{}_i(\thetavec{}_{\textrm r},\thetavec{}_{\textrm b}, \vect{\gamma})}\\
& ~\mathcal{N}(\thetavec{}_{\textrm r})
~\mathcal{N}(\thetavec{}_{\textrm b})
~\pi{}(\Truth{})
~\pi{}(\vect{K}_{\text{norm}})
~\textrm{d}\thetavec{}_{\textrm r}
~\textrm{d}\thetavec{}_{\textrm b},\\
\end{split}
\end{equation}
 
where $\Bckg{} = \Bckg{}(\thetavec{}_{\textrm r},\thetavec{}_{\textrm b}, \vect{\gamma})$ is the total background prediction, the probability densities $\pi{}$ are uniform priors and \Reco{} is the reconstructed signal prediction. Three categories of nuisance parameters are considered:
the normalisation of the background processes ($\thetavec{}_{\textrm b}$), the uncertainties in the background prediction due to limited MC sample sizes ($\vect{\gamma}$), and the uncertainties associated with signal and background modelling, object identification, reconstruction and calibrations ($\thetavec{}_{\textrm r}$). Uncertainties belonging to the last category affect both the reconstructed distribution of the \ttbar{} signal and the background prediction. The $\vect{\gamma}$ nuisance parameters are per-bin variables, allowing the total background prediction to fluctuate within its statistical uncertainty~\cite{Barlow:1993dm}. In each region, four $\vect{\gamma}$ nuisance parameters correspond to four \dYtt bins. In the differential measurements, four $\vect{\gamma}$ nuisance parameters are considered in each differential bin, in each region.

When combining multiple regions in the unfolding procedure, assumptions have to be made about the correlations of systematic uncertainties. The experimental uncertainties are treated as fully correlated across all combined regions in all \Actt measurements. For the \ttbar signal and background modelling uncertainties, several decorrelation schemes are studied. For these uncertainties, constraints may lead to an unjustified reduction of the uncertainty. This is because model comparisons include multiple effects, where implicit assumptions are made about their correlation structure. To limit the number of NPs, the modelling uncertainties are correlated among regions unless they are significantly constrained (by more than about 30\%). For a small number of cases in which decorrelating across the regions is found to increase the total uncertainty by 5\%--20\%, the more conservative scheme is chosen. Decorrelating more uncertainties increases the complexity of the FBU marginalisation but does not lead to a sizeable increase in the total uncertainty. The impact of the decorrelation scheme on the unfolded \Actt mean value is found to be negligible. Overall, the changes to the asymmetry value and uncertainty due to the decorrelation scheme choice are much smaller than the statistical uncertainty in all measurement regions.
 
In the single-lepton channel measurement, the following decorrelation scheme is used:
\begin{itemize}
\item With the exception of the \ttbar Var3c and PDF uncertainties, all of the signal modelling uncertainties are decorrelated between the resolved and boosted regions, to reduce the constraints, given the significantly different kinematics in these regions. The aforementioned exceptions are kept correlated.
\item In order to reduce constraints on the systematic uncertainties, the \ttbar PS+hadronisation uncertainty is decorrelated between all of the combined regions.
\item In addition, to mitigate stronger constraints in the differential \Actt measurements, the \ttbar \hdamp, \ttbar FSR and \ttbar PS+hadronisation systematic uncertainties are decorrelated between differential bins, but the \dYtt bins of the systematic variation are kept correlated within a given differential bin.
\item For the background modelling uncertainties, only the single-top PS+hadronisation, single-top FSR and \Wjets \muR and \muF scales are decorrelated between the resolved and boosted regions. In the differential measurements, single-top $tW$-channel FSR systematic uncertainties are decorrelated between differential bins as well.
\item The fake-lepton shape and normalisation uncertainties are decorrelated among all combined regions, given that the fake-events composition in each region is expected to be different.
\end{itemize}
In the dilepton channel measurement, a different decorrelation scheme was used:
\begin{itemize}
\item All of the \ttbar signal and background modelling uncertainties are fully correlated across all combined regions, with the exception of the \Zjets and fake-lepton normalisation uncertainties. This differs from the single-lepton channel (where the resolved and boosted regions have very different kinematics) because the dilepton channel includes only the resolved topology.
\item The \Zjets and fake-lepton normalisation uncertainties are decorrelated between individual regions as they have either very different contributions or compositions in the respective regions.
\item To reduce the constraints on the systematic uncertainties in the differential measurements, the \ttbar \hdamp and \ttbar PS+hadronisation systematic uncertainties are decorrelated between differential bins, but the \dYtt bins of the systematic variation are kept correlated within a given differential bin.
\end{itemize}
Finally, in the combination, the following  decorrelation scheme is adopted:
\begin{itemize}
\item In the inclusive measurement, the systematic uncertainties for \ttbar \hdamp, \ttbar \muR and \muF scales, \ttbar FSR and single-top $tW$ channel FSR are correlated across single-lepton resolved and dilepton regions, while they are decorrelated from the single-lepton boosted regions, given the significantly different kinematics in these regions. In order to reduce the number of constraints, the \ttbar PS+hadronisation is decorrelated among single-lepton and dilepton channels. A single NP is used for all dilepton regions, while the uncertainty is decorrelated among the individual single-lepton regions. The \ttbar Var3c and PDF uncertainties are kept correlated across all combined regions.
\item In the differential measurements, the \ttbar \muR and \muF scales are correlated among the single-lepton resolved and dilepton regions, while they are uncorrelated with the single-lepton boosted regions. The \ttbar PS+hadronisation, \ttbar \hdamp, \ttbar FSR as well as single-top PS+hadronisation and single-top FSR are decorrelated between single-lepton resolved, single-lepton boosted and dilepton regions. In addition, \ttbar PS+hadronisation, \ttbar \hdamp and single-top FSR are decorrelated between differential bins, with \ttbar PS+hadronisation also decorrelated between 1-$b$-tag exclusive and 2-$b$-tag inclusive regions. The \ttbar FSR is also decorrelated between differential bins, but only in single-lepton regions, where sizeable constraints are observed otherwise.
\item  Similarly to the single-lepton channel measurement, the \Wjets \muR and \muF scales are decorrelated between single-lepton resolved and single-lepton boosted regions.
\item The single-top $tW$ normalisation is correlated across all regions.
\item The fake-lepton shape uncertainty in the single-lepton regions and fake-lepton normalisation uncertainties are decorrelated among all combined regions.
\item For the other background processes, usually their yields are small in one of the combined channels, and therefore the background contributions are combined into a single contribution. The normalisation uncertainties of the other background processes are then uncorrelated between single-lepton and dilepton regions. In addition, the \Zjets normalisation uncertainty is decorrelated  among dilepton regions.
\end{itemize}
 
As shown in the bottom panels of Figure~\ref{fig:prefit_postfit1}, the marginalisation procedure significantly reduces the total uncertainty and results in better overall agreement between data and predictions.

\FloatBarrier
\subsubsection{Comparison with the Standard Model prediction}
 
The SM prediction of \Actt{} is obtained by performing fixed-order calculations at NNLO in the perturbative expansion in the strong coupling constant of QCD, with NLO EW corrections~\cite{Czakon:2017lgo}.
Fully differential predictions are available, including a detailed analysis of the charge asymmetry at the LHC~\cite{Czakon:2017wor,Czakon:2015owf}.
The same calculation is compared with Tevatron results for the forward--backward asymmetry~\cite{Czakon:2016ckf,Czakon:2014xsa}.

The \Actt{} predictions are calculated using $\mtop = 172.5$~\gev{}~\cite{Czakon:2017lgo}. The PDF of the proton is taken from the \text{LUXqed17} NNLO PDF set (LUXqed17\_plus\_PDF4LHC15\_nnlo\_100)~\cite{Czakon:2017lgo}. The central value and uncertainties are based on the PDF4LHC 2015 recommendation. The photon content of the proton is determined following Refs.~\cite{Manohar:2017eqh,Manohar:2016nzj}.
A dynamical renormalisation and factorisation scale~\cite{Czakon:2016dgf} is used, with the nominal value $\mu_0$ chosen as $H_\mathrm{T}/4$, where $H_{\mathrm{T}} = \sqrt{m_{t}^2 + p_{\mathrm{T},t}^2} + \sqrt{m_{\bar{t}}^2 + p_{\mathrm{T},\bar{t}}^2}$.
The scale uncertainty band indicates the maximum and the minimum value of the asymmetry obtained from an independent variation of the renormalisation and factorisation scales by factors of 2 or 0.5 around $\mu_0$. The variations in which one scale is multiplied by 2 while the other scale is divided by 2 are excluded. The MC integration uncertainty is typically smaller than the scale uncertainty. Finally, the scale and MC integration uncertainties are added in quadrature.
 
\begin{figure}[H]
\centering
\subfigure[][]{\includegraphics[width=0.49\textwidth]{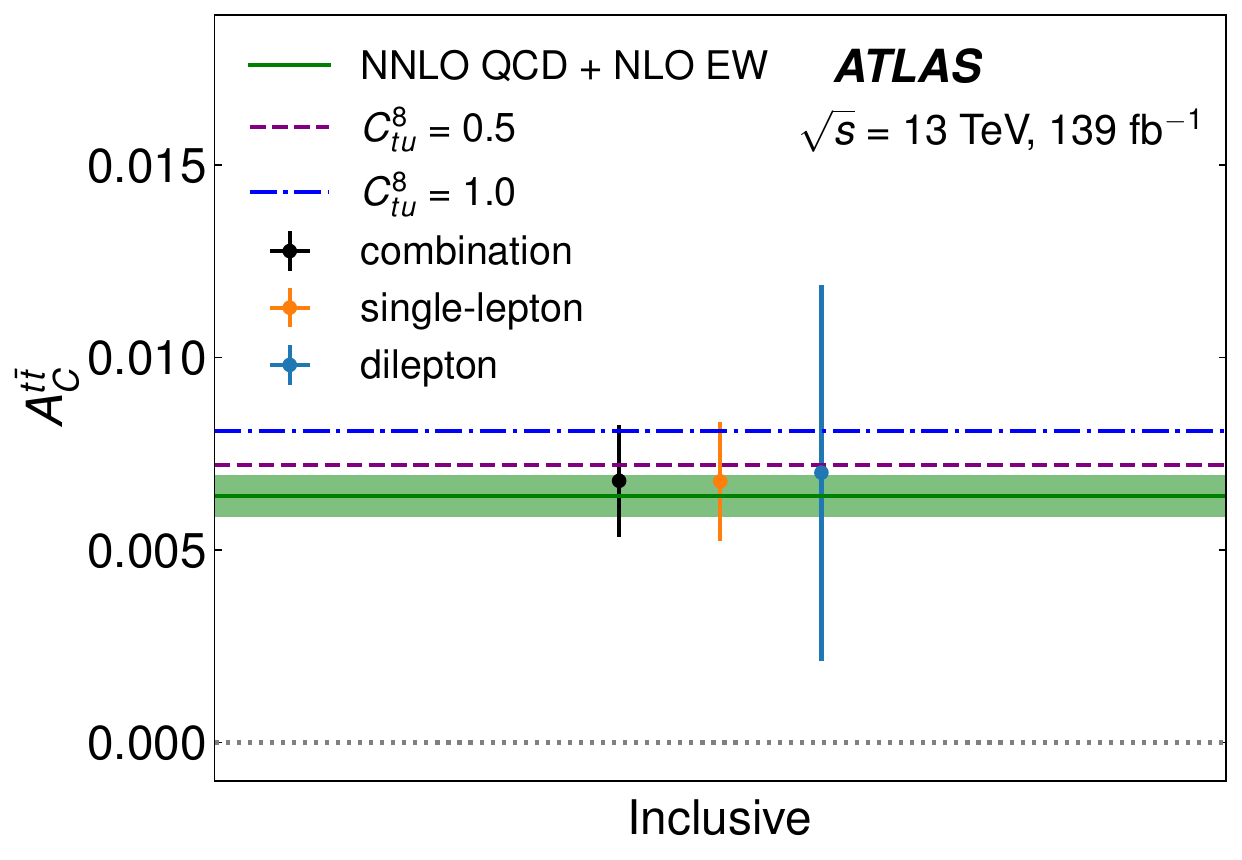}}
\subfigure[][]{\includegraphics[width=0.49\textwidth]{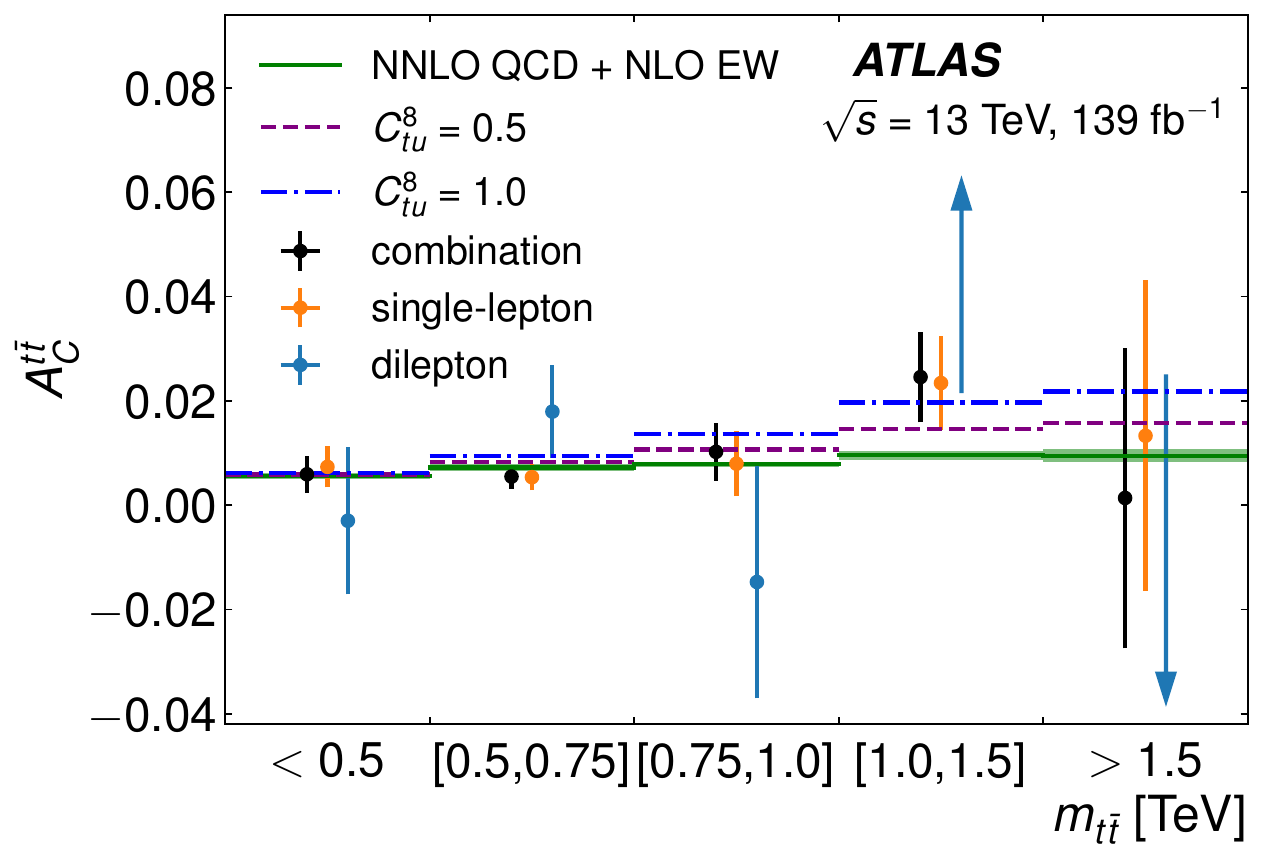}}
\caption{\label{fig:LJ_data1}
The unfolded inclusive (a) and differential charge asymmetries as a function of the invariant mass (b) of the reconstructed top-quark pair in data. Vertical bars correspond to the total uncertainties.
Shaded regions show SM theory predictions calculated at NNLO in QCD and NLO in EW theory.
The impact of the linear term of the $C_{tu}^{8}$ Wilson coefficient on the \Actt{} prediction for two different values is shown as dashed lines. In the dilepton channel, the two points corresponding to the higher invariant masses have large statistical uncertainty. A partial vertical uncertainty bar is shown, while the central point is in the direction of the associated arrow, with a value given in Table~\ref{tab:ljets_summary_data_stat_syst_list_incl_mtt}.}
\end{figure}

\begin{figure}[H]
\centering
\subfigure[][]{\includegraphics[width=0.49\textwidth]{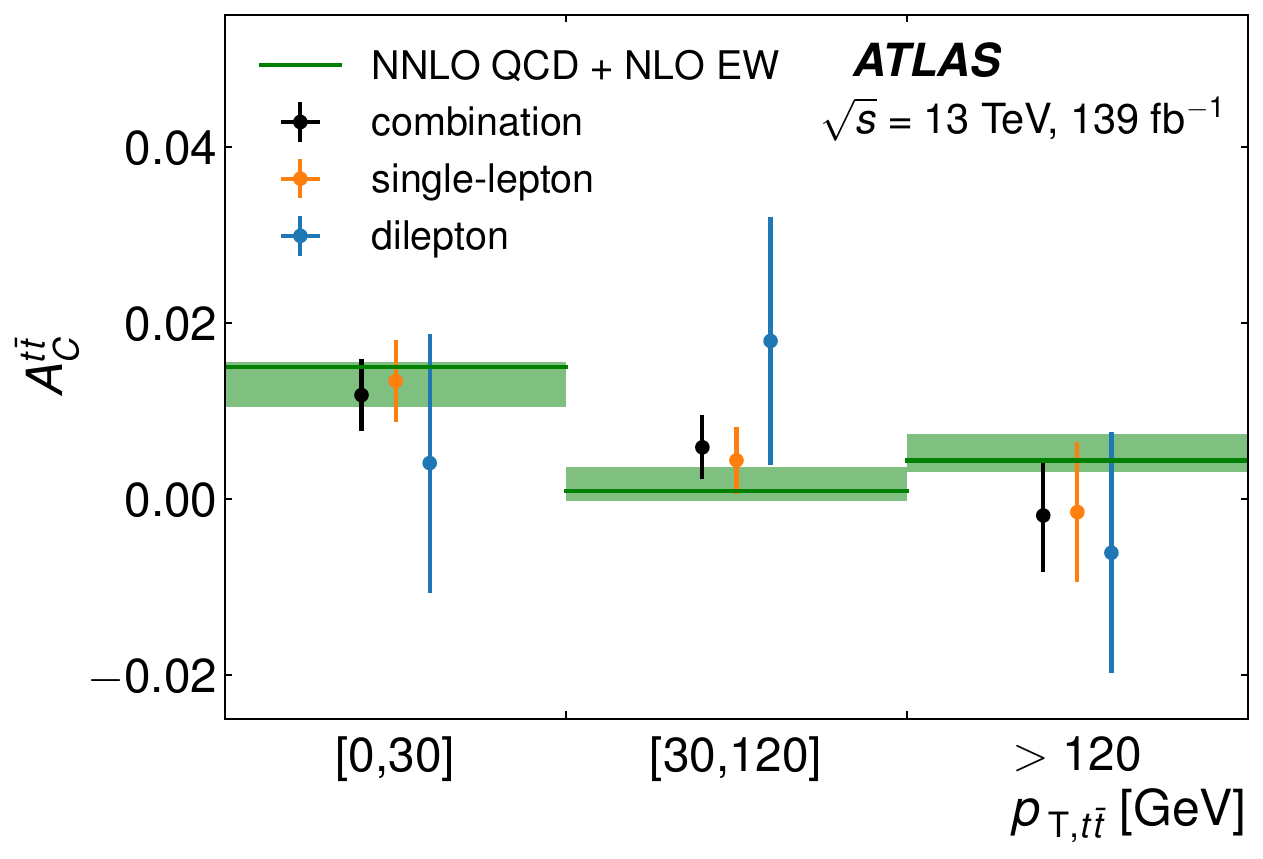}}
\subfigure[][]{\includegraphics[width=0.49\textwidth]{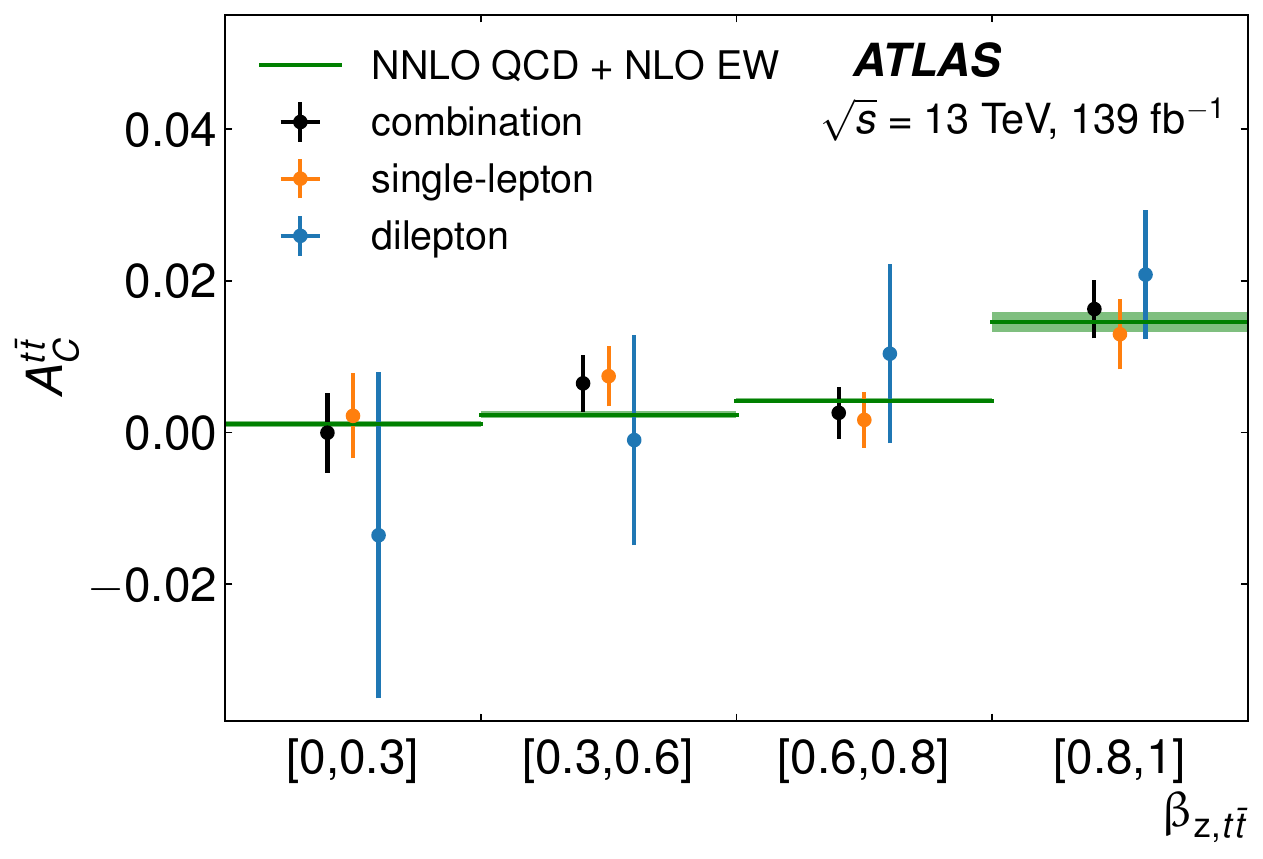}}
\caption{\label{fig:LJ_data2}
The unfolded differential charge asymmetries as a function of the transverse momentum (a)
and the longitudinal boost (b) of the reconstructed top-quark pair in data. Vertical bars correspond to the total uncertainties.
Shaded regions show SM theory predictions calculated at NNLO in QCD and NLO in EW theory.
}
\end{figure}

Tables~\ref{tab:ljets_summary_data_stat_syst_list_incl_mtt}--\ref{tab:ljets_summary_data_stat_syst_list_pttt_betatt} and Figures~\ref{fig:LJ_data1}--\ref{fig:LJ_data2} show a comparison of the measured \Actt asymmetry with the SM prediction. The measured asymmetry is consistent with the NNLO prediction in Ref.~\cite{Czakon:2017lgo} within the uncertainty. In addition, the impact of the linear term of the $C_{tu}^{8}$ Wilson coefficient on the \Actt{} prediction (see Section~\ref{sec_eft}) is shown for the inclusive and \mtt differential measurements.
 
\begin{table}[hbtp]
\caption{Results with statistical, systematic and the total uncertainties for the inclusive and \mtt differential \Actt measurements. The statistical uncertainty is obtained by unfolding the data, but excluding all the nuisance parameters. The squared systematic uncertainty is evaluated as the difference between the squares of the total uncertainty and the statistical uncertainty. The SM predictions are calculated at NNLO in QCD and NLO in EW theory.}
\label{tab:ljets_summary_data_stat_syst_list_incl_mtt}
\begin{center}
\setlength\extrarowheight{2.5pt}
\begin{tabular}{crlccccc}
& & & \multicolumn{4}{c}{Data \lumi} & \multirow{2}{*}{SM prediction} \\
& & Channel & \Actt{} & Stat. & Syst.  & Total unc. & \\
\hline
\multirow{3}{*}{Inclusive}
& & Single-lepton & 0.0068 & 0.0011 & 0.0011 & 0.0015 & \multirow{3}{*}{$0.0064^{+0.0005}_{-0.0006}$} \\
& & Dilepton      & 0.0070 & 0.0034 & 0.0035 & 0.0049 & \\
& & Combination   & 0.0068 & 0.0010 & 0.0010 & 0.0015 & \\
\hline
\multirow{15}{12pt}{\centering \mtt}
& \multirow{3}{*}{$<$ 500~\GeV}
& Single-lepton & 0.0074  &  0.0028 & 0.0028 & 0.0039 & \multirow{3}{*}{$0.0056^{+0.0006}_{-0.0006}$} \\
& & Dilepton    & $-0.0030$~~\,  &  0.0114 & 0.0084 & 0.0141 & \\
& & Combination & 0.0059  &  0.0027 & 0.0024 & 0.0036 & \\
\cline{2-8}
& \multirow{3}{*}{500--750~\GeV}
& Single-lepton & 0.0054  &  0.0020 & 0.0015 & 0.0025 & \multirow{3}{*}{$0.0072^{+0.0006}_{-0.0006}$} \\
& & Dilepton    & 0.0180  &  0.0061 & 0.0066 & 0.0089 & \\
& & Combination & 0.0055  &  0.0019 & 0.0013 & 0.0023 & \\
\cline{2-8}
& \multirow{3}{*}{750--1000~\GeV}
& Single-lepton & 0.0080  &  0.0048 & 0.0040 & 0.0062 & \multirow{3}{*}{$0.0079^{+0.0004}_{-0.0006}$} \\
& & Dilepton    & $-0.0147$~~\,  &  0.0188 & 0.0120 & 0.0223 & \\
& & Combination & 0.0102  &  0.0046 & 0.0030 & 0.0056 & \\
\cline{2-8}
& \multirow{3}{*}{1000--1500~\GeV}
& Single-lepton & 0.0234  &  0.0075 & 0.0050 & 0.0090 & \multirow{3}{*}{$0.0096^{+0.0009}_{-0.0009}$} \\
& & Dilepton    & 0.0663  &  0.0371 & 0.0244 & 0.0444 & \\
& & Combination & 0.0246  &  0.0074 & 0.0045 & 0.0087 & \\
\cline{2-8}
& \multirow{3}{*}{$>$ 1500~\GeV}
& Single-lepton & 0.0133  &  0.0288 & 0.0076 & 0.0298 & \multirow{3}{*}{$0.0094^{+0.0015}_{-0.0011}$}  \\
& & Dilepton    & $-0.1313$~~\,  &  0.1444 & 0.0590 & 0.1560 & \\
& & Combination & 0.0014  &  0.0280 & 0.0068 & 0.0288 & \\
\hline
\end{tabular}
\end{center}
\end{table}
 
\begin{table}[hbtp]
\caption{Results with statistical, systematic and the total uncertainties for the \pttt and \betatt differential \Actt measurements. The statistical uncertainty is obtained by unfolding the data, but excluding all the nuisance parameters. The squared systematic uncertainty is evaluated as the difference between the squares of the total uncertainty and the statistical uncertainty. The SM predictions are calculated at NNLO in QCD and NLO in EW theory.}
\label{tab:ljets_summary_data_stat_syst_list_pttt_betatt}
\begin{center}
\setlength\extrarowheight{2.5pt}
\begin{tabular}{crlccccc}
& & & \multicolumn{4}{c}{Data \lumi} & \multirow{2}{*}{SM prediction} \\
& & Channel & \Actt{} & Stat. & Syst.  & Total unc. & \\
\hline
\multirow{9}{12pt}{\centering \pttt}
& \multirow{3}{*}{$<$ 30~\GeV}
& Single-lepton & 0.0134  &  0.0034 & 0.0031 & 0.0046 & \multirow{3}{*}{$0.0150^{+0.0006}_{-0.0046}$} \\
& & Dilepton    & 0.0041  &  0.0105 & 0.0103 & 0.0147 & \\
& & Combination & 0.0118  &  0.0032 & 0.0025 & 0.0041 & \\
\cline{2-8}
& \multirow{3}{*}{30--120~\GeV}
& Single-lepton & 0.0044  &  0.0027 & 0.0027 & 0.0038 & \multirow{3}{*}{$0.0009^{+0.0028}_{-0.0012}$} \\
& & Dilepton    & 0.0179  &  0.0096 & 0.0103 & 0.0141 & \\
& & Combination & 0.0058  &  0.0026 & 0.0026 & 0.0037 & \\
\cline{2-8}
& \multirow{3}{*}{$>$ 120~\GeV}
& Single-lepton & $-0.0015$~~\,  &  0.0047 & 0.0064 & 0.0080 & \multirow{3}{*}{$0.0044^{+0.0030}_{-0.0014}$} \\
& & Dilepton    & $-0.0061$~~\,  &  0.0122 & 0.0062 & 0.0137 & \\
& & Combination & $-0.0019$~~\,  &  0.0044 & 0.0048 & 0.0065 & \\
\hline
\multirow{12}{12pt}{\centering \betatt}
& \multirow{3}{*}{0--0.3}
& Single-lepton & 0.0022  &  0.0040 & 0.0039 & 0.0056 & \multirow{3}{*}{$0.0011^{+0.0005}_{-0.0004}$} \\
& & Dilepton    & $-0.0135$~~\,  &  0.0179 & 0.0119 & 0.0215 & \\
& & Combination & ${<}0.0001$~~\,  &  0.0039 & 0.0035 & 0.0052 & \\
\cline{2-8}
& \multirow{3}{*}{0.3--0.6}
& Single-lepton & 0.0074  &  0.0031 & 0.0025 & 0.0040 & \multirow{3}{*}{$0.0023^{+0.0006}_{-0.0004}$} \\
& & Dilepton    & $-0.0010$~~\,  &  0.0116 & 0.0076 & 0.0139 & \\
& & Combination & 0.0065  &  0.0029 & 0.0023 & 0.0037 & \\
\cline{2-8}
& \multirow{3}{*}{0.6--0.8}
& Single-lepton & 0.0017  &  0.0028 & 0.0023 & 0.0037 & \multirow{3}{*}{$0.0042^{+0.0003}_{-0.0003}$} \\
& & Dilepton    & 0.0104  &  0.0090 & 0.0076 & 0.0118 & \\
& & Combination & 0.0026  &  0.0027 & 0.0021 & 0.0034 & \\
\cline{2-8}
& \multirow{3}{*}{0.8--1.0}
& Single-lepton & 0.0130  &  0.0026 & 0.0038 & 0.0046 & \multirow{3}{*}{$0.0146^{+0.0012}_{-0.0014}$} \\
& & Dilepton    & 0.0208  &  0.0069 & 0.0050 & 0.0085 & \\
& & Combination & 0.0163  &  0.0024 & 0.0030 & 0.0039 & \\
\end{tabular}
\end{center}
\end{table}
 
\FloatBarrier
\subsubsection{Uncertainties}
In order to determine the relative importance of systematic uncertainties, a ranking of individual nuisance parameters is performed. For each uncertainty, the unfolding is repeated with the corresponding NP  fixed to $+/-1\sigma$ around the NP post-marginalisation mean value. The difference between the nominal unfolded \Ac{} result and the result with the shifted NP is considered as the pre-marginalisation ranking estimate of the uncertainty. Similarly, a ranking where the NP is varied by $+/-c$, where $c$ is the post-marginalisation constraint, is performed and the post-marginalisation uncertainty ranking is obtained.
This approach hence takes into account post-marginalisation correlations between the ranked systematic uncertainty and other uncertainties.
The rankings of the leading systematic uncertainties are summarised in Tables~\ref{tab:Actt_inlc_ranking} and~\ref{tab:Actt_mtt_ranking} for the inclusive and \mtt differential \Actt combined measurements, respectively. Tables for the \pttt and \betatt differential measurements can be found in Appendix~\ref{app:comb_Actt_ranking}.
In the tables, the information about constraints can be obtained by comparing the pre-marginalisation and post-marginalisation impacts. The strongest constraints reduce the uncertainties to about 60\%--70\% of the pre-marginalisation value, with the exception of the third \mtt bin. In this bin, the constraint related to the \ttbar modelling impact reduces the pre-marginalisation uncertainty by 50\%.
A larger reduction in the uncertainty can be seen in Figure~\ref{fig:prefit_postfit1}, due to the inclusion of systematic uncertainties with mostly normalisation effect on \dYtt distribution, but with negligible effect on \Actt.
 
\begin{table}[H]
\caption{Impact of systematic uncertainties on the measurement of inclusive \Actt{}, grouped into categories. The impact of each category is obtained by summing the symmetrised ranking impact of the respective individual uncertainties in quadrature. The total uncertainty differs from the  sum-in-quadrature of the groups due to post-marginalisation nuisance parameter correlations in the unfolding.}
\label{tab:Actt_inlc_ranking}
\centering
\begin{tabular}{ll}
\toprule
& Post-marg. (pre-marg.) impact $\times 100$ \\
\midrule
$t\bar{t}$ modelling & 0.06 (0.08)\\
$t\bar{t}$ normalisation (flat prior) & 0.02\\
Background modelling & 0.04 (0.05)\\
Monte Carlo statistics & 0.05\\
Small-$R$ JES & 0.03 (0.03)\\
Small-$R$ JER & 0.03 (0.03)\\
Large-$R$ JES, JER & 0.01 (0.01)\\
Leptons, \MET & 0.02 (0.03)\\
$b$-tagging eff. & 0.01 (0.01)\\
Pile-up, JVT, luminosity & 0.01 (0.01)\\
\midrule
Statistical uncertainty & 0.10\\
\midrule
Total uncertainty & 0.15\\
\bottomrule
\end{tabular}
\end{table}
 
\begin{table}[H]
\caption{Impact of systematic uncertainties on the measurement of \Actt{} vs \mtt{}, grouped into categories. The impact of each category is obtained by summing the symmetrised ranking impact of the respective individual uncertainties in quadrature. The total uncertainty differs from the  sum-in-quadrature of the groups due to post-marginalisation nuisance parameter correlations in the unfolding.}
\label{tab:Actt_mtt_ranking}
\centering
\begin{tabular}{llllll}
\toprule
& \multicolumn{5}{c}{Post-marg. (pre-marg.) impact $\times 100$} \\
\cmidrule{2-6}
\mtt{} bin [\GeV] & $< 500$ & 500--750 & 750--1000 & 1000--1500 & $> 1500$ \\
\midrule
$t\bar{t}$ modelling & 0.15 (0.22) & 0.07 (0.10) & 0.14 (0.27) & 0.31 (0.45) & 0.64 (0.80)\\
$t\bar{t}$ normalisation (flat prior) & 0.06 & 0.02 & 0.08 & 0.06 & 0.18\\
Background modelling & 0.09 (0.11) & 0.06 (0.08) & 0.13 (0.15) & 0.22 (0.28) & 0.84 (1.02)\\
Monte Carlo statistics & 0.13 & 0.08 & 0.20 & 0.30 & 1.02\\
Small-$R$ JES & 0.09 (0.10) & 0.05 (0.06) & 0.13 (0.15) & 0.15 (0.17) & 0.43 (0.52)\\
Small-$R$ JER & 0.11 (0.16) & 0.03 (0.05) & 0.09 (0.15) & 0.21 (0.29) & 0.42 (0.65)\\
Large-$R$ JES, JER & 0.02 (0.02) & 0.02 (0.02) & 0.02 (0.03) & 0.05 (0.05) & 0.19 (0.28)\\
Leptons, \MET & 0.04 (0.06) & 0.03 (0.04) & 0.08 (0.10) & 0.06 (0.08) & 0.26 (0.39)\\
$b$-tagging eff. & 0.02 (0.02) & 0.01 (0.01) & 0.03 (0.03) & 0.03 (0.04) & 0.27 (0.30)\\
Pile-up, JVT, luminosity & 0.01 (0.01) & 0.01 (0.01) & 0.01 (0.01) & 0.04 (0.04) & 0.02 (0.02)\\
\midrule
Statistical uncertainty & 0.27 & 0.19 & 0.46 & 0.74 & 2.80\\
\midrule
Total uncertainty & 0.36 & 0.23 & 0.56 & 0.87 & 2.88\\
\bottomrule
\end{tabular}
\end{table}
 
\FloatBarrier
\subsection{\Acll{} measurement}
 
Similarly to \Actt measurements, the measured \Acll{} is inferred from the posterior distribution of \detall{} according to Eq.~(\ref{intro:eq:acll}), where the \detall{} posterior is obtained by marginalising the extended likelihood in Eq.~(\ref{eq:results:like}). The \Acll value is measured in the dilepton channel only, combining four  dilepton regions. The \detall distribution is split into four bins in all regions and in each differential bin of all differential measurements.
The \mll, \ptll and \betall differential bins are chosen in a way that compromises between data-event counts and event migration due to poor reconstruction. For the systematic uncertainties, the same correlation scheme as in the dilepton channel \Actt measurement is used.
 
\begin{figure}[htb]
\centering
\vspace{-0.2cm}
\subfigure[][]{\includegraphics[width=0.48\textwidth]{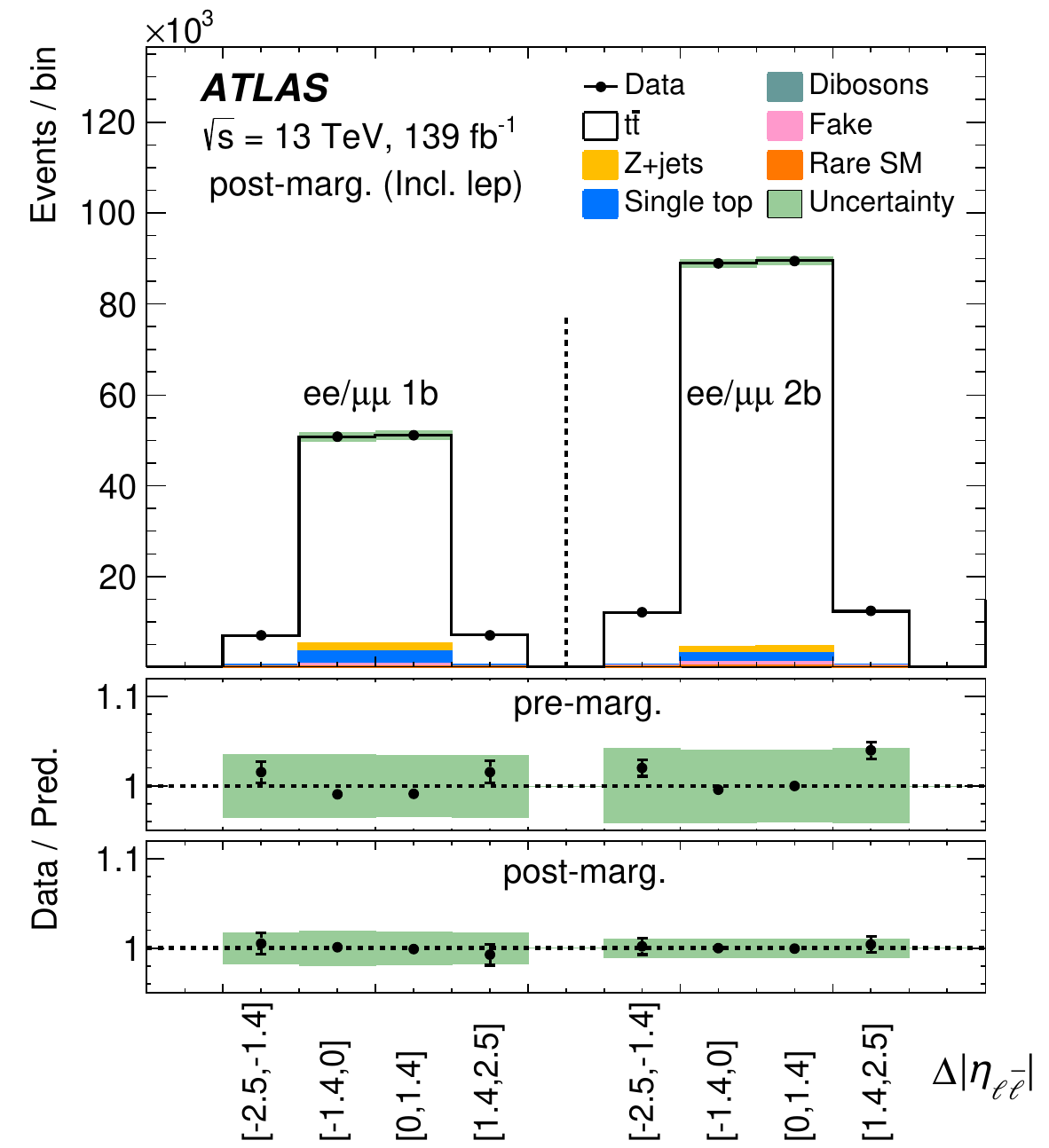}}
\subfigure[][]{\includegraphics[width=0.48\textwidth]{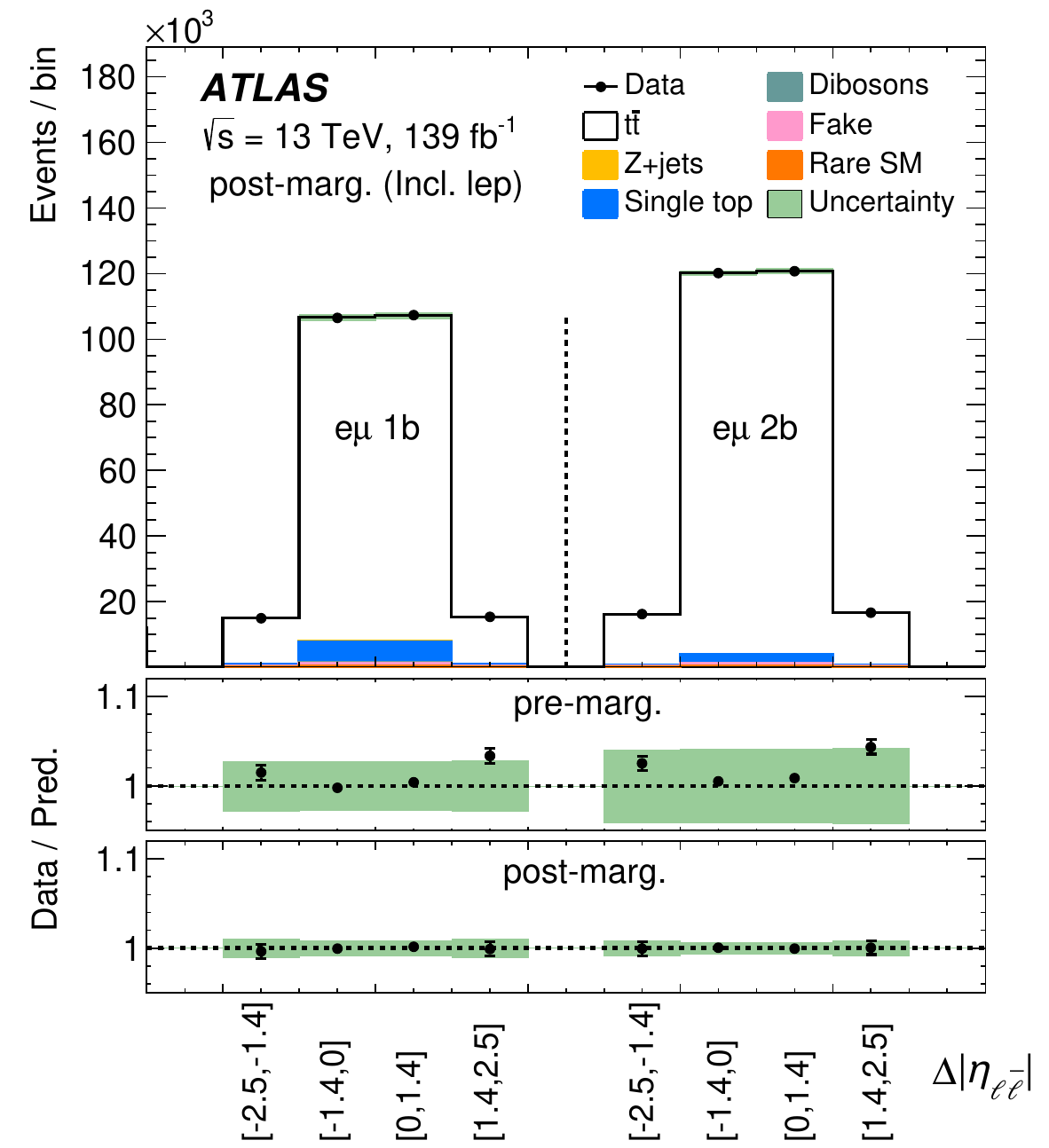}}
\vspace{-0.1cm}
\caption{
Comparison between the data and the prediction for the reconstructed \detall in the inclusive \Acll measurement in the (a) dilepton $ee+\mu\mu$ and (b) dilepton $e\mu$ region. The ratio of data to prediction  is shown before and after marginalisation within FBU. Shaded bands correspond to the total uncertainty of the prediction. The signal modelling uncertainties do not include normalisation effects because the overall acceptance difference cancels out in the \Acll observable.
}
\label{fig:prefit_postfit_ll_1}
\end{figure}

As shown in Figure \ref{fig:prefit_postfit_ll_1}, the marginalisation procedure reduces the total uncertainty significantly and results in better agreement between the data and the predictions.

\FloatBarrier
\subsubsection{Comparison with the Standard Model prediction}
The SM prediction of \Acll{} has been calculated at NLO in SM gauge couplings, including corrections up to order $\alphas^3$ in the QCD coupling and the mixed QCD--QED and mixed QCD--weak-interaction corrections~\cite{Bern:reuther2012sx}.
Fully differential predictions are available, including a detailed analysis of the charge asymmetry at the LHC~\cite{Bern:reuther2012sx}.
 
The \Acll{} predictions are calculated using $\mtop = 173.34$~\gev{}, $\alpha_{\text{QED}}(m_Z) = 0.008$ and the weak mixing angle $\sin^2 \theta_\text{W} = 0.23$~\cite{Bern:reuther2012sx}. The PDF of the proton is taken from the \CT[10] set~\cite{Lai:2010vv}.
The same value of $\mu=\mtop$ is used for both the renormalisation scale and factorisation scale.
The scale uncertainty band indicates the minimum and the maximum value of the asymmetry obtained by varying both scales by a factor of 0.5 or 2.0 respectively.
 
\begin{figure}[H]
\centering
\subfigure[][]{\includegraphics[width=0.48\textwidth]{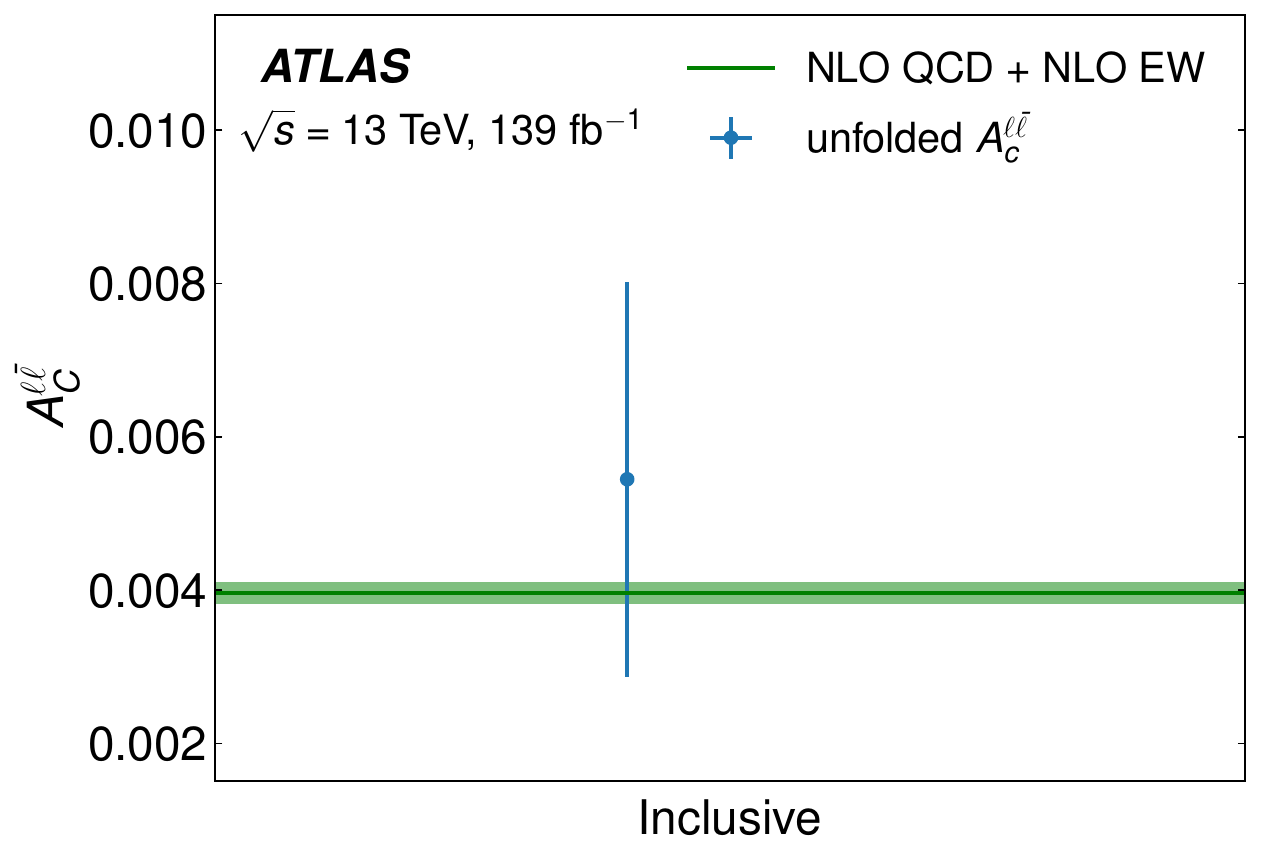}}
\subfigure[][]{\includegraphics[width=0.48\textwidth]{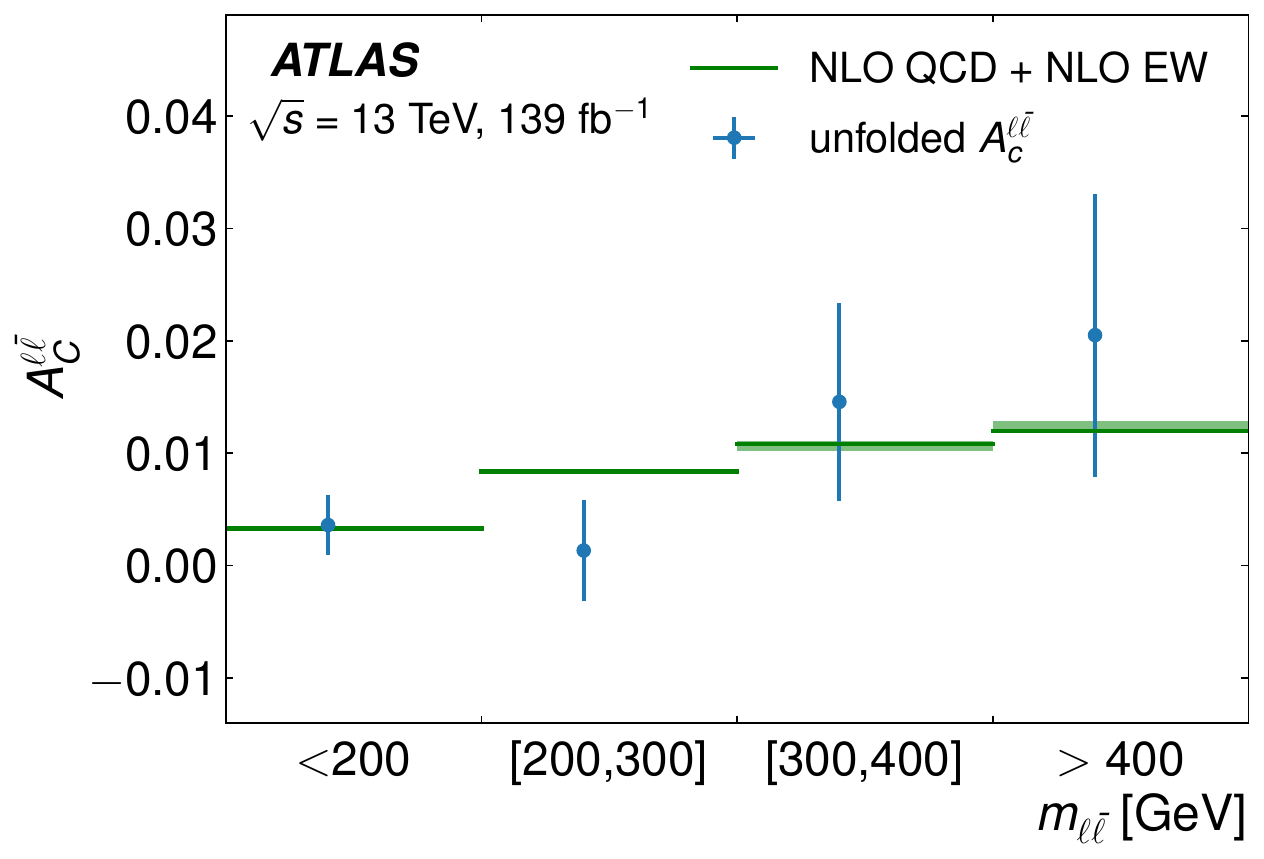}}\\
\subfigure[][]{\includegraphics[width=0.48\textwidth]{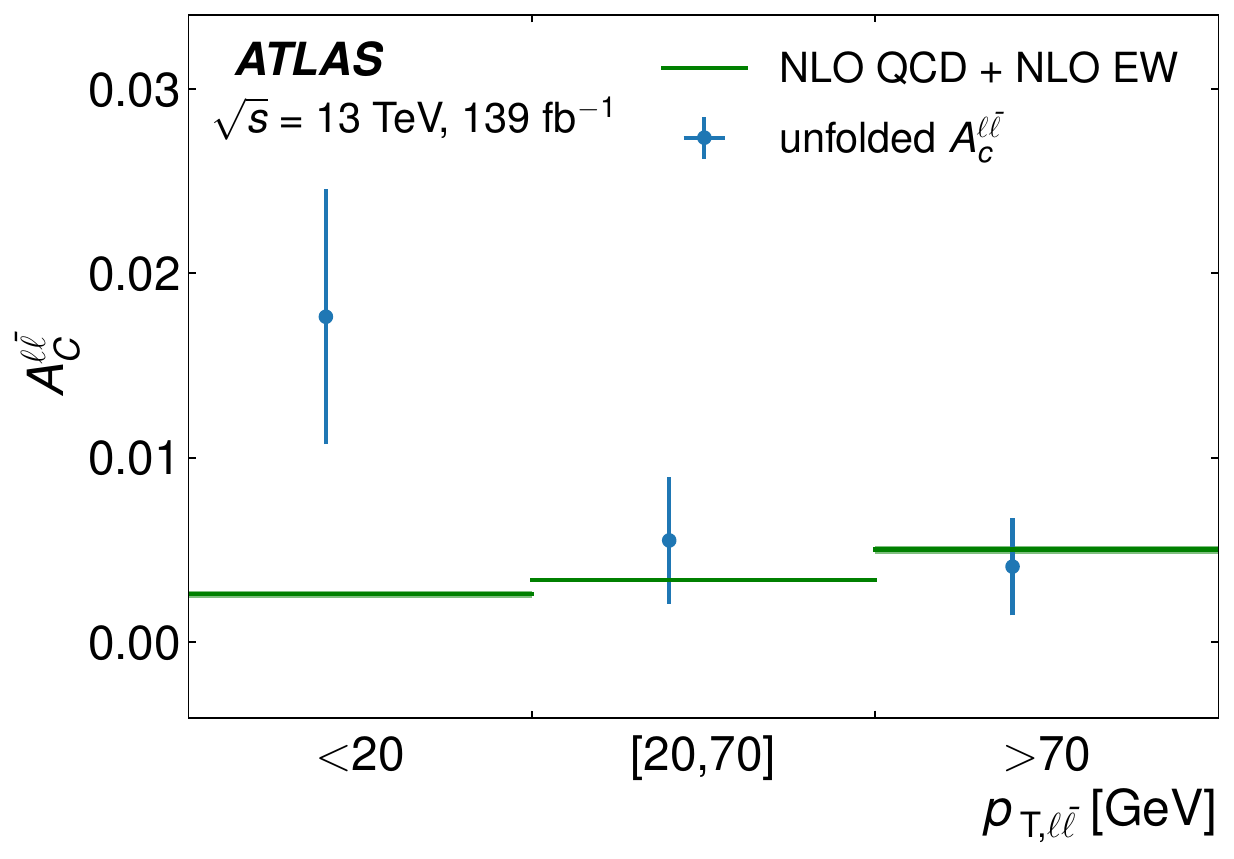}}
\subfigure[][]{\includegraphics[width=0.48\textwidth]{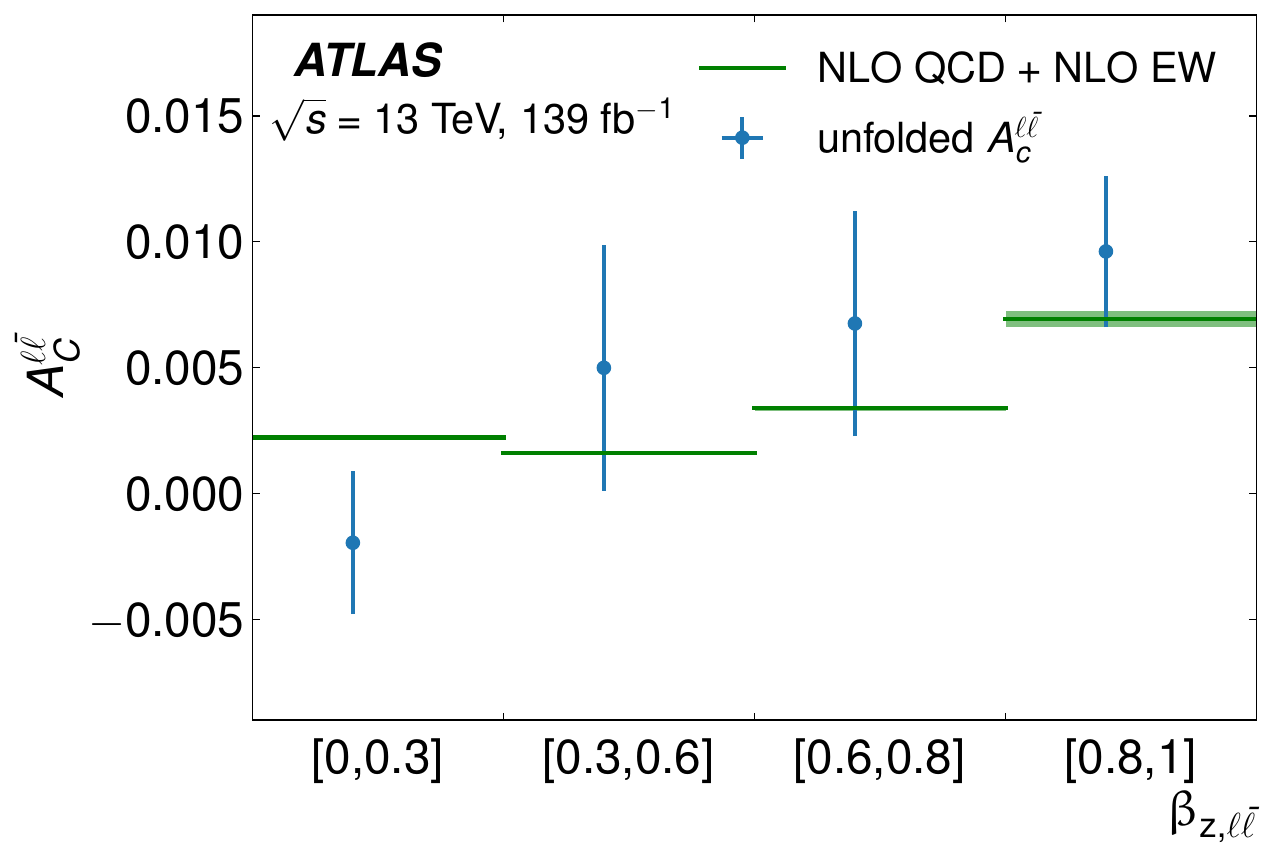}}
\caption{\label{fig:Acll_data}
The unfolded inclusive (a) and differential leptonic asymmetries as a function of the invariant mass (b), transverse momentum (c),
and the longitudinal boost (d) of the reconstructed lepton pair in dilepton channel data.
Shaded regions show SM theory predictions calculated at NLO in QCD and NLO in EW theory.
Vertical bars correspond to the total uncertainties.}
\end{figure}

Figure~\ref{fig:Acll_data} and Table~\ref{tab:Acll_summary_data_stat_syst_list} show a comparison of the measured \Acll asymmetry with the SM prediction. The measured asymmetry is consistent with NLO calculations~\cite{Bern:reuther2012sx}.
 
\begin{table}[hbtp]
\caption{Results of the inclusive and differential \Acll measurements, including statistical, systematic and total uncertainties. The statistical uncertainty is obtained by unfolding the data, excluding all the nuisance parameters. The squared systematic uncertainty is evaluated as the difference between the squares of the total uncertainty and the statistical uncertainty.  The SM predictions are calculated at NLO in QCD and NLO in EW theory.}
\label{tab:Acll_summary_data_stat_syst_list}
\begin{center}
\setlength\extrarowheight{2.5pt}
\begin{tabular}{crccccc}
\toprule
& & \multicolumn{4}{c}{Data \lumi} & \multirow{2}{*}{SM prediction} \\
& & \Acll{} & Stat. & Syst.  & Total unc. & \\
\midrule
Inclusive & &     0.0054  &  0.0012 & 0.0023 & 0.0026 & $0.0040^{+0.0002}_{-0.0001}$ \\
\midrule
\multirow{4}{12pt}{\centering \mll}
& $<$ 200~\GeV &   0.0036  &  0.0014 & 0.0023 & 0.0027 & $0.0033^{+0.0001}_{-0.0001}$ \\
& 200--300~\GeV &   0.0013  &  0.0041 & 0.0020 & 0.0046 & $0.0084^{+0.0002}_{-0.0001}$ \\
& 300--400~\GeV &   0.0146  &  0.0085 & 0.0030 & 0.0090 & $0.0108^{+0.0003}_{-0.0006}$ \\
& $>$ 400~\GeV &   0.0205  &  0.0123 & 0.0036 & 0.0128 & $0.0120^{+0.0009}_{-0.0002}$ \\
\midrule
\multirow{3}{12pt}{\centering \ptll}
& $<$ 20~\GeV  &   0.0176  &  0.0063 & 0.0031 & 0.0070 & $0.0026^{+0.0002}_{-0.0002}$ \\
& 20--70~\GeV   &   0.0055  &  0.0020 & 0.0028 & 0.0034 & $0.0034^{+0.0001}_{<0.0001}$ \\
& $>$ 70~\GeV  &   0.0041  &  0.0018 & 0.0020 & 0.0027 & $0.0050^{+0.0002}_{-0.0003}$ \\
\midrule
\multirow{4}{12pt}{\centering \betall}
& 0--0.3     &  $-0.0020$~~\,  &  0.0028 & 0.0008 & 0.0029 & $0.0022^{+0.0001}_{-0.0001}$ \\
& 0.3--0.6   &   0.0050  &  0.0024 & 0.0043 & 0.0049 & $0.0016^{+0.0001}_{-0.0001}$ \\
& 0.6--0.8   &   0.0068  &  0.0026 & 0.0037 & 0.0045 & $0.0034^{<0.0001}_{-0.0001}$ \\
& 0.8--1.0   &   0.0096  &  0.0028 & 0.0014 & 0.0031 & $0.0069^{+0.0003}_{-0.0003}$ \\
\bottomrule
\end{tabular}
\end{center}
\end{table}

\FloatBarrier
\subsubsection{Uncertainties}
 
The rankings of the leading systematic uncertainties are summarised in Tables~\ref{tab:Acll_inlc_ranking} and~\ref{tab:Acll_mll_ranking} for the inclusive and \mll differential \Acll measurements, respectively.  Tables for the \ptll and \betall differential measurements can be found in Appendix~\ref{app:Acll_ranking}.
In the tables, the information about constraints can be obtained by comparing the pre-marginalisation and post-marginalisation impacts. The strongest constraints  reduce the uncertainties to about 80\%--90\% of the pre-marginalisation value (e.g.\ for \ttbar signal modelling). For some systematic uncertainties with smaller impacts (e.g.\ $b$-tagging efficiency in the \mll differential measurement), constraints reducing the uncertainties to 65\% of the pre-marginalisation values can be found.
A larger reduction in the uncertainty can be seen in Figure~\ref{fig:prefit_postfit_ll_1}, due to the inclusion of systematic uncertainties with mostly normalisation effect on \detall distribution, but with negligible effect on \Acll.
 
\begin{table}[H]
\caption{Impact of systematic uncertainties on the measurement of inclusive \Acll{}, grouped into categories. The impact of each category is obtained by summing the symmetrised ranking impact of the respective individual uncertainties in quadrature. The total uncertainty differs from the  sum-in-quadrature of the groups due to post-marginalisation nuisance parameter correlations in the unfolding.}
\label{tab:Acll_inlc_ranking}
\sisetup{input-symbols = ()}
\centering
\begin{tabular}{ll}
\toprule
& Post-marg. (pre-marg.) impact $\times 100$ \\
\midrule
$t\bar{t}$ modelling & ~~\,0.20 ~~~(0.22)\\
$t\bar{t}$ normalisation (flat prior) & ~~\,0.06\\
Background modelling & ~~\,0.02 ~~~(0.02)\\
Monte Carlo statistics & ~~\,0.04\\
Small-$R$ JES & ~~\,0.03 ~~~(0.03)\\
Small-$R$ JER & ~~\,0.06 ~~~(0.06)\\
Leptons, \MET & ~~\,0.01 ~~~(0.01)\\
$b$-tagging eff. & ${<}0.01$ (${<}0.01$)\\
Pile-up, JVT, luminosity & ~~\,0.01 ~~~(0.01)\\
\midrule
Statistical uncertainty & ~~\,0.12\\
\midrule
Total uncertainty & ~~\,0.26\\
\bottomrule
\end{tabular}
\end{table}
 
\begin{table}[H]
\caption{Impact of systematic uncertainties on the measurement of \Acll{} vs \mll{}, grouped into categories. The impact of each category is obtained by summing the symmetrised ranking impact of the respective individual uncertainties in quadrature. The total uncertainty differs from the  sum-in-quadrature of the groups due to post-marginalisation nuisance parameter correlations in the unfolding.}
\label{tab:Acll_mll_ranking}
\centering
\begin{tabular}{lllll}
\toprule
& \multicolumn{4}{c}{Post-marg. (pre-marg.) impact $\times 100$} \\
\cmidrule{2-5}
\mll{} bin [\GeV] & $< 200$ & 200--300 & 300--400 & $> 400$ \\
\midrule
$t\bar{t}$ modelling & 0.18 (0.21) & 0.17 (0.21) & 0.20 (0.23) & 0.15 (0.15)\\
$t\bar{t}$ normalisation (flat prior) & 0.03 & 0.03 & 0.03 & 0.01\\
Background modelling & 0.02 (0.02) & 0.05 (0.06) & 0.06 (0.07) & 0.06 (0.06)\\
Monte Carlo statistics & 0.05 & 0.09 & 0.19 & 0.26\\
Small-$R$ JES & 0.04 (0.05) & 0.02 (0.02) & 0.03 (0.03) & 0.08 (0.09)\\
Small-$R$ JER & 0.03 (0.03) & 0.05 (0.05) & 0.07 (0.07) & 0.07 (0.06)\\
Leptons, \MET & 0.02 (0.02) & 0.03 (0.02) & 0.06 (0.06) & 0.09 (0.08)\\
$b$-tagging eff. & 0.01 (0.01) & 0.02 (0.02) & 0.02 (0.02) & 0.02 (0.03)\\
Pile-up, JVT, luminosity & 0.01 (0.01) & 0.01 (0.01) & 0.02 (0.01) & 0.01 (0.02)\\
\midrule
Statistical uncertainty & 0.14 & 0.41 & 0.85 & 1.23\\
\midrule
Total uncertainty & 0.27 & 0.46 & 0.90 & 1.28\\
\bottomrule
\end{tabular}
\end{table}


\FloatBarrier
\subsection{SMEFT interpretation of the \Actt{} results}
\label{sec_eft}

\FloatBarrier
 
The combined results of the \Actt{} measurement are interpreted in the SMEFT.
In this formalism the SM Lagrangian is extended with higher-dimensional operators that encode the effect of new-physics phenomena at a scale beyond the direct reach of the experiment. A general effective Lagrangian is written as follows:
\begin{equation*}
\mathcal{L}_\text{eff} = \mathcal{L}_\text{SM} + \frac{1}{\Lambda^2} \sum\limits_i C_i O_i + \mathcal{O}\left(\Lambda^{-4} \right),
\end{equation*}
where $\mathcal{L}_\text{SM}$ represents the SM Lagrangian, $\Lambda$ represents the characteristic new-physics scale and $C_i$ are the Wilson coefficients of the dimension-six operators $O_i$. Any observable receives contributions proportional to $\Lambda^{-2}$, from the interference of dimension-six operators with SM amplitudes. The squares of dimension-six operators are also computed, and these give rise to contributions of order $\Lambda^{-4}$. However, further terms of the same order, from the interference of dimension-eight operators with the SM, are ignored. For this expansion to provide meaningful results, $C_i/\Lambda^2$ must be small. For typical values of the coupling strength $C_i$ of the dimension-six operators, that implies that $\Lambda$ must exceed several~\tev. An indication of the validity of the expansion is provided by comparing the results obtained from a \textit{linear} parameterisation that includes only $\Lambda^{-2}$ terms with the results of a \textit{quadratic} fit that also takes into account the known $\Lambda^{-4}$ terms from dimension-six operators squared.
 
In the following, the left-handed (L) quark doublets of the first two generations are represented by $q_i = (u_\text{L},d_\text{L}), (c_\text{L},s_\text{L})$, and the doublet of the third generation is given by $Q = (t_\text{L},b_\text{L})$. Right-handed (R) quarks of the first two generations are denoted by $u_i$, $d_i$ and the right-handed top quark by $t$.
 
Using this notation, our interpretation considers 15 dimension-six operators in the Warsaw basis~\cite{Grzadkowski:2010es, AguilarSaavedra:2018nen}:\\
\begin{itemize}
\item There are eight $q\bar{q}t\bar{t}$ operators with LL and RR chiral structures:
\end{itemize}
\begin{align}\nonumber
O_{Qq}^{1,8} & = (\bar{Q}\gamma_\mu T^A Q)(\bar{q}_i\gamma^\mu T^A q_i),  & O_{Qq}^{1,1} & = (\bar{Q}\gamma_\mu Q)(\bar{q}_i\gamma^\mu q_i), \\\nonumber
O_{Qq}^{3,8} & = (\bar{Q}\gamma_\mu T^A\tau^I Q)(\bar{q}_i\gamma^\mu T^A\tau^I q_i), & O_{Qq}^{3,1} & = (\bar{Q}\gamma_\mu \tau^I Q)(\bar{q}_i\gamma^\mu \tau^I q_i),\\\nonumber
O_{tu}^{8} & = (\bar{t}\gamma_\mu T^A t)(\bar{u}_i\gamma^\mu T^A u_i) & O_{tu}^{1} & = (\bar{t}\gamma_\mu t)(\bar{u}_i\gamma^\mu u_i)\\\nonumber
O_{td}^{8} & = (\bar{t}\gamma^\mu T^A t) (\bar{d}_i\gamma_\mu T^A d_i) & O_{td}^{1} & = (\bar{t}\gamma_\mu t)(\bar{d}_i\gamma^\mu d_i)\,.
\end{align}
\begin{itemize}
\item There are six further $q\bar{q}t\bar{t}$ operators with LR structures:
\end{itemize}
\begin{align}\nonumber
O_{Qu}^{8} & = (\bar{Q}\gamma_\mu T^A Q)(\bar{u}_i\gamma^\mu T^A u_i) & O_{Qu}^{1} & = (\bar{Q}\gamma_\mu Q)(\bar{u}_i\gamma^\mu u_i)\\\nonumber
O_{Qd}^{8} & = (\bar{Q}\gamma_\mu T^A Q)(\bar{d}_i\gamma^\mu T^A d_i) & O_{Qd}^{1} & = (\bar{Q}\gamma_\mu Q)(\bar{d}_i\gamma^\mu d_i)\\\nonumber
O_{tq}^{8} & = (\bar{t}\gamma^\mu T^A t)(\bar{q}_i\gamma_\mu T^A q_i) & O_{tq}^{1} & = (\bar{t}\gamma^\mu t)(\bar{q}_i\gamma_\mu q_i)\,.
\end{align}
\begin{itemize}
\item There is one tensor operator that modifies the top--gluon interaction:
\begin{align}\nonumber
O_{tG} = (\bar{t} \sigma^{\mu\nu} T^A t) \tilde{\varphi} G^A_{\mu\nu}.
\end{align}
\end{itemize}
 
The indices indicate the strong and weak structures:  colour-singlet operators are indicated with $(1)$ and colour-octet operators with $(8)$, weak-singlet operators with (1) and weak-triplet operators with $(3)$. Also, $T^A \equiv \lambda^A/2$ where $\lambda^A$ are Gell-Mann matrices; $\tau^I$ represents Pauli matrices, $\tilde{\varphi}$ represents the Higgs doublet $\varphi$ with the antisymmetric SU(2) tensor by $\epsilon = i \tau^2$ and $\tilde{\varphi} = \epsilon \varphi^*$. The CP-violating imaginary parts of operator coefficients are not considered.
 
The charge asymmetry is affected by a particular linear combination of operator coefficients~\cite{Brivio:2019ius,Rosello:2015sck,Zhang:2010dr}. This measurement is therefore complementary to cross-section measurements and very valuable in disentangling blind directions in global SMEFT fits~\cite{Brivio:2019ius}.

The operators are implemented in the SMEFT@NLO UFO model~\cite{Degrande:2020evl} in the \MGNLO package~\cite{Alwall:2014hca}. The relation between the operator coefficients and the charge asymmetry is determined at NLO accuracy in the strong coupling constant and is parameterised with an analytic function~\cite{Ethier:2021bye}\footnote{We thank E. Vryonidou and C. Zhang for providing us the NLO predictions in the SMEFT, and for the parameterisation of the dependence of the differential charge asymmetry measurement on the Wilson coefficients of the SMEFT that form the basis of the EFT interpretation.}. If the `positive' and `negative' cross-sections are defined as $\sigma^{+}=\sigma(\dYtt>0)$ and $\sigma^{-}=\sigma(\dYtt<0)$, respectively, they can be expressed as a function of the relevant operator coefficients:
\begin{equation*}
\sigma^{\pm} = \sigma^{\pm}_\text{SM} + \sum\limits_i \sigma^{\pm}_{1,i} \times C_i/\Lambda^2 + \sum\limits_i \sigma^{\pm}_{2,i} \times C^2_i/\Lambda^4 + \sum\limits_{i,j} \sigma^{\pm}_{3,i,j} \times C_i C_j/\Lambda^4
\end{equation*}
Fits with two different parameterisations are performed: one that retains only the \textit{linear} terms proportional to $\Lambda^{-2}$ due to the interference of the dimension-six operators with the SM, and a second \textit{quadratic} parameterisation that also retains the dimension-six-squared terms proportional to $\Lambda^{-4}$. In the quadratic fit the cross-terms $C_i C_j$ involving two different operator coefficients are taken into account. The $\sigma^{\pm}_\text{SM}$ term in the parameterisation -- determined at NLO accuracy with SMEFT@NLO~\cite{Degrande:2020evl} -- is replaced by the NNLO prediction from Ref.~\cite{Czakon:2017lgo}.
 
\begin{figure}
\centering
\includegraphics[width=0.8\textwidth]{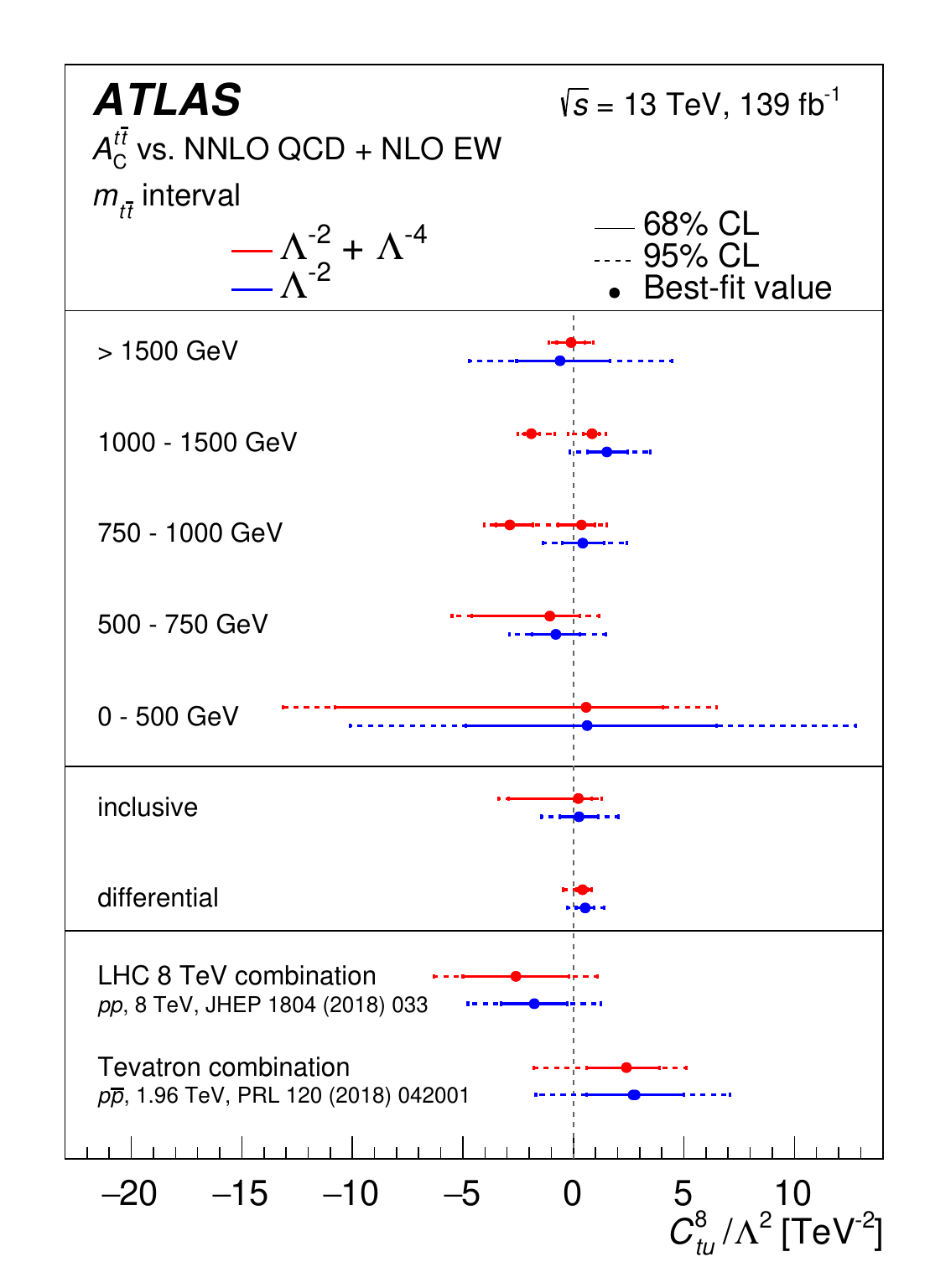}
\caption{Individual 68\% (solid line) and 95\% (dashed line) CL limits on the Wilson coefficient $C_{tu}^{8}/ \Lambda^2$ in the SMEFT. The bounds are derived from the charge asymmetry measurement presented in this paper, combining the single-lepton and dilepton channels. For the quadratic fit, two solutions were found by the $\chi^2$ minimisation in some cases. The theory uncertainty corresponds to the NNLO QCD + NLO EW calculation.
The impact of dimension-six operators is parameterised at NLO accuracy in QCD using the SMEFT@NLO package in \MGNLO. Bounds are also shown from the forward--backward asymmetry measurements in $\sqrt{s}= 1.96$~\TeV{} $p\bar{p}$ collisions at the Tevatron and the charge asymmetry measurements in 8~\TeV{} $pp$ collisions in LHC Run~1.}
\label{fig:combined_constraints_LHC_Tevatron_nlo}
\end{figure}

Limits on individual Wilson coefficients are derived by $\chi^2$ minimisation. Just one Wilson coefficient is varied, while the others are fixed to the SM prediction, i.e.\ zero.
As an example, the 68\% and 95\% CL bounds on $C_{tu}^{8}/\Lambda^2$ from the \Actt{} measurement are shown in Figure~\ref{fig:combined_constraints_LHC_Tevatron_nlo}. The different points correspond to the inclusive \Actt{} measurement, the individual bins of the differential \Actt{} measurement as a function of \mtt{}, and the combined fit of the bins of the \Actt{} vs \mtt{} measurement. Limits derived from the LHC combination of charge asymmetry measurements at $\sqrt{s}=$ 8~\tev{} and from the Tevatron combination of forward--backward asymmetry measurements in 1.96~\tev{} $p\bar{p}$ collisions are shown in the lower panel, for reference.
 
The nominal result, obtained with a linear parameterisation of the $C_{tu}^{8}/\Lambda^{2}$ dependence, is shown in blue. The result of the fit that also takes into account the $\Lambda^{-4}$ term is shown in red. The central values returned by the fits are quite similar, but in many cases the 95\% CL intervals differ considerably and $\Lambda^{-4}$ terms are not negligible. The inclusive measurement, with an uncertainty of 0.15\%, yields tight bounds: $-1.44 < C_{tu}^{8}/\Lambda^2 < 2.05~\TeV^{-2}$ at 95\% CL in the linear fit. Despite the large dilution by gluon-initiated $t\bar{t}$ production at the LHC at $\sqrt{s}=$ 13~\tev, these bounds based on a single measurement improve considerably on the limits derived from the LHC 8~\TeV{} combination~\cite{TOPQ-2016-16-updated} and from the Tevatron combination~\cite{CDF:2017cvy}.
 
The bounds from the differential measurements reflect the interplay between the sensitivity, which increases rapidly at higher $m_{t\bar{t}}$, and the uncertainty, which grows from 0.2\%--0.3\% in the lowest mass bin to 2.9\% in the highest bin. For the linear fit, the tightest limit is obtained in the mass bin from 1 to 1.5~\TeV.
 
The result labelled `differential' combines all results of the differential measurement versus $m_{\ttbar}$ in a $\chi^2$ minimisation. The covariance matrix that keeps track of bin-to-bin correlations in the differential result includes experimental uncertainties from the measurement and the uncertainties from the fixed-order prediction. The experimental covariance matrix includes the uncertainties from experimental and modelling uncertainties in the measurement, as returned by the unfolding procedure. The covariance matrix for the SM prediction includes the effect of different choices for the renormalisation and factorisation scales and of the difference between the LUXqed and NNPDF3.1 PDF sets. The covariance matrices due to experimental and theoretical uncertainties are added together. The differential analysis yields a tight set of 95\% CL bounds. For example, the linear 95\% CL bound on $C_{tu}^{8}/\Lambda^2$ improves from [$-1.44$, 2.05]~\TeV$^{-2}$ in the inclusive measurement to [$-0.28$, 1.38]~\TeV$^{-2}$.
 
\begin{figure}
\centering
\includegraphics[width=0.7\textwidth]{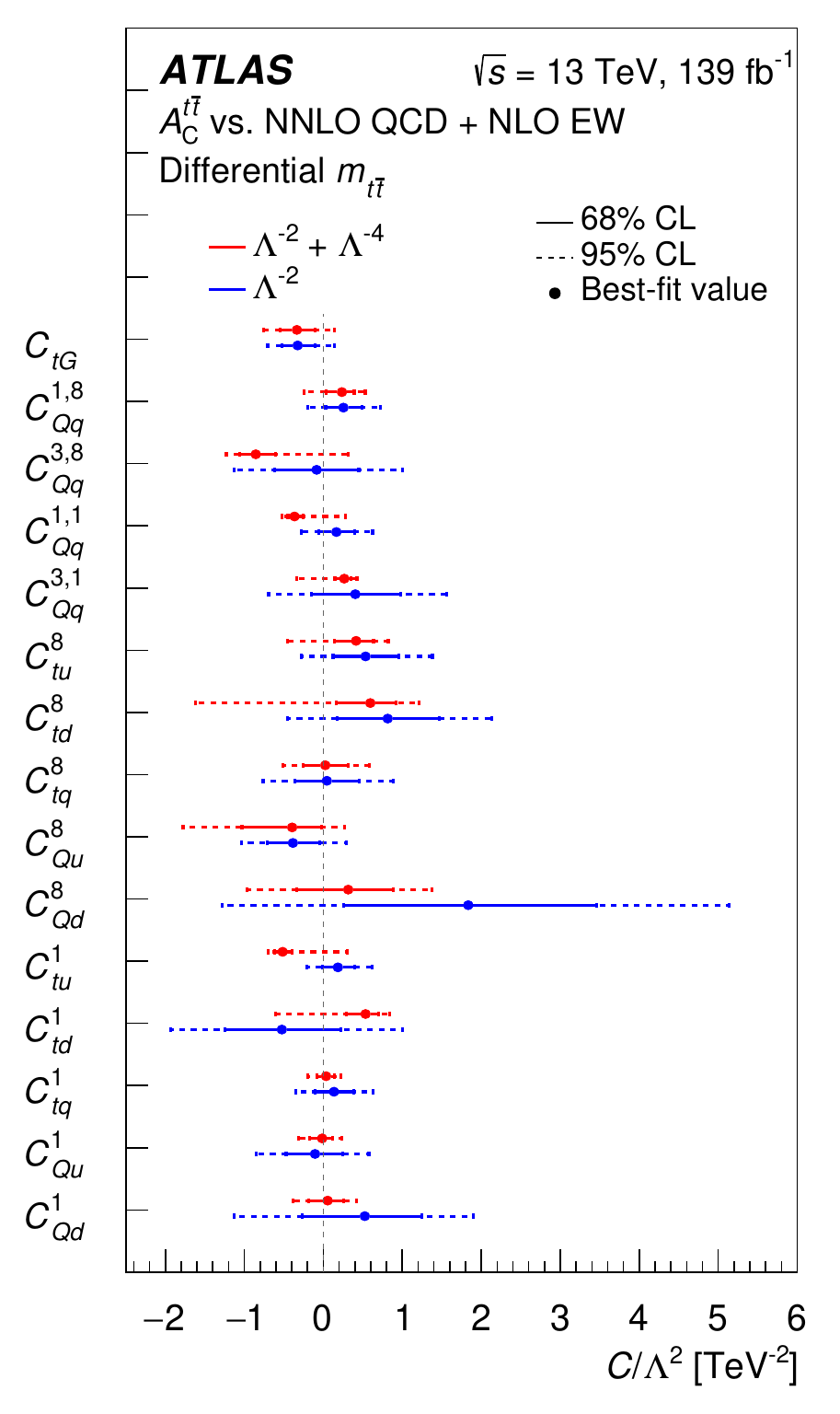}
\caption{Individual 68\% (solid line) and 95\% (dashed line) CL limits on the Wilson coefficients $C_i/ \Lambda^2$ in the SMEFT from the differential charge asymmetry measurements versus $m_{\ttbar}$. Only one Wilson coefficient is varied at a time while the others are kept at 0. The bounds are derived from the charge asymmetry measurement presented in this paper, combining the single-lepton and dilepton channels. The theory uncertainty corresponds to the NNLO QCD + NLO EW calculation.
The impact of dimension-six operators is parameterised at NLO accuracy in QCD using the SMEFT@NLO package in \MGNLO.}
\label{fig:combined_constraints_all_operators}
\end{figure}

The individual 68\% CL and 95\% CL bounds on the Wilson coefficients of all relevant $q\bar{q}t\bar{t}$ operators and $C_{tG}$ obtained from the combination of \mtt differential results are presented in Figure~\ref{fig:combined_constraints_all_operators}.
All intervals are also given in Table~\ref{tab:eftbounds_all_wilson_coefficients}.
The 95\% CL bounds typically range from $\cal{O}$$(1~\tev^{-2})$ to over $\cal{O}$$(10~\tev^{-2})$.
For the inclusive measurement, the linear fit yields tighter bounds on the Wilson coefficients for octet operators, which enter at tree level, than for the coefficients of singlet operators that only affect \ttbar{} production at NLO~\cite{Brivio:2019ius}.
The quadratic fit, however, provides tighter constraints for most of the singlet operators.
The 95\% CL bounds from the differential measurement are generally about a factor of two stronger than the bounds derived from the inclusive measurement, thanks to the rapidly increasing sensitivity at high $m_{t\bar{t}}$.
 
\begin{table}[ht!]
\caption{Individual 68\% and 95\% CL bounds on the relevant Wilson coefficients of the SM Effective Field Theory in units of \TeV$^{-2}$. The bounds are derived from the charge asymmetry measurement presented in this paper, combining the single-lepton and dilepton channels. The upper block corresponds to the bounds derived from the inclusive measurement, the second block to the bounds from the differential charge asymmetry measurements versus $m_{\ttbar}$. In both cases, the experimental uncertainties are accounted for, in the form of the complete covariance matrix that keeps track of correlations between bins for the differential measurement. The theory uncertainty from the NNLO QCD + NLO EW calculation is included by explicitly varying the renormalisation and factorisation scales, or the parton distribution functions, in the calculation and registering the variations in the intervals.}
\label{tab:eftbounds_all_wilson_coefficients}
\begin{tabular}{lcccc}
\hline
\multicolumn{5}{c}{Individual bounds (in units of \TeV$^{-2}$) from the inclusive $A_C^{\ttbar}$ measurement.} \\
& \multicolumn{2}{c}{\textit{Linear} fit including terms $\propto \Lambda^{-2}$} & \multicolumn{2}{c}{\textit{Quadratic} fit adding $(D6)^2$ terms $\propto \Lambda^{-4}$} \\
Operator coefficient  & 68\% CL & 95\% CL & 68\% CL & 95\% CL \\ \hline
$C_{tG} / \Lambda^2$       & $[-0.54,0.37]$ & $[-0.89,1.03]$ & $[-0.56,0.37]$ & $[-0.97,0.99]$ \\
$C_{Qq}^{1,8} / \Lambda^2$ & $[-0.32,0.61]$ & $[-0.78,1.10]$ & $[-0.37,0.51]$ & $[-3.47,0.84]$ \\
$C_{Qq}^{3,8} / \Lambda^2$ & $[-0.88,1.63]$ & $[-2.10,2.94]$ & $[-1.97,0.90]$ & $[-2.41,1.33]$ \\
$C_{Qq}^{1,1} / \Lambda^2$ & $[-1.24,2.29]$ & $[-2.97,4.13]$ & $[-0.60,0.48]$ & $[-0.79,0.67]$ \\
$C_{Qq}^{3,1} / \Lambda^2$ & $[-6.74,3.73]$ & $[-12.1,8.99]$ & $[-0.51,0.57]$ & $[-0.70,0.75]$ \\
$C_{tu}^{8}  / \Lambda^2$  & $[-0.60,1.13]$ & $[-1.44,2.05]$ & $[-2.93,0.82]$ & $[-3.38,1.28]$ \\
$C_{td}^{8}  / \Lambda^2$  & $[-0.97,1.80]$ & $[-2.30,3.26]$ & $[-4.34,1.28]$ & $[-5.05,1.99]$ \\
$C_{tq}^{8}  / \Lambda^2$  & $[-0.96,0.54]$ & $[-1.73,1.29]$ & $[-2.77,0.45]$ & $[-3.23,0.93]$ \\
$C_{Qu}^{8}  / \Lambda^2$  & $[-1.06,0.59]$ & $[-1.90,1.41]$ & $[-3.63,0.51]$ & $[-4.17,1.06]$ \\
$C_{Qd}^{8}  / \Lambda^2$  & $[-3.71,2.08]$ & $[-6.61,5.02]$ & $[-3.36,1.29]$ & $[-4.41,2.35]$ \\
$C_{tu}^{1}  / \Lambda^2$  & $[-0.81,1.49]$ & $[-1.93,2.69]$ & $[-0.80,0.53]$ & $[-1.02,0.75]$ \\
$C_{td}^{1}  / \Lambda^2$  & $[-17.6,32.7]$ & $[-41.8,59.4]$ & $[-0.98,0.96]$ & $[-1.32,1.30]$ \\
$C_{tq}^{1}  / \Lambda^2$  & $[-1.25,2.31]$ & $[-3.00,4.16]$ & $[-0.34,0.47]$ & $[-0.55,0.68]$ \\
$C_{Qu}^{1}  / \Lambda^2$  & $[-2.40,4.40]$ & $[-5.76,7.92]$ & $[-0.43,0.53]$ & $[-0.68,0.78]$ \\
$C_{Qd}^{1}  / \Lambda^2$  & $[-53.0,88.0]$ & $[-134,152]$   & $[-0.72,0.74]$ & $[-1.10,1.12]$ \\
\hline
\multicolumn{5}{c}{Individual bounds (in units of \TeV$^{-2}$) from the differential $A_C^{\ttbar}$ measurement versus $m_{\ttbar}$.} \\
& \multicolumn{2}{c}{\textit{Linear} fit including terms $\propto \Lambda^{-2}$} & \multicolumn{2}{c}{\textit{Quadratic} fit adding $(D6)^2$ terms $\propto \Lambda^{-4}$} \\
Operator coefficient  & 68\% CL & 95\% CL & 68\% CL & 95\% CL \\ \hline
$C_{tG}      / \Lambda^2$ & $[-0.53,-0.11]$  & $[-0.70,0.14]$ & $[-0.55,-0.11]$ & $[-0.75,0.14]$ \\
$C_{Qq}^{1,8} / \Lambda^2$& $[~~\,0.03,~~\,0.49]$  & $[-0.20,0.73]$ & $[~~0.04,~~\,0.39]$ & $[-0.25,0.53]$ \\
$C_{Qq}^{3,8} / \Lambda^2$& $[-0.61,~~\,0.45]$  & $[-1.13,1.00]$ & $[-1.06,-0.61]$ & $[-1.23,0.31]$ \\
$C_{Qq}^{1,1} / \Lambda^2$& $[-0.06,~~\,0.39]$  & $[-0.28,0.62]$ & $[-0.45,-0.25]$ & $[-0.52,0.28]$ \\
$C_{Qq}^{3,1} / \Lambda^2$& $[-0.15,~~\,0.98]$  & $[-0.69,1.56]$ & $[~~0.15,~~\,0.35]$ & $[-0.34,0.43]$ \\
$C_{tu}^{8}  / \Lambda^2$ & $[~~\,0.12,~~\,0.96]$  & $[-0.28,1.38]$ & $[~~0.14,~~\,0.63]$ & $[-0.45,0.82]$ \\
$C_{td}^{8}  / \Lambda^2$ & $[~~\,0.18,~~\,1.47]$  & $[-0.45,2.13]$ & $[~~0.17,~~\,0.92]$ & $[-1.62,1.21]$ \\
$C_{tq}^{8}  / \Lambda^2$ & $[-0.36,~~\,0.46]$  & $[-0.76,0.88]$ & $[-0.25,~~\,0.31]$ & $[-0.51,0.58]$ \\
$C_{Qu}^{8}  / \Lambda^2$ & $[-0.71,-0.05]$  & $[-1.03,0.29]$ & $[-1.03,-0.02]$ & $[-1.78,0.27]$ \\
$C_{Qd}^{8}  / \Lambda^2$ & $[~~\,0.25,~~\,3.46]$  & $[-1.28,5.14]$ & $[-0.33,~~\,0.89]$ & $[-0.96,1.37]$ \\
$C_{tu}^{1}  / \Lambda^2$ & $[-0.02,~~\,0.40]$  & $[-0.21,0.61]$ & $[-0.62,-0.39]$ & $[-0.70,0.31]$ \\
$C_{td}^{1}  / \Lambda^2$ & $[-1.24,~~\,0.22]$  & $[-1.94,1.00]$ & $[~~0.29,~~\,0.70]$ & $[-0.60,0.84]$ \\
$C_{tq}^{1}  / \Lambda^2$ & $[-0.10,~~\,0.38]$  & $[-0.35,0.63]$ & $[-0.08,~~\,0.14]$ & $[-0.20,0.22]$ \\
$C_{Qu}^{1}  / \Lambda^2$ & $[-0.47,~~\,0.24]$  & $[-0.85,0.58]$ & $[-0.18,~~\,0.12]$ & $[-0.31,0.23]$ \\
$C_{Qd}^{1}  / \Lambda^2$ & $[-0.27,~~\,1.24]$  & $[-1.13,1.90]$ & $[-0.19,~~\,0.26]$ & $[-0.39,0.42]$ \\
\hline
\end{tabular}
\end{table}

Compared to global fits of the top-quark sector~\cite{AguilarSaavedra:2018nen,Hartland:2019bjb,Brivio:2019ius} and fits including top, Higgs and EW data~\cite{Ellis:2020unq,Ethier:2021bye} the bounds found in this analysis are of the same order of magnitude.
Often, the bounds from the differential analysis are significantly tighter than the global bounds, indicating that inclusion of these results in future global fits can improve the global result, by disentangling some of the poorly constrained combinations of operator coefficients.

A further asymmetry measurement was proposed in Ref.~\cite{Westhoff:2013, Basan:2020btr}. The energy asymmetry in boosted \ttbar\ production in association with an energetic jet provides complementary information that can eliminate a blind direction of the charge asymmetry measurement. Such a measurement was performed in $t\bar{t}j$ production by the ATLAS Collaboration~\cite{TOPQ-2019-28-updated}.
 
For the energy asymmetry, the SM prediction was obtained from simulations of $t\bar{t}j$ production at NLO with \MGNLOPY[8], while the impact of dimension-six operators is parameterised at LO accuracy in QCD using the SMEFT@NLO package in \MGNLO.
The covariance matrix used in the $\chi^2$ fit  contains the experimental uncertainties as well as the statistical uncertainty of the Monte Carlo prediction and the theory uncertainty due to its scale dependence. The renormalisation and factorisation scales \muR\ and \muF\ are varied from their nominal values by factors of 2 and 0.5, yielding nine different pairs of scale values, and the effect of the scale uncertainty is assessed from the envelope of the resulting asymmetries.

\begin{figure}
\includegraphics[width=.45\textwidth]{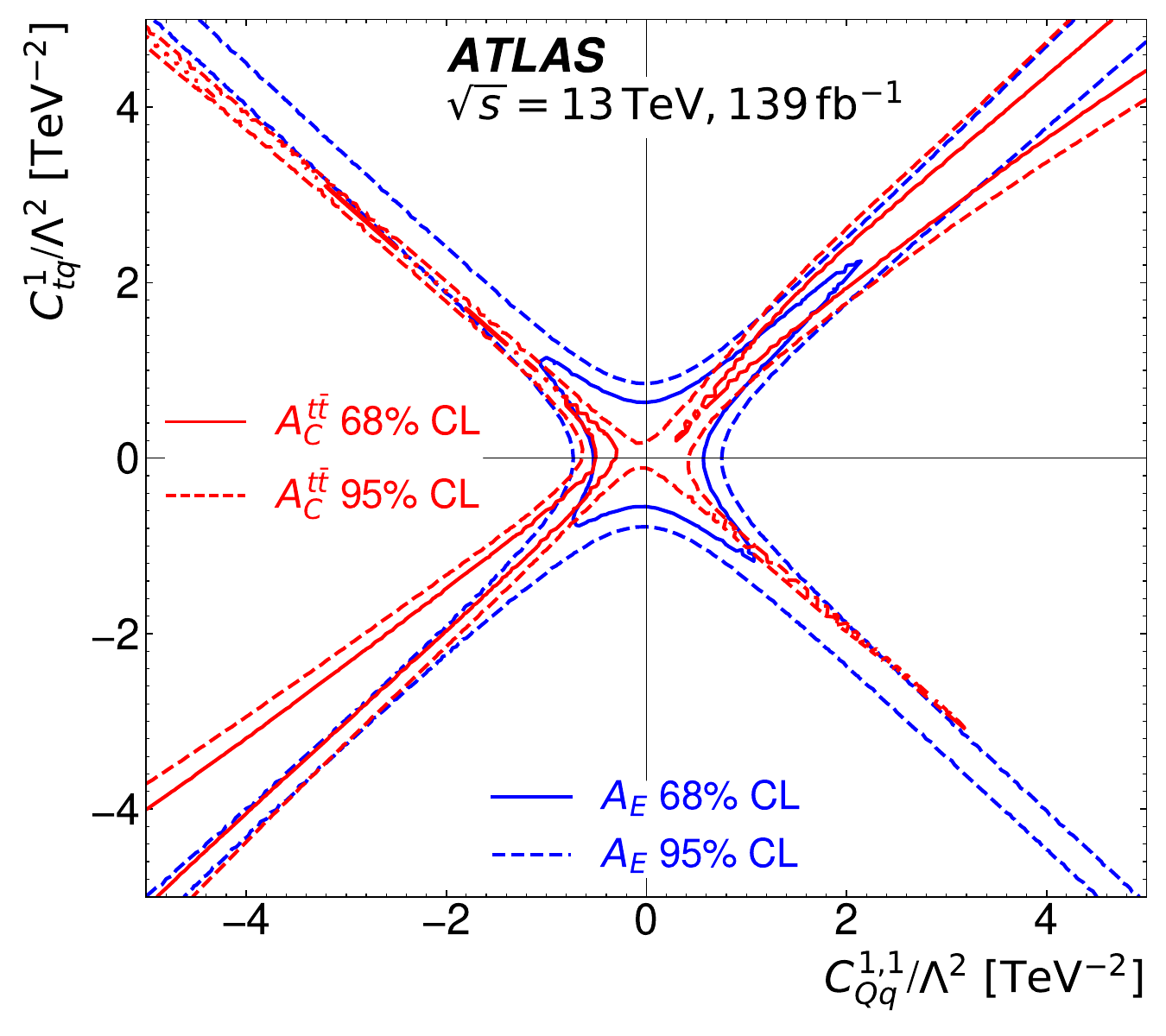}
\includegraphics[width=.45\textwidth]{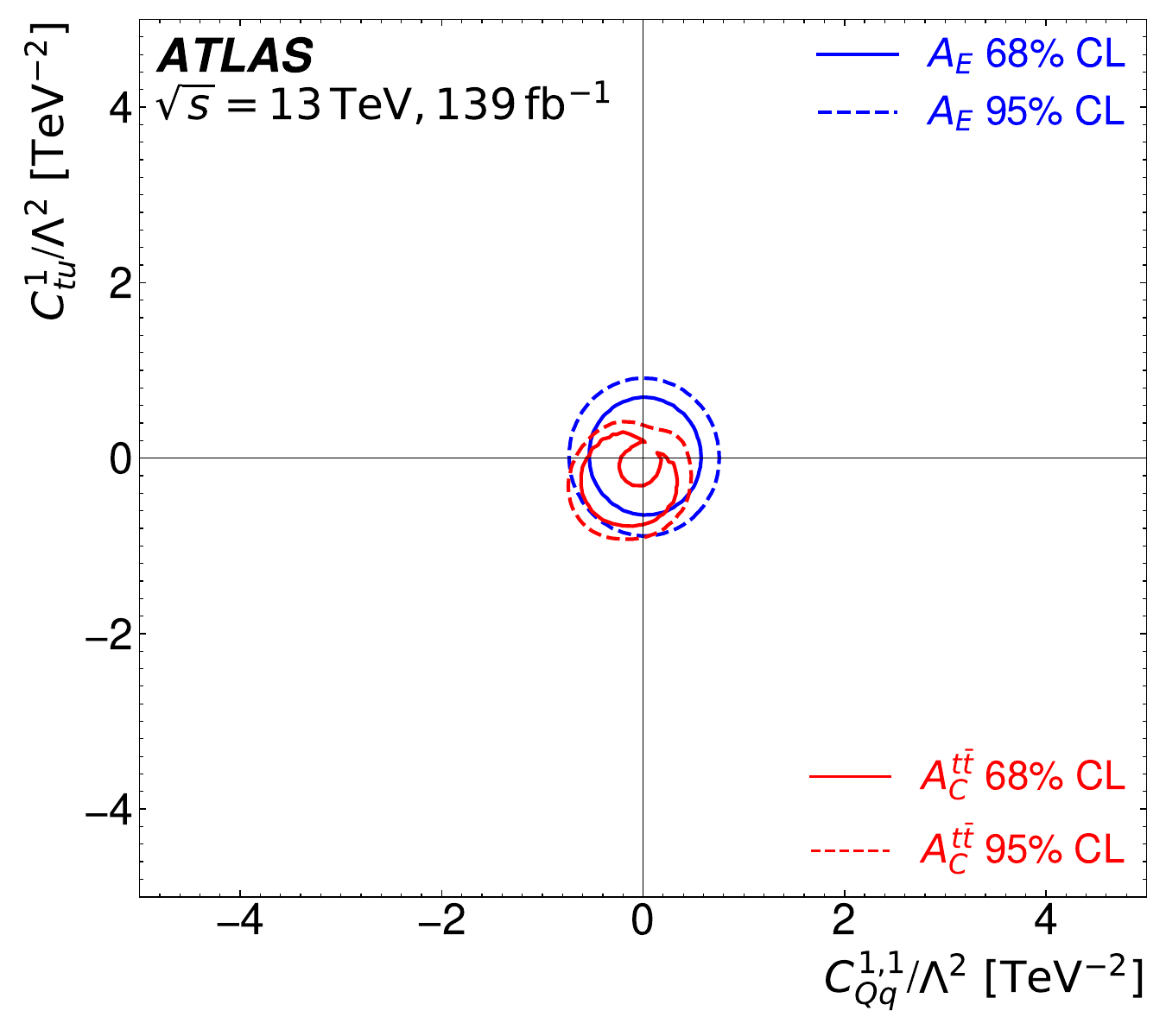}\\
\includegraphics[width=.45\textwidth]{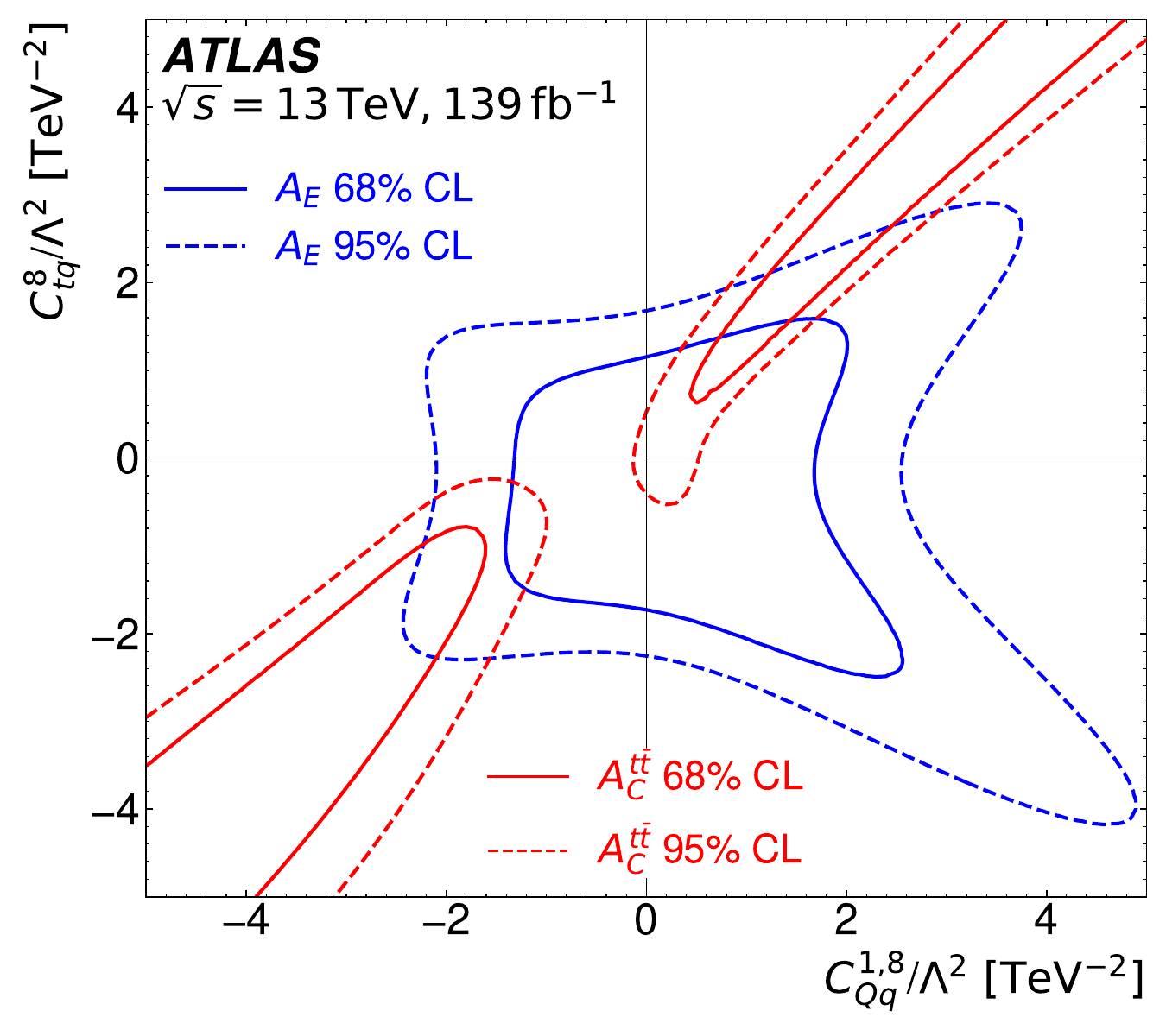}
\includegraphics[width=.45\textwidth]{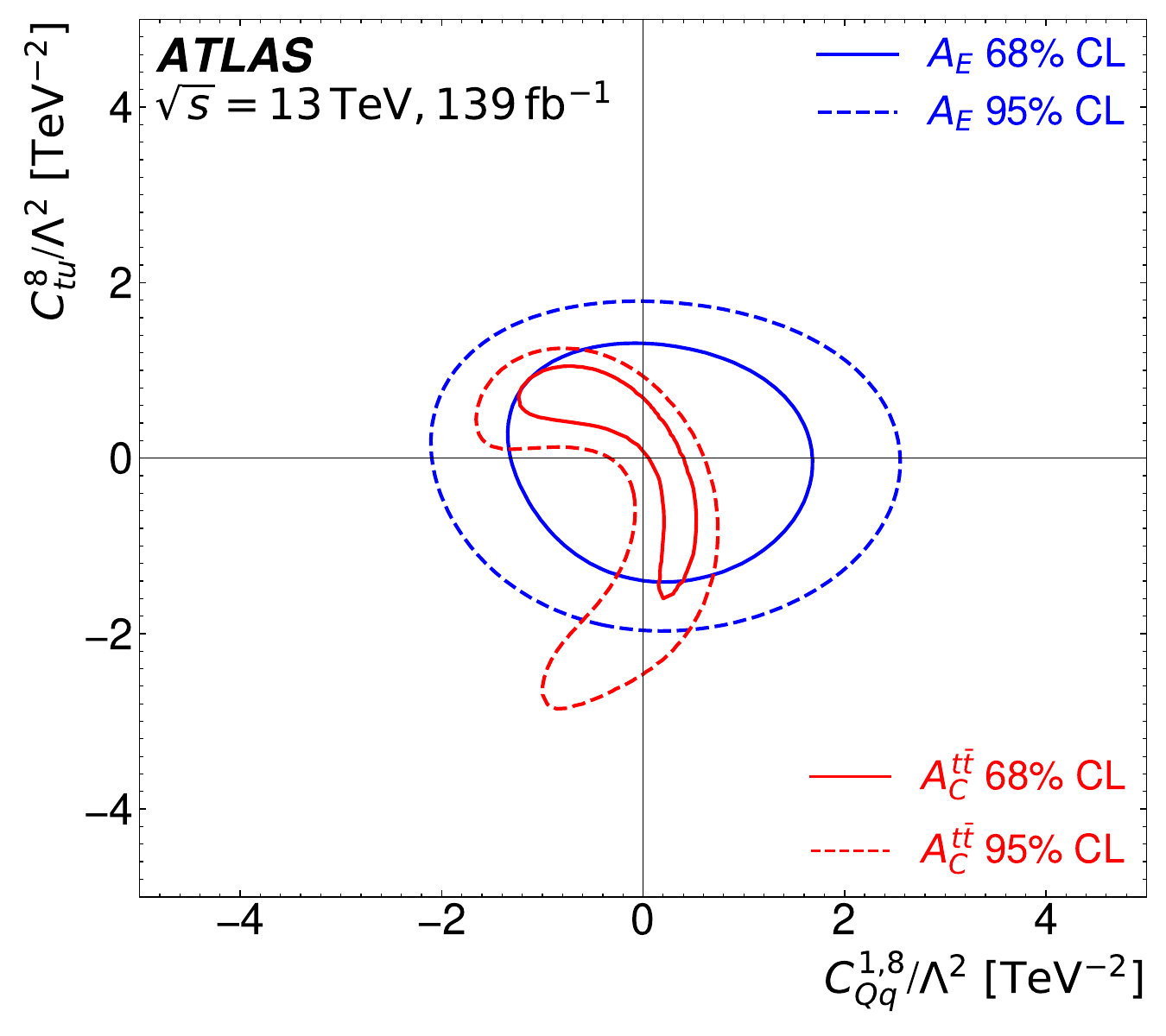}\\
\includegraphics[width=.45\textwidth]{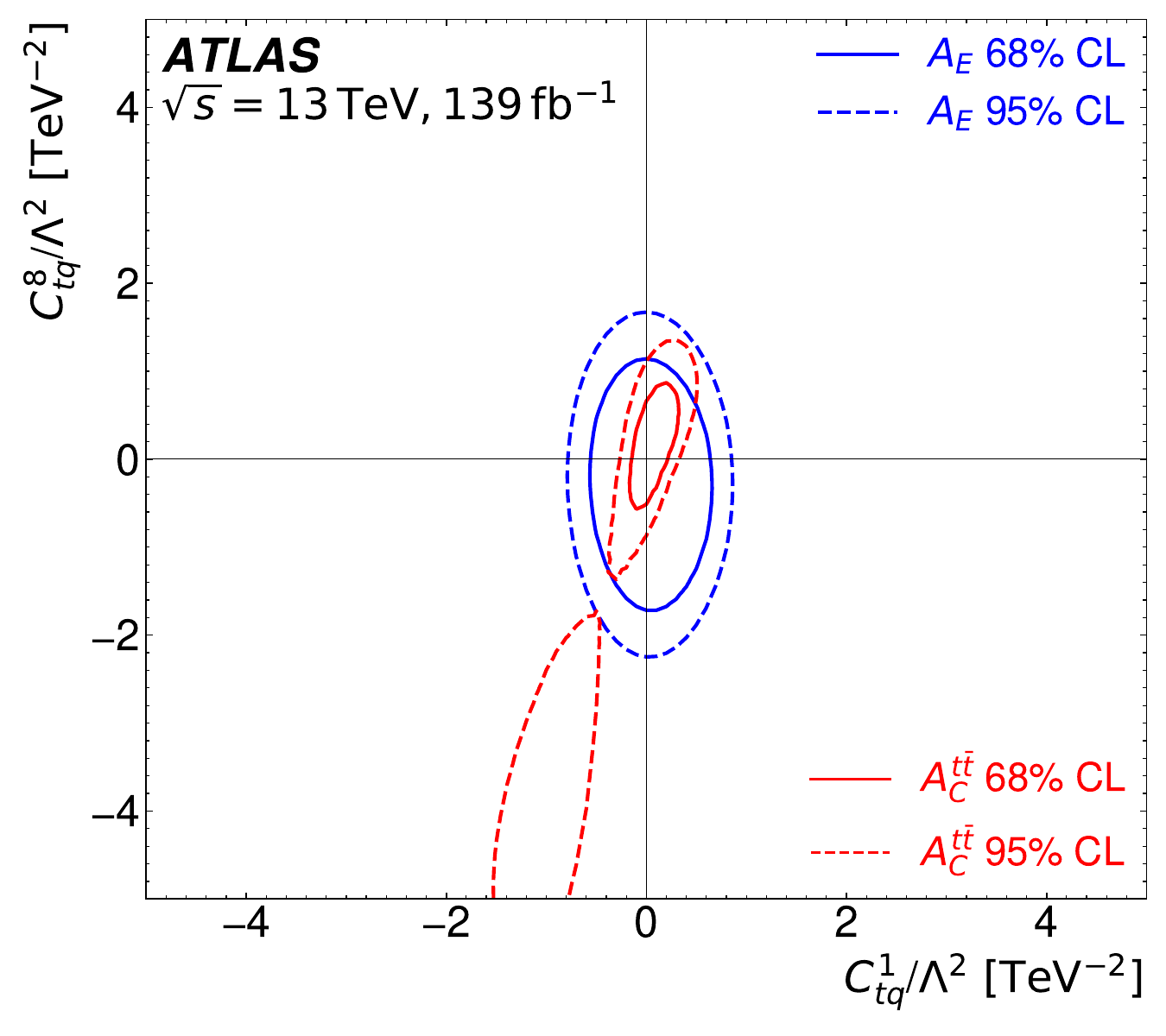}
\includegraphics[width=.45\textwidth]{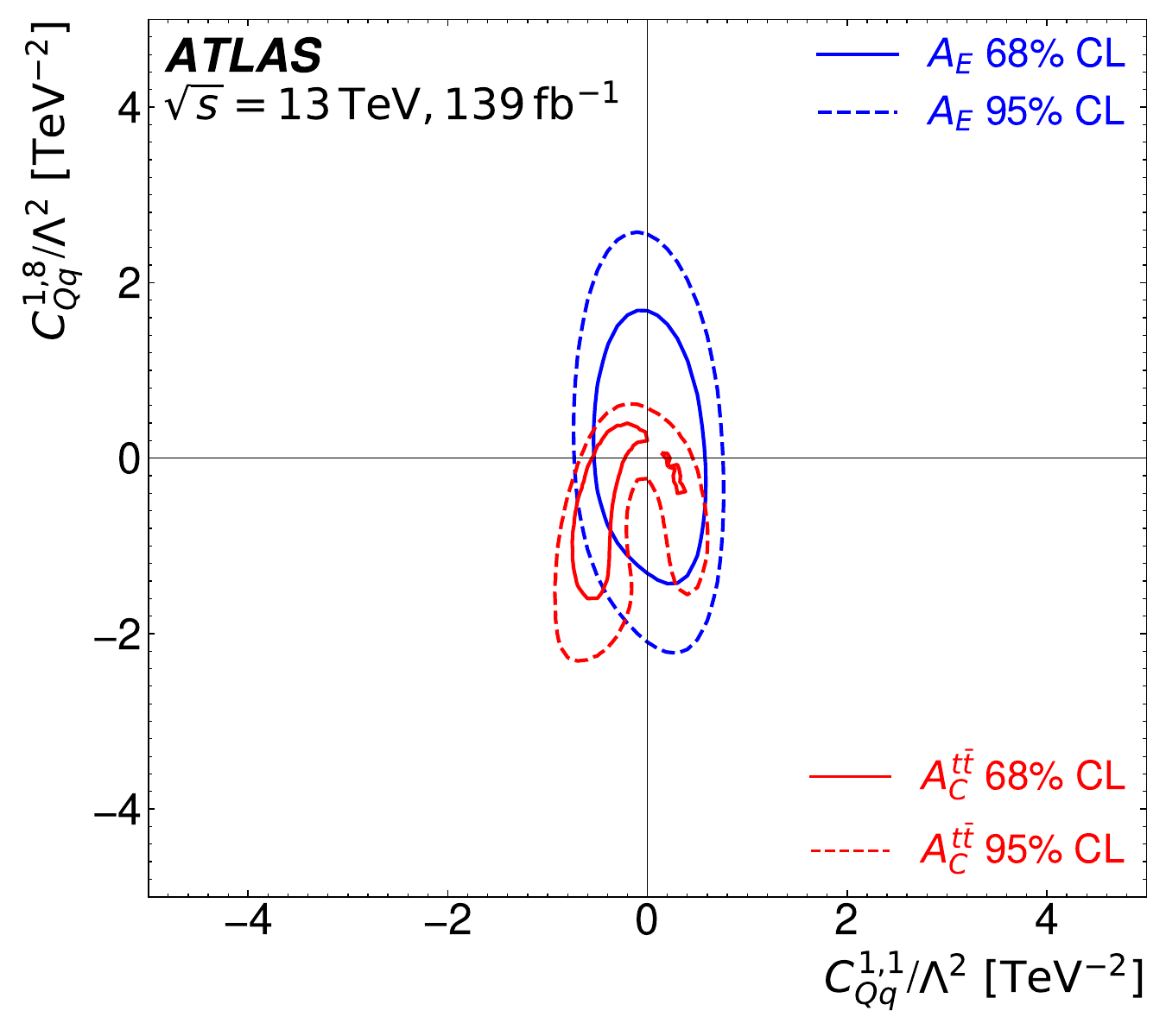}\\
\caption{Bounds on Wilson coefficients from two-parameter fits of the energy asymmetry $A_E$ (blue) and \Actt{} vs \mtt{} (red) measurements. The full and dashed lines show the 68\% and 95\% CL limits, respectively. The bounds are derived from a comparison of the measured asymmetries with the SMEFT predictions using the quadratic parameterisation. The SM prediction of $A_E$ is obtained from simulations of $t\bar{t}j$ production at NLO with \MGNLOPY[8], while the impact of dimension-six operators is parameterised at LO accuracy in QCD using the SMEFT@NLO package in \MGNLO. For \Actt{} the SM prediction is calculated up to NNLO QCD and NLO EW accuracy and the impact of the dimension-six operators is parameterised at NLO accuracy in QCD using the SMEFT@NLO package in \MGNLO.
}
\label{fig:combination:EnergyAsymmetryComparison}
\end{figure}

Due to the extra jet in $t\bar t j$ production, the QCD structure of the energy asymmetry is not the same as for the charge asymmetry in $t\bar t$ production, so the two asymmetries probe different directions in chiral and colour space. This complementarity is illustrated in Figure~\ref{fig:combination:EnergyAsymmetryComparison}, which shows the bounds from \mtt{} differential charge asymmetry measurements compared with the bounds from the energy asymmetry measurement. For colour-singlet operators with different quark chiralities (top row), the two asymmetries probe similar areas in the parameter space. For colour-octet operators with the same chirality scenarios (middle row), however, the shapes of the bounds look very different. In particular, the charge asymmetry leaves a blind direction in the $(C_{Qq}^{1,8},C_{tq}^8)$ plane (left panel), which is broken by the energy asymmetry due to operator interference with the QCD amplitude.
The bottom row shows colour-singlet versus colour-octet operators with the same quark chiralities.
Here, the different shapes of the bounds are due to the different colour-singlet and colour-octet contributions to $t\bar{t}$ and $t\bar t j$ production, which is probed with high sensitivity by the asymmetries. The EFT contributions include dimension-six-squared terms proportional to $\Lambda^{-4}$ for both the charge asymmetry and energy asymmetry.
A comparison for $C^{1}_{tu}$ versus $C^{8}_{tu}$ is not included, because the different sensitivity to colour-singlet and colour-octet operators has already been illustrated with $C^{1}_{tq}$ versus $C^{8}_{tq}$, and the difference between L and R light quarks has been illustrated with $C^{1,1}_{Qq}$ versus $C^{1}_{tu}$.

\FloatBarrier
\section{Conclusion}
\label{sec:conclusion}

The charge asymmetry in top-quark pair production is measured in the single-lepton and dilepton decay channels using 139~\ifb\ of $\sqrt{s} = 13$~\TeV\ proton--proton collision data collected by the ATLAS experiment at the LHC.  A fully Bayesian unfolding method is used to correct for detector resolution and acceptance effects. The resolved and boosted topologies of the single-lepton channels are unfolded simultaneously with the same- and opposite-flavour dilepton channels.
 
Differential measurements are performed as a function of the invariant mass, transverse momentum and longitudinal boost of the \ttbar system. In comparison with previous results obtained at 8\,\TeV, the differential variables benefit from a finer binning at larger values which are sensitive to possible enhancements of the charge asymmetry due to new-physics phenomena beyond the Standard Model. The leptonic charge asymmetry is measured inclusively and differentially as a function of invariant mass, transverse momentum and longitudinal boost of the dilepton pair. The results are compatible with the Standard Model predictions calculated at next-to-leading order in quantum chromodynamics and next-to-leading order in electroweak theory.
 
The asymmetry obtained in the inclusive measurement, combining the single-lepton and dilepton results, is $A_{\text{C}} = 0.0068 \pm 0.0015~\textrm{(stat.+syst.)}$. The result differs from zero by 4.7~standard deviations and thus provides strong evidence for charge asymmetry in \ttbar{} production at the LHC. The precision of the combination is dominated by the lepton+jets channel, primarily due to a much smaller statistical uncertainty than in the dilepton channel. The inclusive and differential measurements are consistent with the Standard Model predictions calculated at next-to-next-to-leading order in quantum chromodynamics and next-to-leading order in electroweak theory.
 
The results are interpreted in the Standard Model Effective Field Theory. Individual limits on the Wilson coefficients related to two-light-quark--two-heavy-quark operators and to the operator that modifies the top--gluon interaction are derived from the inclusive charge asymmetry measurement and from the differential measurement versus \mtt. The 95\% CL bound on $C_{tu}^{8}/\Lambda^2$ from the inclusive \Actt{} measurement improves considerably on the limits derived from the LHC 8~\TeV{} combination and from the Tevatron $A_\textrm{FB}$ combination. The combined bound on $C_{tu}^{8}/\Lambda^2$ from the \mtt differential measurement, $[-0.28,1.38]$~\TeV$^{-2}$ using a linear fit, is more than a factor of two better than the bound derived from the inclusive measurement, due to the rapid increase in sensitivity with \mtt. The charge asymmetry presented in this paper complements the energy asymmetry measured recently in $\ttbar j$ production by ATLAS, since the two observables probe different directions in chiral and colour space.


\section*{Acknowledgements}


We thank CERN for the very successful operation of the LHC, as well as the
support staff from our institutions without whom ATLAS could not be
operated efficiently.
 
We acknowledge the support of
ANPCyT, Argentina;
YerPhI, Armenia;
ARC, Australia;
BMWFW and FWF, Austria;
ANAS, Azerbaijan;
SSTC, Belarus;
CNPq and FAPESP, Brazil;
NSERC, NRC and CFI, Canada;
CERN;
ANID, Chile;
CAS, MOST and NSFC, China;
Minciencias, Colombia;
MEYS CR, Czech Republic;
DNRF and DNSRC, Denmark;
IN2P3-CNRS and CEA-DRF/IRFU, France;
SRNSFG, Georgia;
BMBF, HGF and MPG, Germany;
GSRI, Greece;
RGC and Hong Kong SAR, China;
ISF and Benoziyo Center, Israel;
INFN, Italy;
MEXT and JSPS, Japan;
CNRST, Morocco;
NWO, Netherlands;
RCN, Norway;
MEiN, Poland;
FCT, Portugal;
MNE/IFA, Romania;
JINR;
MES of Russia and NRC KI, Russian Federation;
MESTD, Serbia;
MSSR, Slovakia;
ARRS and MIZ\v{S}, Slovenia;
DSI/NRF, South Africa;
MICINN, Spain;
SRC and Wallenberg Foundation, Sweden;
SERI, SNSF and Cantons of Bern and Geneva, Switzerland;
MOST, Taiwan;
TENMAK, T\"urkiye;
STFC, United Kingdom;
DOE and NSF, United States of America.
In addition, individual groups and members have received support from
BCKDF, CANARIE, Compute Canada and CRC, Canada;
COST, ERC, ERDF, Horizon 2020 and Marie Sk{\l}odowska-Curie Actions, European Union;
Investissements d'Avenir Labex, Investissements d'Avenir Idex and ANR, France;
DFG and AvH Foundation, Germany;
Herakleitos, Thales and Aristeia programmes co-financed by EU-ESF and the Greek NSRF, Greece;
BSF-NSF and GIF, Israel;
Norwegian Financial Mechanism 2014-2021, Norway;
NCN and NAWA, Poland;
La Caixa Banking Foundation, CERCA Programme Generalitat de Catalunya and PROMETEO and GenT Programmes Generalitat Valenciana, Spain;
G\"{o}ran Gustafssons Stiftelse, Sweden;
The Royal Society and Leverhulme Trust, United Kingdom.
 
The crucial computing support from all WLCG partners is acknowledged gratefully, in particular from CERN, the ATLAS Tier-1 facilities at TRIUMF (Canada), NDGF (Denmark, Norway, Sweden), CC-IN2P3 (France), KIT/GridKA (Germany), INFN-CNAF (Italy), NL-T1 (Netherlands), PIC (Spain), ASGC (Taiwan), RAL (UK) and BNL (USA), the Tier-2 facilities worldwide and large non-WLCG resource providers. Major contributors of computing resources are listed in Ref.~\cite{ATL-SOFT-PUB-2021-003}.


\clearpage
\appendix
\part*{Appendix}
\addcontentsline{toc}{part}{Appendix}
\FloatBarrier
\section{Ranking}

\subsection{\Actt combination measurement}
\label{app:comb_Actt_ranking}
The ranking of the leading systematic uncertainties is summarised in Tables~\ref{tab:Actt_betatt_ranking} and~\ref{tab:Actt_pttt_ranking} for the \betatt and \pttt differential \Actt combined measurements, respectively.
Figures~\ref{fig:NP_rankpull_dy_comb_incl_dybetatt_bin0}--\ref{fig:NP_rankpull_dypttt_comb_bin2} show the ten highest-ranking nuisance parameters.  Comparisons of pre-marginalisation and post-marginalisation ranking are presented together with the pulls and constraints of the ranked nuisance parameters obtained from data. In the figures, the $\gamma_1$, $\gamma_2$, $\gamma_3$ and $\gamma_4$ nuisance parameters correspond to $[-5,-x]$, $[-x,0]$, $[0,x]$ and $[x,5]$ \dYtt bins, respectively, where $x$ stands for the \dYtt bin edge.
 
\begin{table}[H]
\caption{Impact of systematic uncertainties on the measurement of \Actt{} vs \betatt{}, grouped into categories. The impact of each category is obtained by summing the symmetrised ranking impact of the respective individual uncertainties in quadrature. The total uncertainty differs from the  sum-in-quadrature of the groups due to post-marginalisation nuisance parameter correlations in the unfolding.}
\label{tab:Actt_betatt_ranking}
\centering
\begin{tabular}{lllll}
\toprule
& \multicolumn{4}{c}{Post-marg. (pre-marg.) impact $\times 100$} \\
\cmidrule{2-5}
\betatt{} bin & 0--0.3 & 0.3--0.6 & 0.6--0.8 & 0.8--1.0 \\
\midrule
$t\bar{t}$ modelling & 0.18 (0.25) & 0.12 (0.16) & 0.11 (0.14) & 0.23 (0.34)\\
$t\bar{t}$ normalisation (flat prior) & 0.07 & 0.05 & 0.04 & 0.07\\
Background modelling & 0.15 (0.18) & 0.07 (0.08) & 0.07 (0.07) & 0.12 (0.13)\\
Monte Carlo statistics & 0.19 & 0.15 & 0.12 & 0.12\\
Small-$R$ JES & 0.15 (0.18) & 0.10 (0.11) & 0.09 (0.10) & 0.10 (0.14)\\
Small-$R$ JER & 0.10 (0.16) & 0.06 (0.09) & 0.07 (0.10) & 0.12 (0.17)\\
Large-$R$ JES, JER & 0.05 (0.05) & 0.02 (0.03) & 0.01 (0.02) & 0.02 (0.03)\\
Leptons, \MET & 0.03 (0.04) & 0.03 (0.04) & 0.05 (0.06) & 0.06 (0.08)\\
$b$-tagging eff. & 0.03 (0.03) & 0.02 (0.02) & 0.02 (0.02) & 0.03 (0.03)\\
Pile-up, JVT, luminosity & 0.01 (0.01) & 0.01 (0.01) & 0.02 (0.02) & 0.01 (0.01)\\
\midrule
Statistical uncertainty & 0.39 & 0.29 & 0.27 & 0.24\\
\midrule
Total uncertainty & 0.52 & 0.37 & 0.34 & 0.39\\
\bottomrule
\end{tabular}
\end{table}
 
\begin{table}[H]
\caption{Impact of systematic uncertainties on the measurement of \Actt{} vs \pttt{}, grouped into categories. The impact of each category is obtained by summing the symmetrised ranking impact of the respective individual uncertainties in quadrature. The total uncertainty differs from the  sum-in-quadrature of the groups due to post-marginalisation nuisance parameter correlations in the unfolding.}
\label{tab:Actt_pttt_ranking}
\centering
\begin{tabular}{llll}
\toprule
& \multicolumn{3}{c}{Post-marg. (pre-marg.) impact $\times 100$} \\
\cmidrule{2-4}
\pttt{} bin [\GeV] & 0--30 & 30--120 & $> 120$ \\
\midrule
$t\bar{t}$ modelling & 0.13 (0.19) & 0.11 (0.16) & 0.40 (0.57)\\
$t\bar{t}$ normalisation (flat prior) & 0.03 & 0.05 & 0.12\\
Background modelling & 0.11 (0.12) & 0.12 (0.13) & 0.14 (0.16)\\
Monte Carlo statistics & 0.15 & 0.13 & 0.19\\
Small-$R$ JES & 0.09 (0.10) & 0.08 (0.09) & 0.16 (0.18)\\
Small-$R$ JER & 0.08 (0.12) & 0.08 (0.11) & 0.13 (0.20)\\
Large-$R$ JES, JER & 0.02 (0.03) & 0.02 (0.03) & 0.05 (0.07)\\
Leptons, \MET & 0.12 (0.14) & 0.05 (0.07) & 0.08 (0.08)\\
$b$-tagging eff. & 0.03 (0.03) & 0.03 (0.03) & 0.04 (0.04)\\
Pile-up, JVT, luminosity & 0.01 (0.01) & 0.02 (0.02) & 0.03 (0.03)\\
\midrule
Statistical uncertainty & 0.32 & 0.26 & 0.44\\
\midrule
Total uncertainty & 0.41 & 0.37 & 0.65\\
\bottomrule
\end{tabular}
\end{table}
 
\begin{figure}[H]
\centering
\subfigure[][Inclusive]{\includegraphics[width=0.7\textwidth]{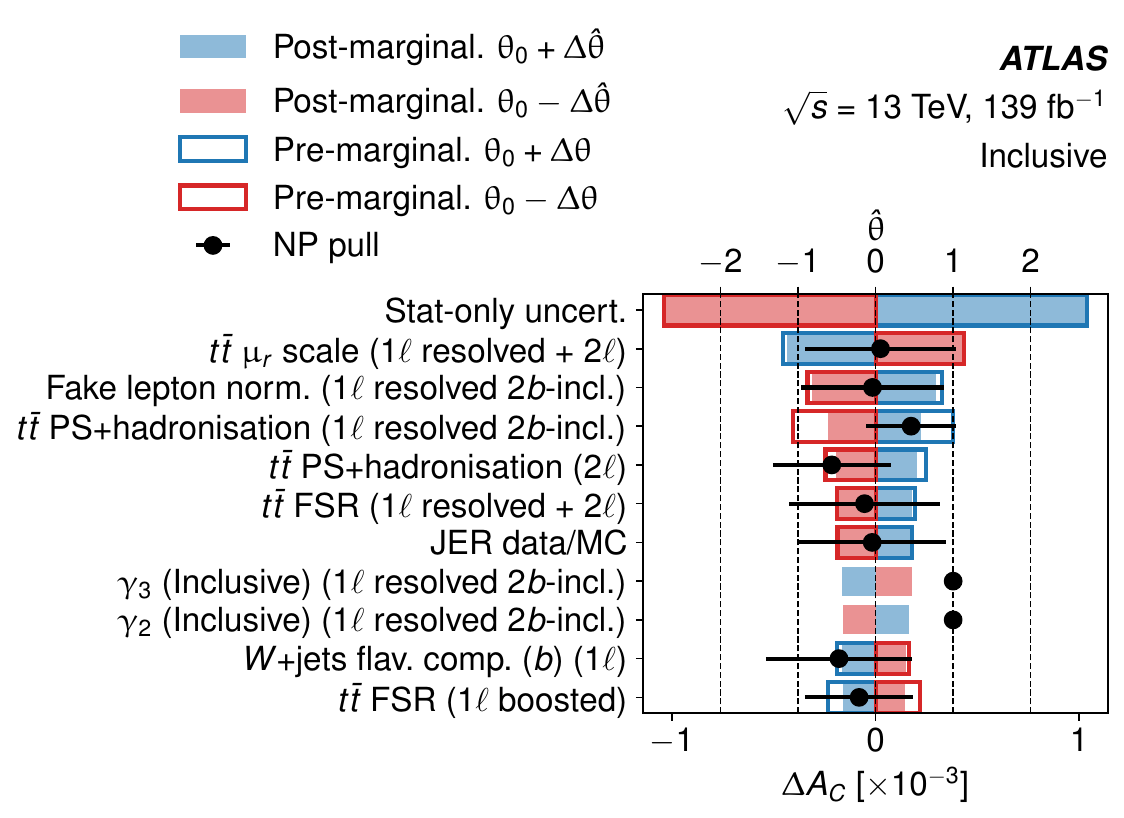}}\\
\subfigure[][0.0 < \betatt < 0.3]{\includegraphics[width=0.78\textwidth]{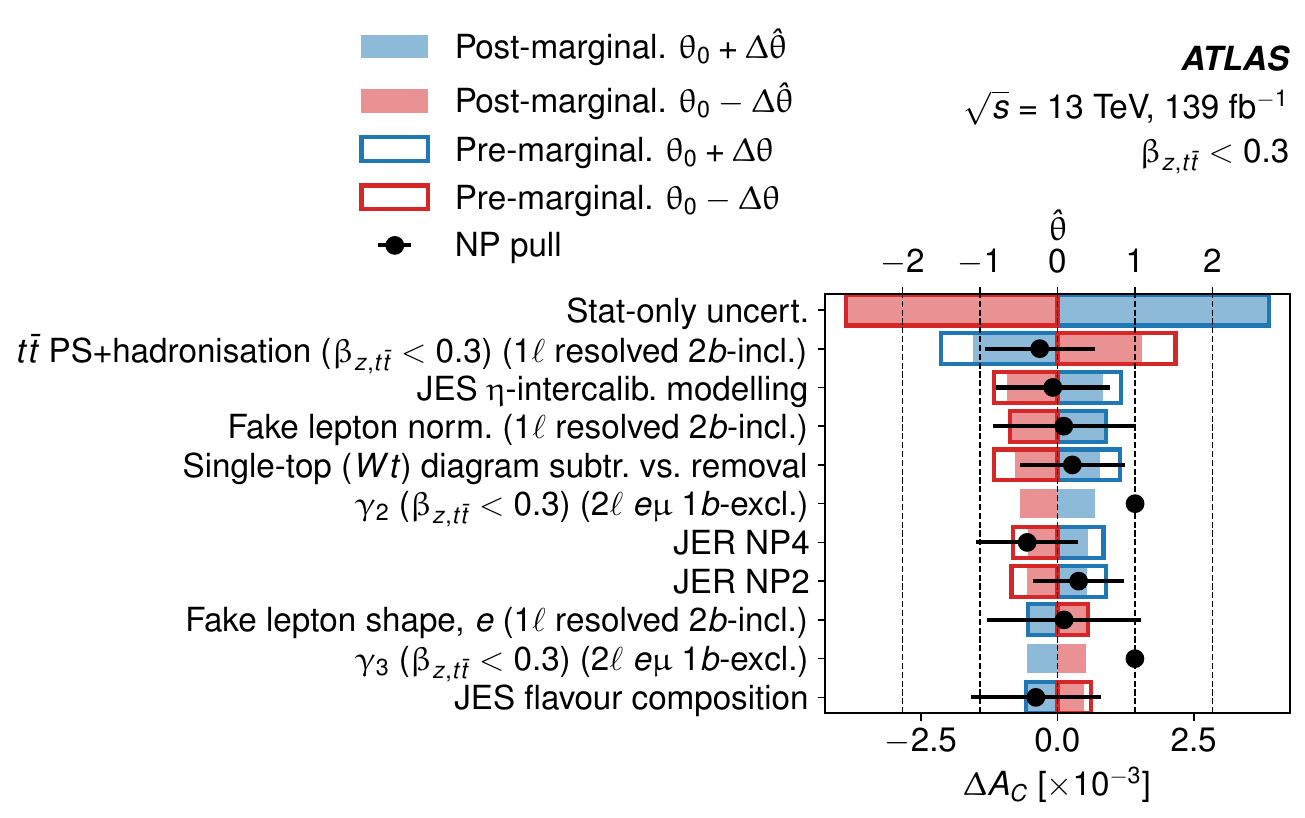}}
\caption{Ranking of the systematic uncertainties with marginalisation (a) for the inclusive \Ac measurement and (b) for $\betatt \in [0.0, 0.3]$. Only the ten highest-ranking uncertainties are shown, ordered according to their post-marginalisation ranking, which considers the constraining power of the marginalisation for the data. The red and blue bars show the effect on unfolded \Ac of downward and upward variation of the systematic uncertainty, respectively. The points show the pulls and constraints of the ranked NPs obtained from data.}
\label{fig:NP_rankpull_dy_comb_incl_dybetatt_bin0}
\end{figure}
 
\begin{figure}[H]
\centering
\subfigure[][0.3 < \betatt <  0.6]{\includegraphics[width=0.8\textwidth]{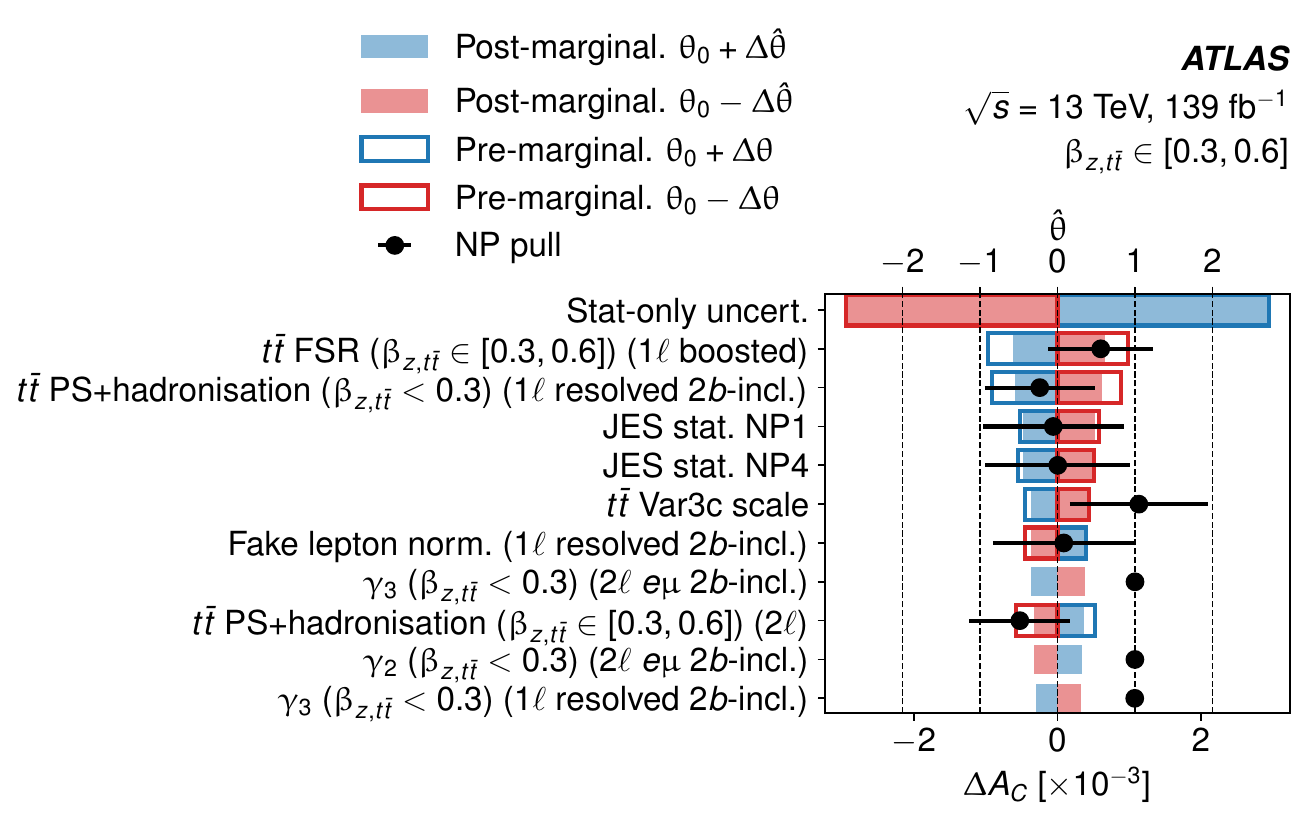}}\\
\subfigure[][0.6 < \betatt <  0.8]{\includegraphics[width=0.76\textwidth]{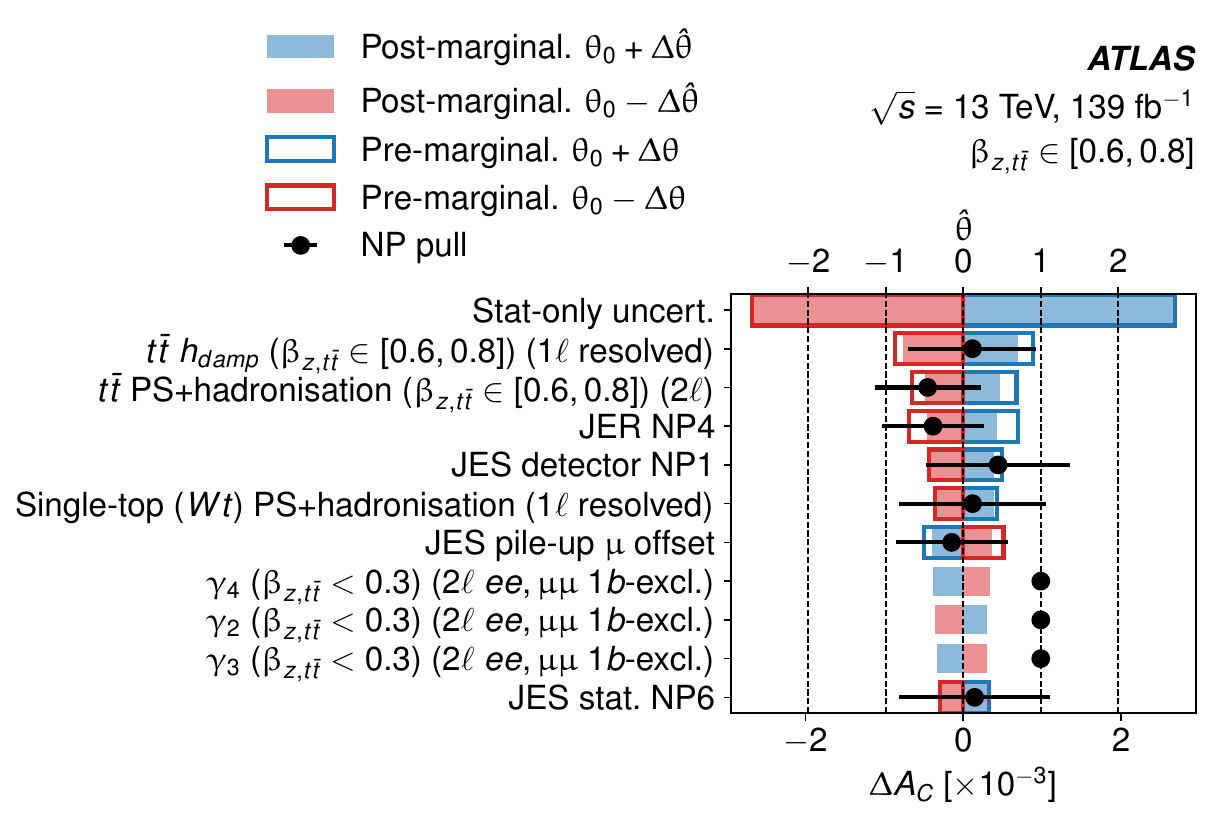}}
\caption{Ranking of the systematic uncertainties with marginalisation (a) for $\betatt \in [0.3, 0.6]$ and (b) for $\betatt \in [0.6, 0.8]$. Only the ten highest-ranking uncertainties are shown, ordered according to their post-marginalisation ranking, which considers the constraining power of the marginalisation for the data. The red and blue bars show the effect on unfolded \Ac of downward and upward variation of the systematic uncertainty, respectively. The points show the pulls and constraints of the ranked NPs obtained from data.}
\end{figure}
 
\begin{figure}[H]
\centering
\subfigure[][0.8 < \betatt <  1.0]{\includegraphics[width=0.85\textwidth]{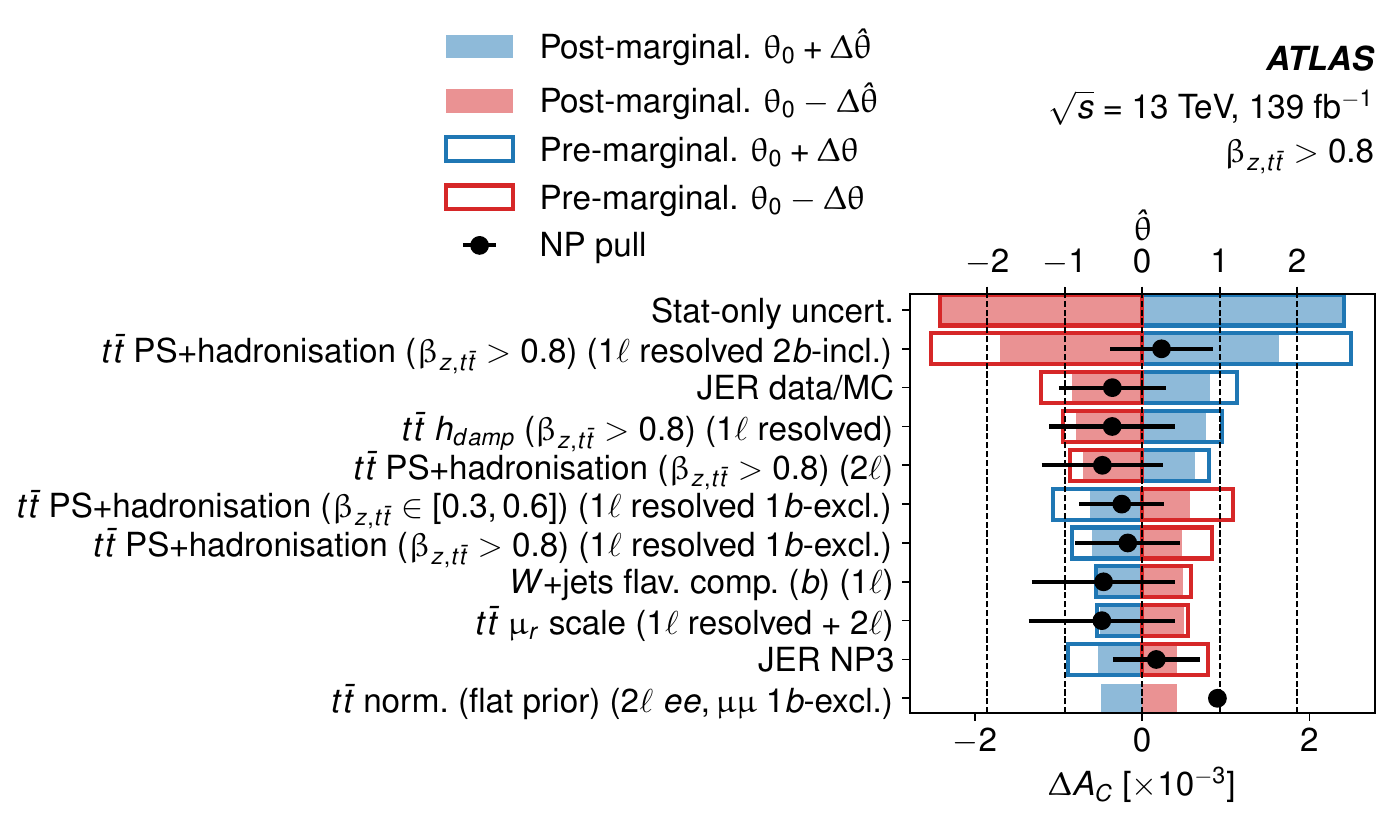}}\\
\subfigure[][\mtt < 500 GeV]{\includegraphics[width=0.9\textwidth]{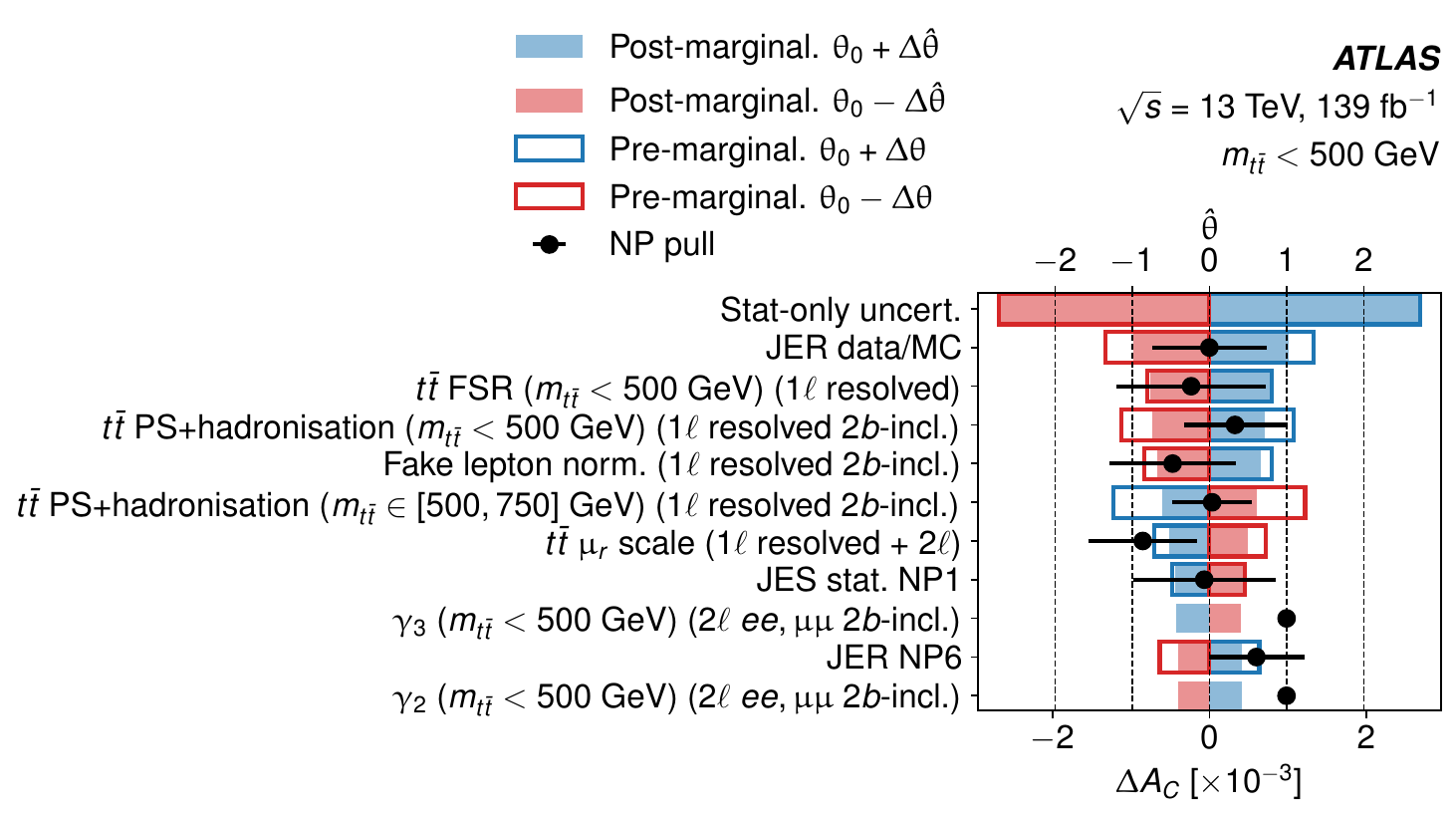}}
\caption{Ranking of the systematic uncertainties with marginalisation (a) for $\betatt \in [0.8, 1.0]$ and (b) for $\mtt \in [0, 500]$~\GeV. Only the ten highest-ranking uncertainties are shown, ordered according to their post-marginalisation ranking, which considers the constraining power of the marginalisation for the data. The red and blue bars show the effect on unfolded \Ac of downward and upward variation of the systematic uncertainty, respectively. The points show the pulls and constraints of the ranked NPs obtained from data.}
\end{figure}
 
\begin{figure}[H]
\centering
\subfigure[][500 < \mtt < 750 GeV]{\includegraphics[width=0.75\textwidth]{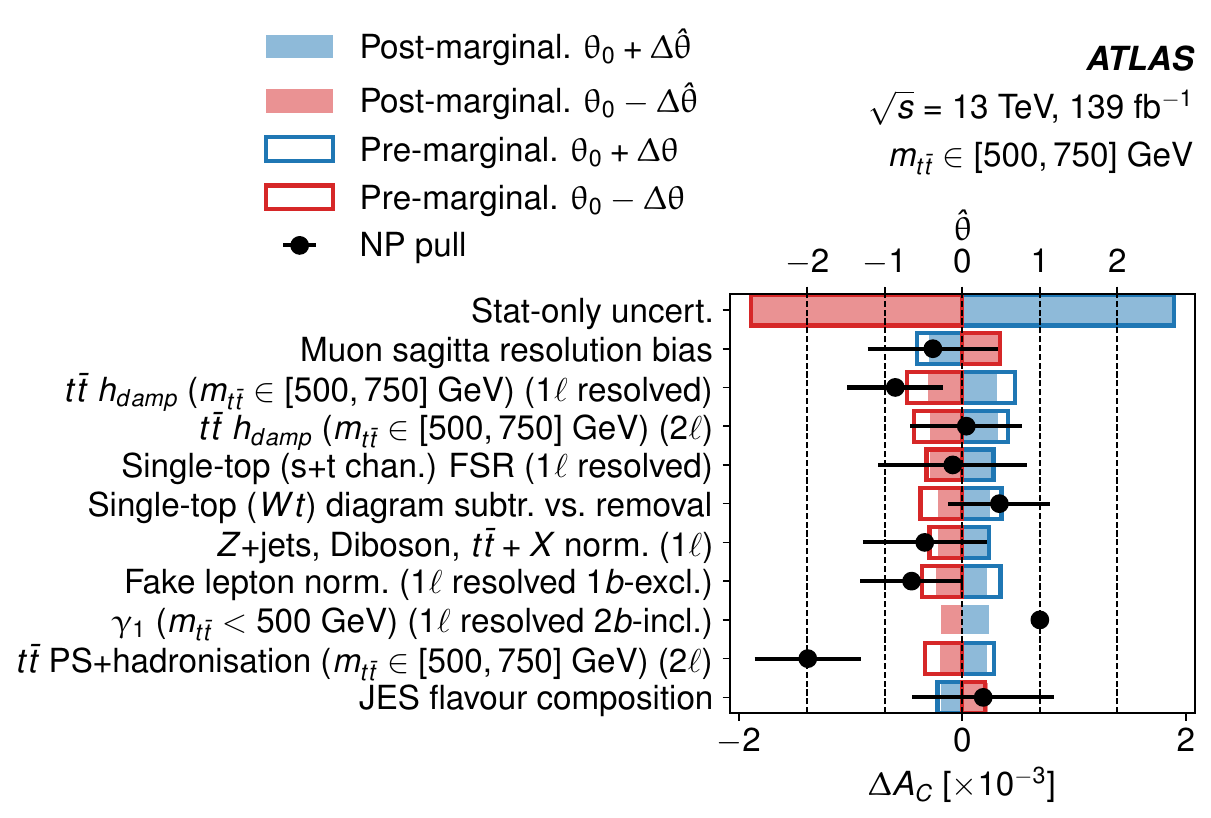}} \\
\subfigure[][750 < \mtt < 1000 GeV]{\includegraphics[width=0.92\textwidth]{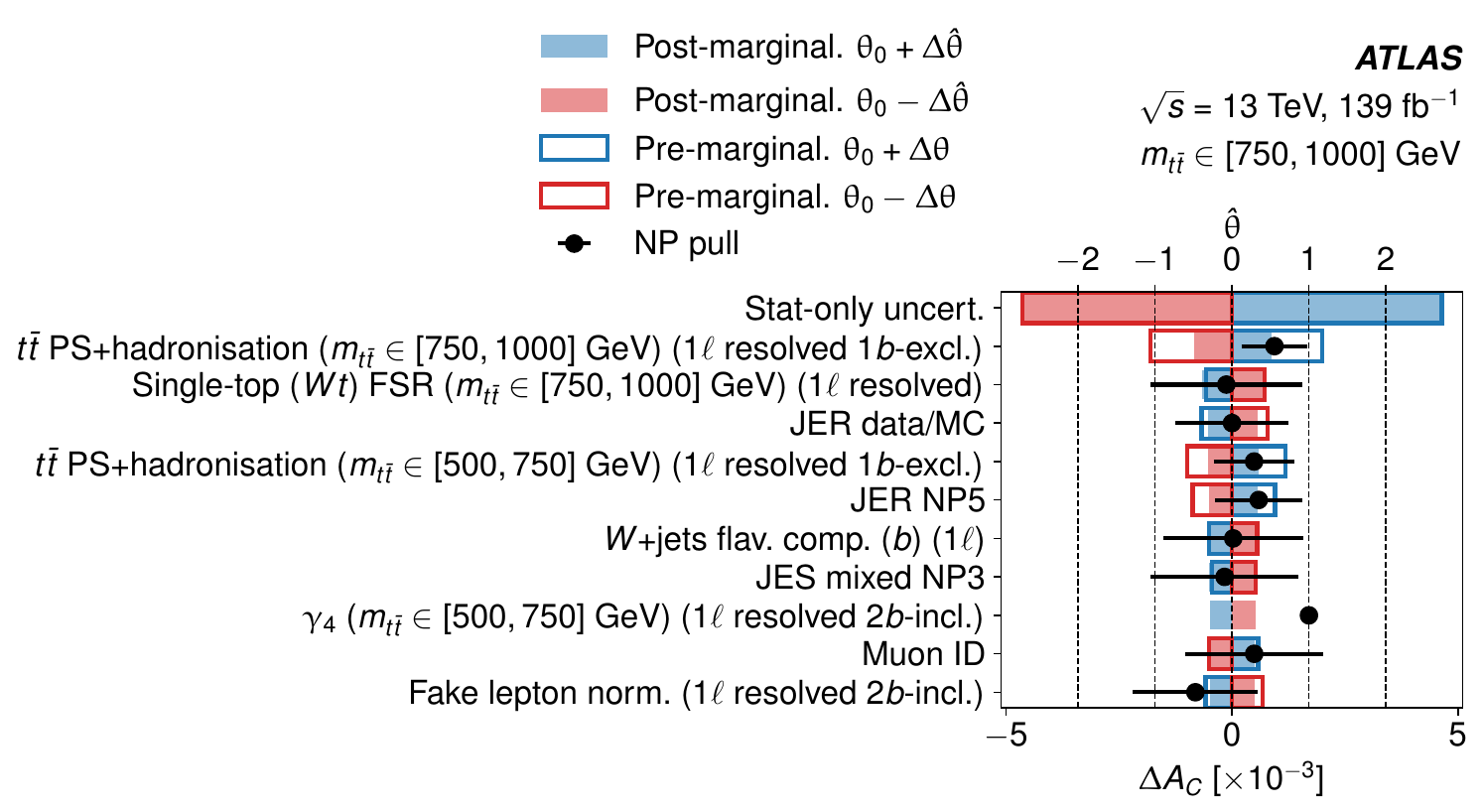}}
\caption{Ranking of the systematic uncertainties with marginalisation (a) for $\mtt \in [500, 750]$~\GeV\ and (b) for $\mtt \in [750, 1000]$~\GeV. Only the ten highest-ranking uncertainties are shown, ordered according to their post-marginalisation ranking, which considers the constraining power of the marginalisation for the data. The red and blue bars show the effect on unfolded \Ac of downward and upward variation of the systematic uncertainty, respectively. The points show the pulls and constraints of the ranked NPs obtained from data.}
\end{figure}
 
\begin{figure}[H]
\centering
\subfigure[][1000 < \mtt < 1500 GeV]{\includegraphics[width=0.92\textwidth]{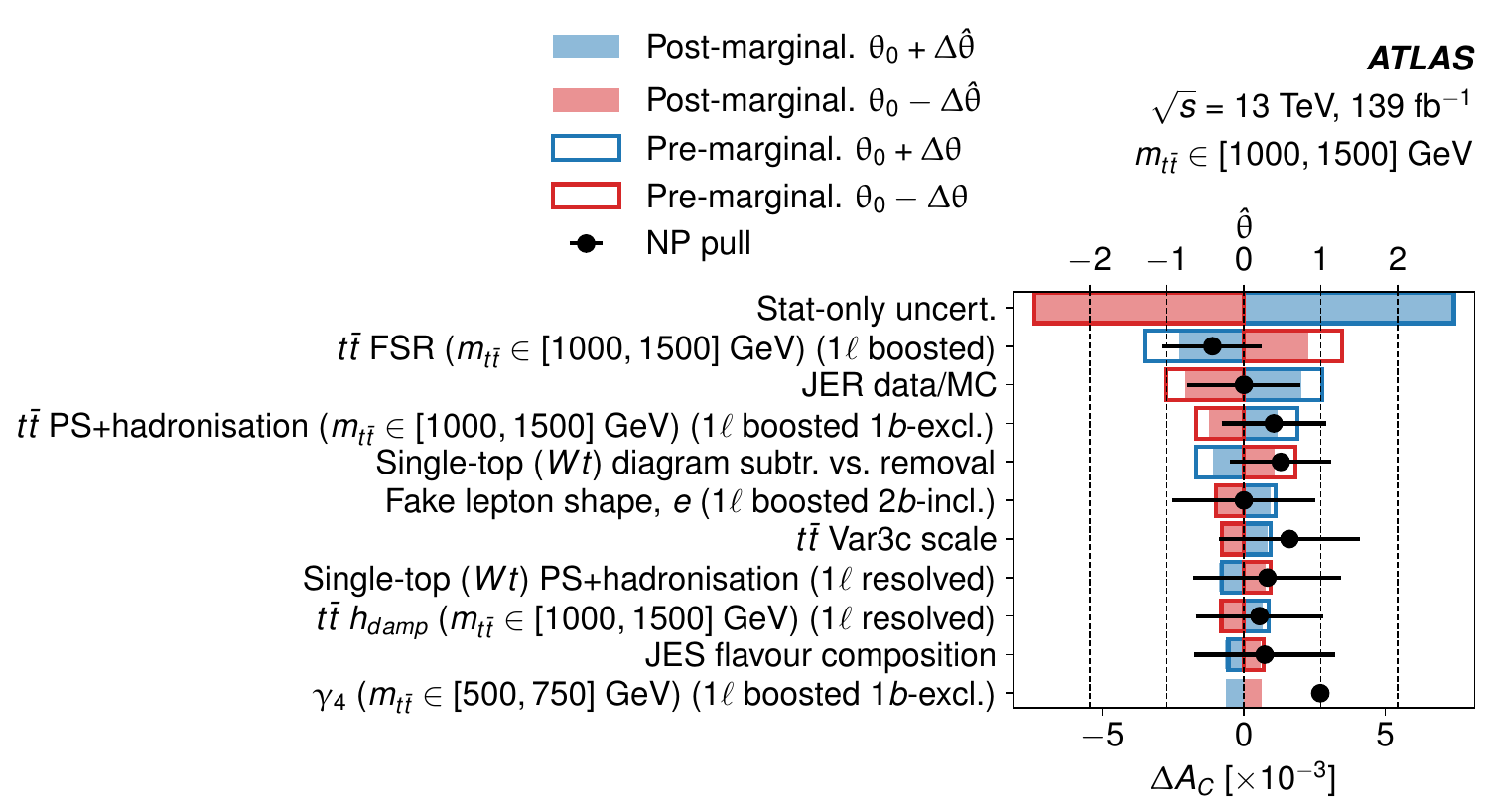}}\\
\subfigure[][\mtt > 1500 GeV]{\includegraphics[width=0.78\textwidth]{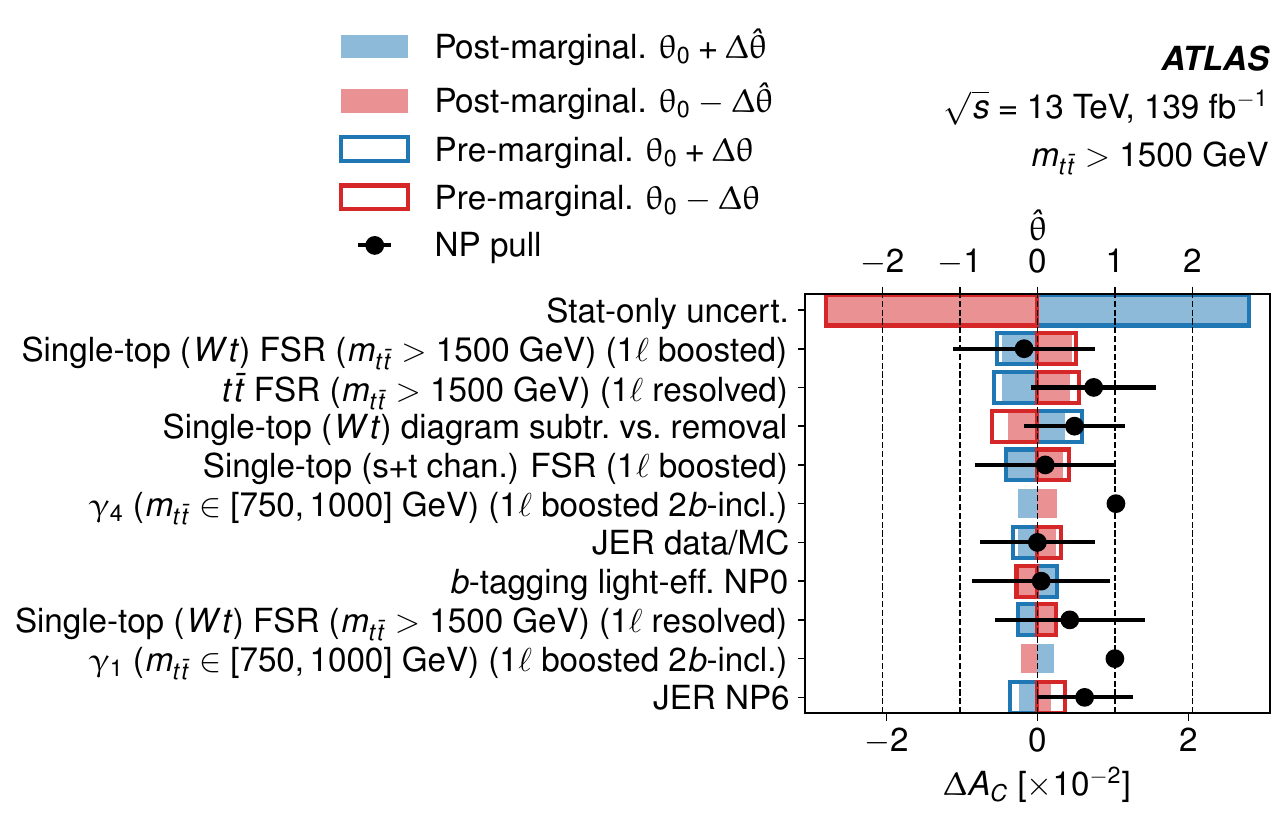}}
\caption{Ranking of the systematic uncertainties with marginalisation (a) for $\mtt \in [1000, 1500]$~\GeV\ and (b) for $\mtt > 1500$~\GeV. Only the ten highest-ranking uncertainties are shown, ordered according to their post-marginalisation ranking, which considers the constraining power of the marginalisation for the data. The red and blue bars show the effect on unfolded \Ac of downward and upward variation of the systematic uncertainty, respectively. The points show the pulls and constraints of the ranked NPs obtained from data.}
\end{figure}
\begin{figure}[H]
\centering
\subfigure[][ \pttt < 30 GeV]{\includegraphics[width=0.76\textwidth]{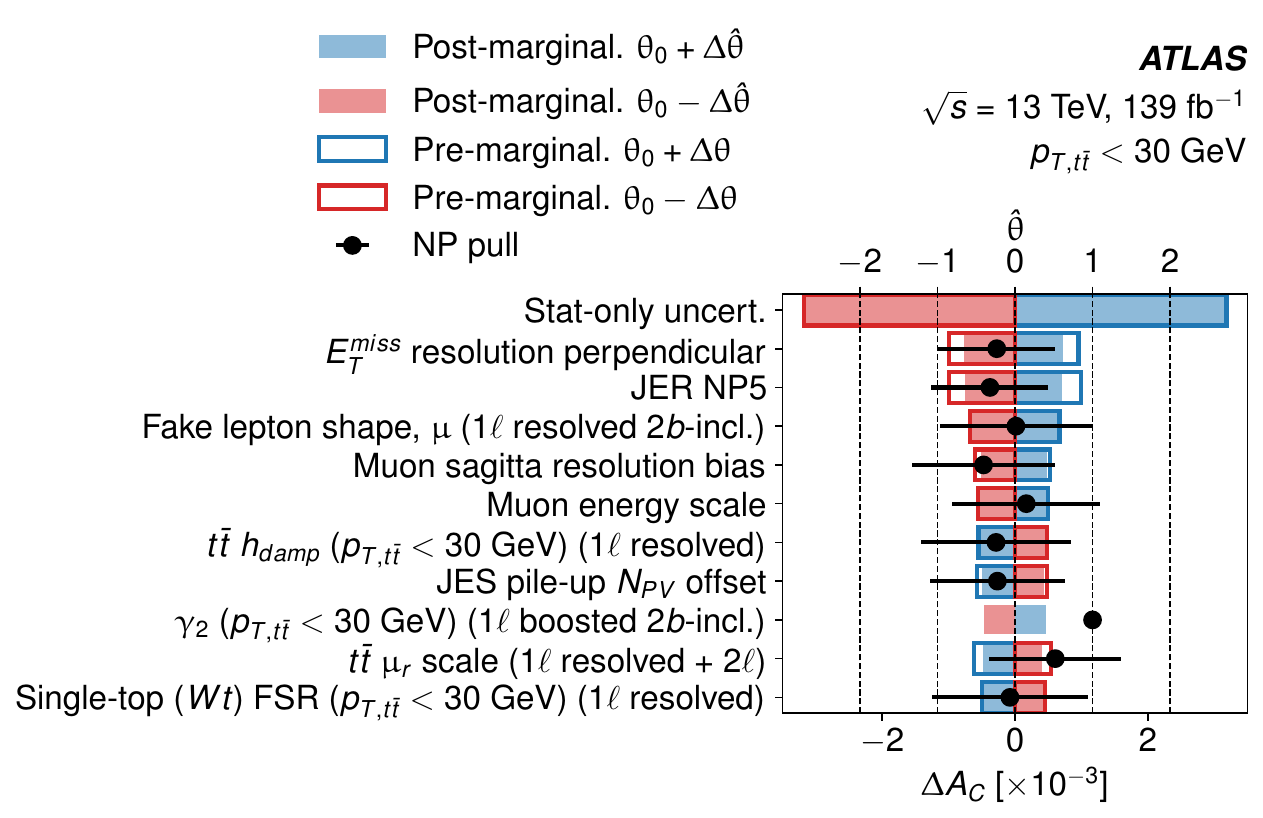}} \\
\subfigure[][30 < \pttt < 120 GeV]{\includegraphics[width=0.7\textwidth]{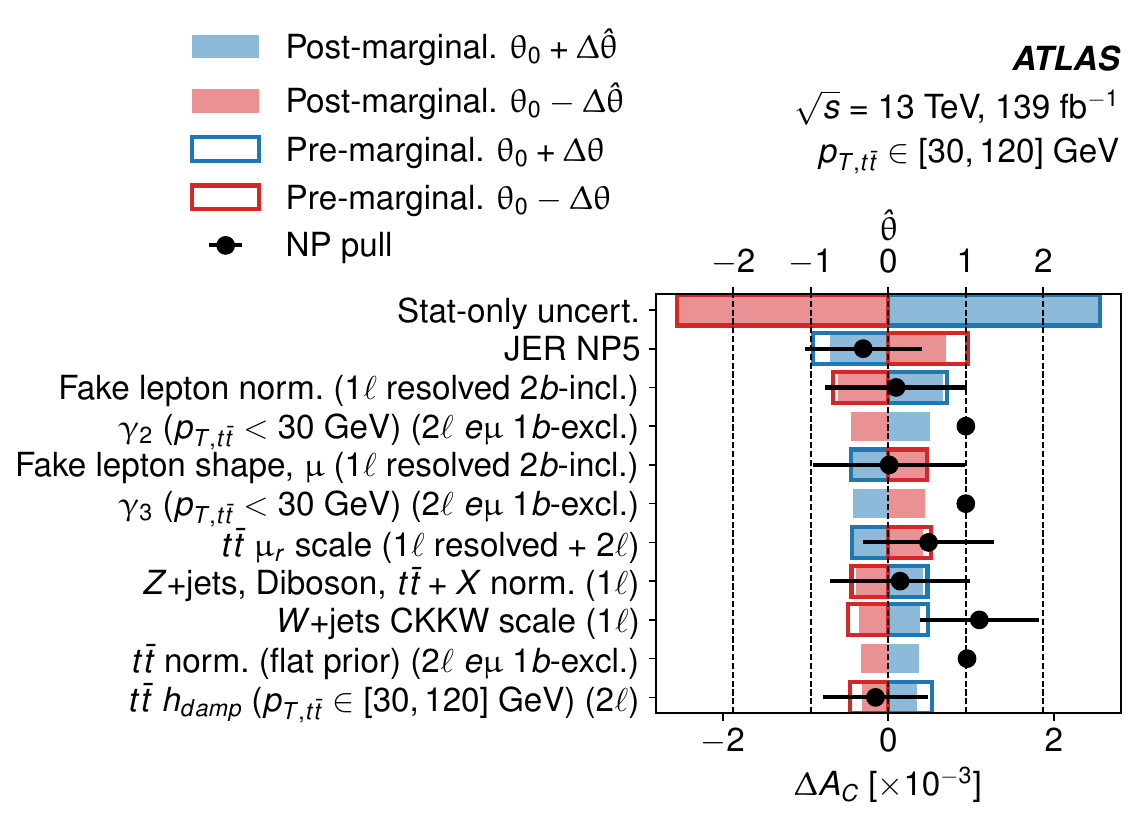}}
\caption{Ranking of the systematic uncertainties with marginalisation (a) for $\pttt \in [0, 30]$~\GeV\ and (b) for $\pttt \in [30, 120]$~\GeV. Only the ten highest-ranking uncertainties are shown, ordered according to their post-marginalisation ranking, which considers the constraining power of the marginalisation for the data. The red and blue bars show the effect on unfolded \Ac of downward and upward variation of the systematic uncertainty, respectively. The points show the pulls and constraints of the ranked NPs obtained from data.}
\end{figure}
 
\begin{figure}[H]
\centering
\subfigure[][ \pttt > 120 GeV]{\includegraphics[width=0.86\textwidth]{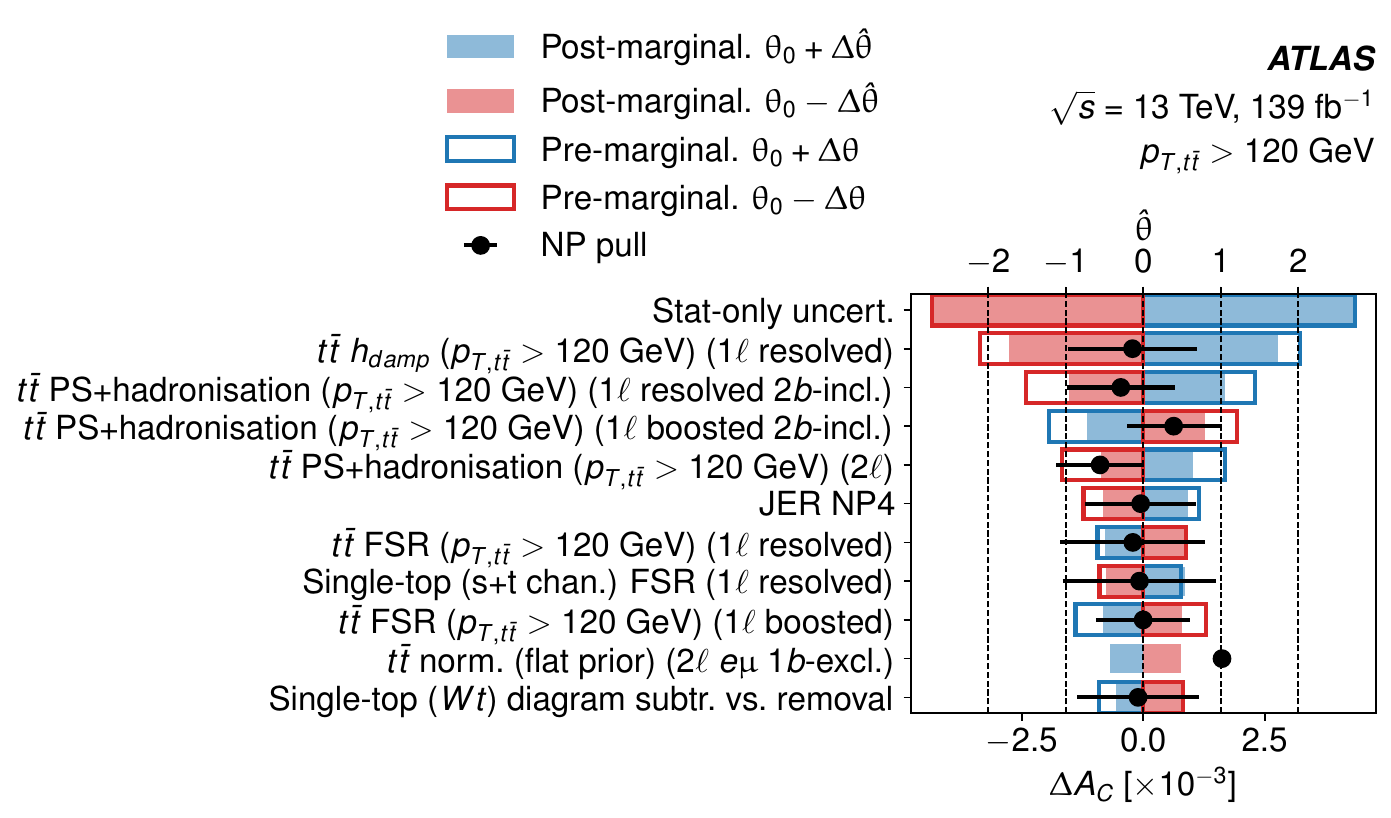}}\\
\caption{Ranking of the systematic uncertainties with marginalisation for $\pttt > 120$~\GeV. Only the ten highest-ranking uncertainties are shown, ordered according to their post-marginalisation ranking, which considers the constraining power of the marginalisation for the data. The red and blue bars show the effect on unfolded \Ac of downward and upward variation of the systematic uncertainty, respectively. The points show the pulls and constraints of the ranked NPs obtained from data.}
\label{fig:NP_rankpull_dypttt_comb_bin2}
\end{figure}


\subsection{\Acll measurement}
\label{app:Acll_ranking}
The ranking of the leading systematic uncertainties is summarised in Tables~\ref{tab:Acll_betall_ranking} and~\ref{tab:Acll_ptll_ranking} for the \betall and \ptll differential \Acll combined measurements, respectively.
Figures~\ref{fig:NP_rankpull_dy_leptonic}--\ref{fig:NP_rankpull_dyptll_leptonic_bin2} show the ten highest-ranking nuisance parameters.  Comparisons of pre-marginalisation and post-marginalisation ranking are presented together with the pulls and constraints of the ranked nuisance parameters obtained from data. In the figures, the $\gamma_1$, $\gamma_2$, $\gamma_3$ and $\gamma_4$ nuisance parameters correspond to $[-2.5,-x]$, $[-x,0]$, $[0,x]$ and $[x,2.5]$ \detall bins, respectively, where $x$ stands for the \detall bin edge.
 
\begin{table}[H]
\caption{Impact of systematic uncertainties on the measurement of \Acll{} vs \betall{}, grouped into categories. The impact of each category is obtained by summing the symmetrised ranking impact of the respective individual uncertainties in quadrature. The total uncertainty differs from the  sum-in-quadrature of the groups due to post-marginalisation nuisance parameter correlations in the unfolding.}
\label{tab:Acll_betall_ranking}
\centering
\begin{tabular}{lllll}
\toprule
& \multicolumn{4}{c}{Post-marg. (pre-marg.) impact $\times 100$} \\
\cmidrule{2-5}
\betall{} bin & ~~\,0--0.3 & 0.3--0.6 & 0.6--0.8 & 0.8--1.0 \\
\midrule
$t\bar{t}$ modelling & ~~\,0.04 ~~~(0.04) & 0.37 (0.41) & 0.27 (0.32) & 0.09 (0.10)\\
$t\bar{t}$ normalisation (flat prior) & ~~\,0.03 & 0.05 & 0.02 & 0.05\\
Background modelling & ~~\,0.04 ~~~(0.05) & 0.05 (0.06) & 0.06 (0.07) & 0.02 (0.03)\\
Monte Carlo statistics & ~~\,0.06 & 0.07 & 0.11 & 0.07\\
Small-$R$ JES & ~~\,0.01 ~~~(0.01) & 0.02 (0.02) & 0.06 (0.07) & 0.02 (0.02)\\
Small-$R$ JER & ~~\,0.03 ~~~(0.03) & 0.07 (0.08) & 0.04 (0.05) & 0.05 (0.06)\\
Leptons, \MET & ~~\,0.02 ~~~(0.02) & 0.02 (0.01) & 0.02 (0.02) & 0.01 (0.01)\\
$b$-tagging eff. & ~~\,0.01 ~~~(0.01) & 0.01 (0.01) & 0.01 (0.01) & 0.01 (0.01)\\
Pile-up, JVT, luminosity & ${<}0.01$ (${<}0.01$) & 0.01 (0.01) & 0.02 (0.02) & 0.01 (0.01)\\
\midrule
Statistical uncertainty & ~~\,0.28 & 0.24 & 0.26 & 0.28\\
\midrule
Total uncertainty & ~~\,0.29 & 0.49 & 0.45 & 0.31\\
\bottomrule
\end{tabular}
\end{table}
 
\begin{table}[H]
\caption{Impact of systematic uncertainties on the measurement of \Acll{} vs \ptll{}, grouped into categories. The impact of each category is obtained by summing the symmetrised ranking impact of the respective individual uncertainties in quadrature. The total uncertainty differs from the  sum-in-quadrature of the groups due to post-marginalisation nuisance parameter correlations in the unfolding.}
\label{tab:Acll_ptll_ranking}
\centering
\begin{tabular}{llll}
\toprule
& \multicolumn{3}{c}{Post-marg. (pre-marg.) impact $\times 100$} \\
\cmidrule{2-4}
\ptll{} bin [\GeV] & 0--20 & 20--70 & $> 70$ \\
\midrule
$t\bar{t}$ modelling & 0.19 (0.23) & 0.24 (0.26) & ~~\,0.16 (0.18)\\
$t\bar{t}$ normalisation (flat prior) & 0.05 & 0.03 & ~~\,0.05\\
Background modelling & 0.06 (0.07) & 0.03 (0.04) & ~~\,0.02 (0.03)\\
Monte Carlo statistics & 0.18 & 0.06 & ~~\,0.05\\
Small-$R$ JES & 0.03 (0.03) & 0.02 (0.03) & ~~\,0.02 (0.02)\\
Small-$R$ JER & 0.05 (0.06) & 0.06 (0.06) & ~~\,0.03 (0.04)\\
Leptons, \MET & 0.04 (0.04) & 0.03 (0.03) & ~~\,0.01 (0.01)\\
$b$-tagging eff. & 0.05 (0.05) & 0.01 (0.01) & ${<}0.01$ (0.01)\\
Pile-up, JVT, luminosity & 0.04 (0.04) & 0.01 (0.01) & ~~\,0.01 (0.01)\\
\midrule
Statistical uncertainty & 0.63 & 0.20 & ~~\,0.18\\
\midrule
Total uncertainty & 0.70 & 0.34 & ~~\,0.27\\
\bottomrule
\end{tabular}
\end{table}
 
\begin{figure}[H]
\centering
\subfigure[][Inclusive]{\includegraphics[width=0.64\textwidth]{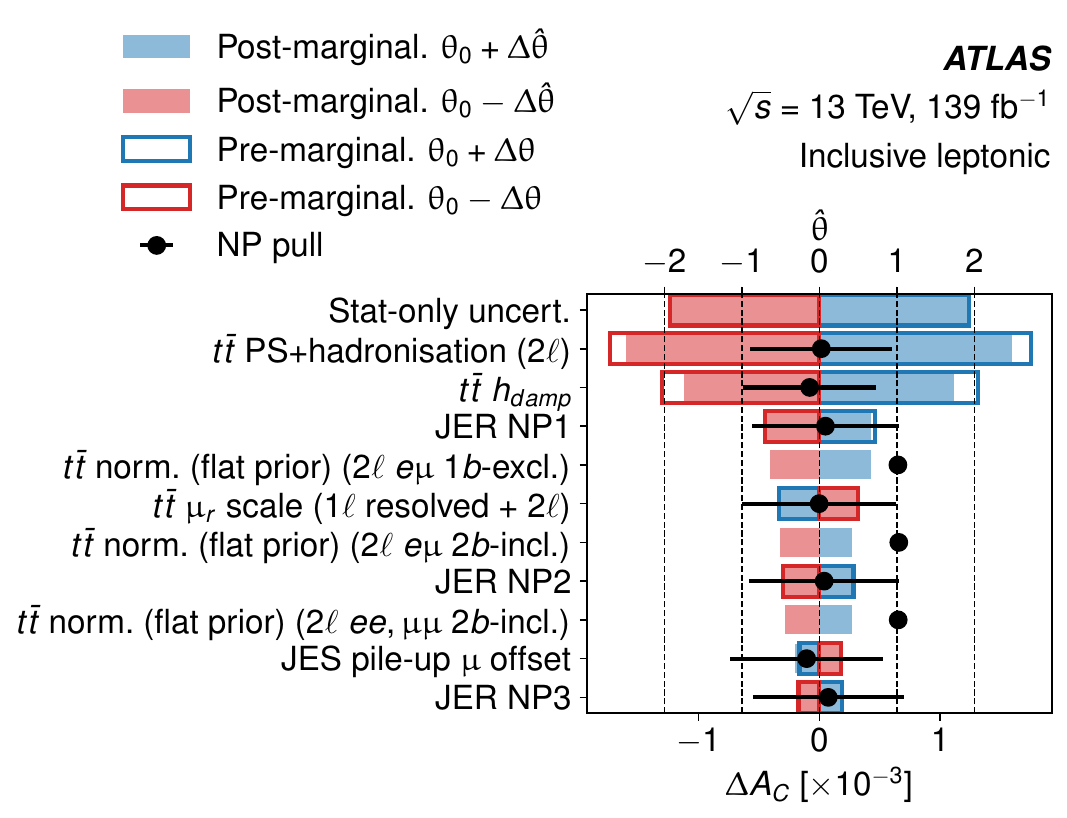}}\\
\subfigure[][0.0 < \betall < 0.3]{\includegraphics[width=0.7\textwidth]{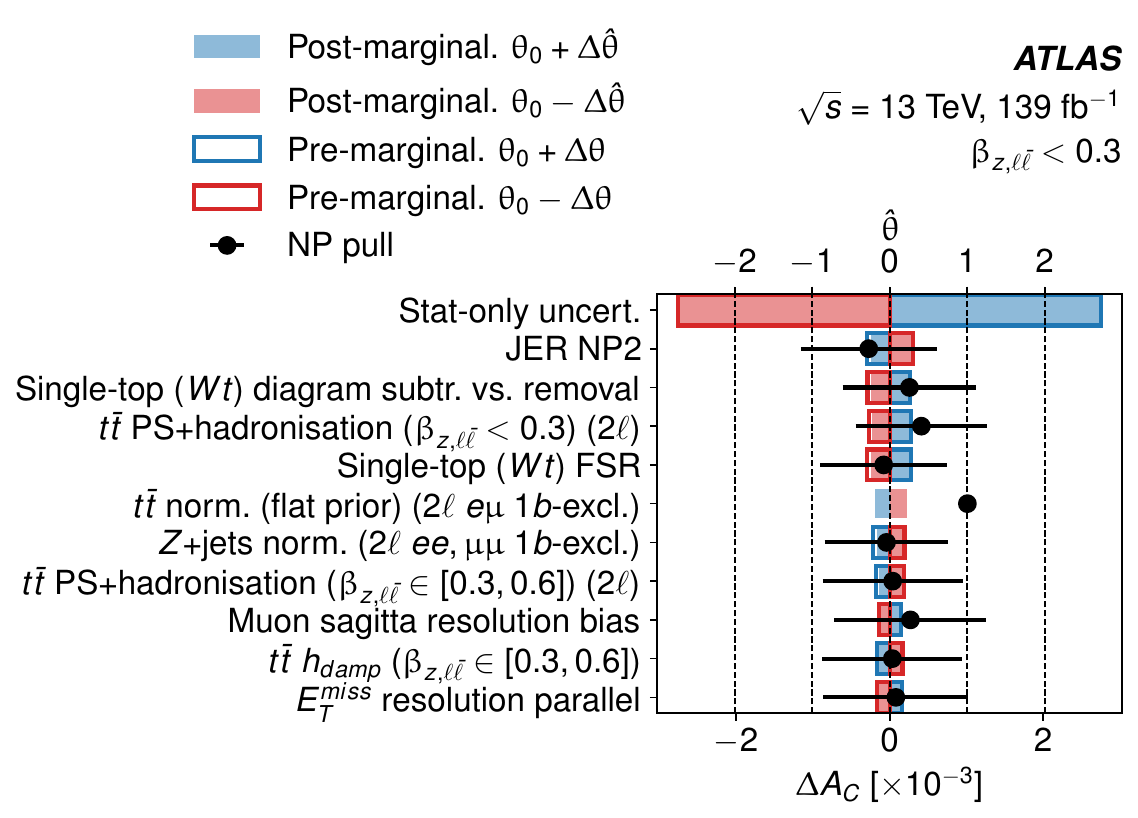}}
\caption{Ranking of the systematic uncertainties with marginalisation (a) for the inclusive \Ac measurement and (b) for $\betall \in [0.0, 0.3]$. Only the ten highest-ranking uncertainties are shown, ordered according to their post-marginalisation ranking, which considers the constraining power of the marginalisation for the data. The red and blue bars show the effect on unfolded \Ac of downward and upward variation of the systematic uncertainty, respectively. The points show the pulls and constraints of the ranked NPs obtained from data.}
\label{fig:NP_rankpull_dy_leptonic}
\end{figure}
 
\begin{figure}[H]
\centering
\subfigure[][0.3 < \betall <  0.6]{\includegraphics[width=0.7\textwidth]{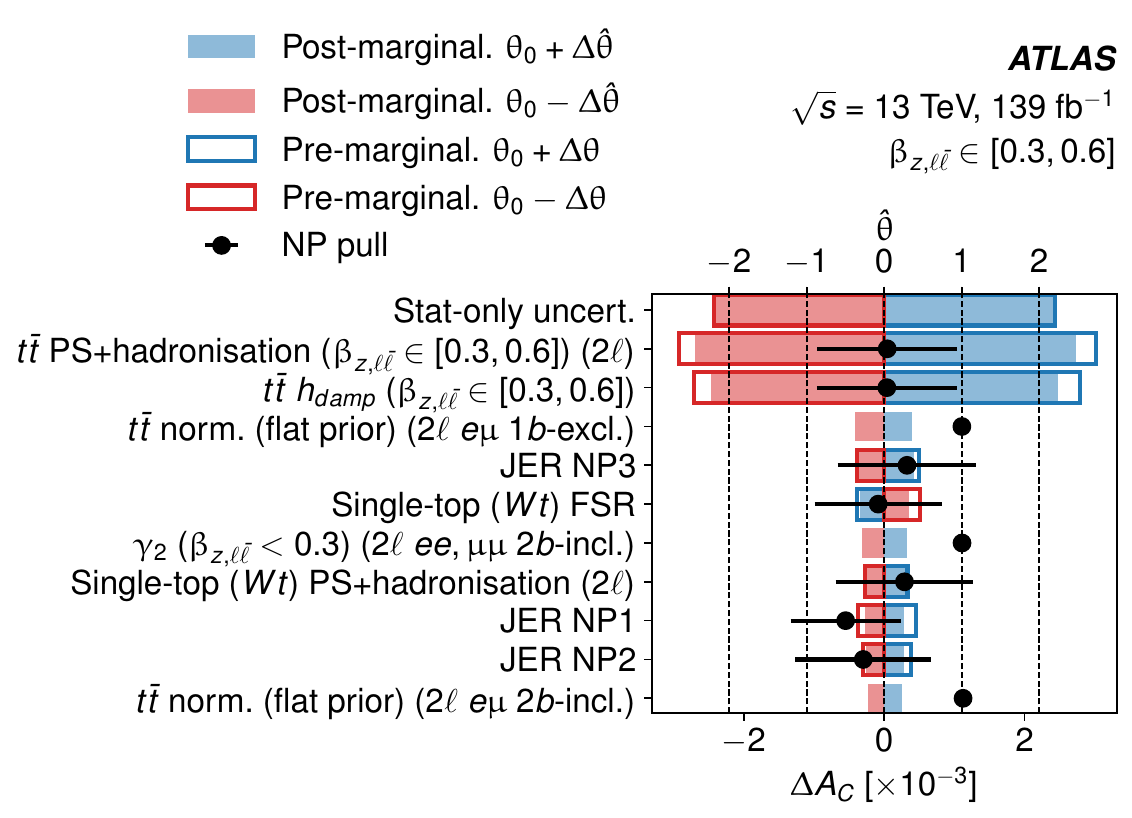}}\\
\subfigure[][0.6 < \betall <  0.8]{\includegraphics[width=0.72\textwidth]{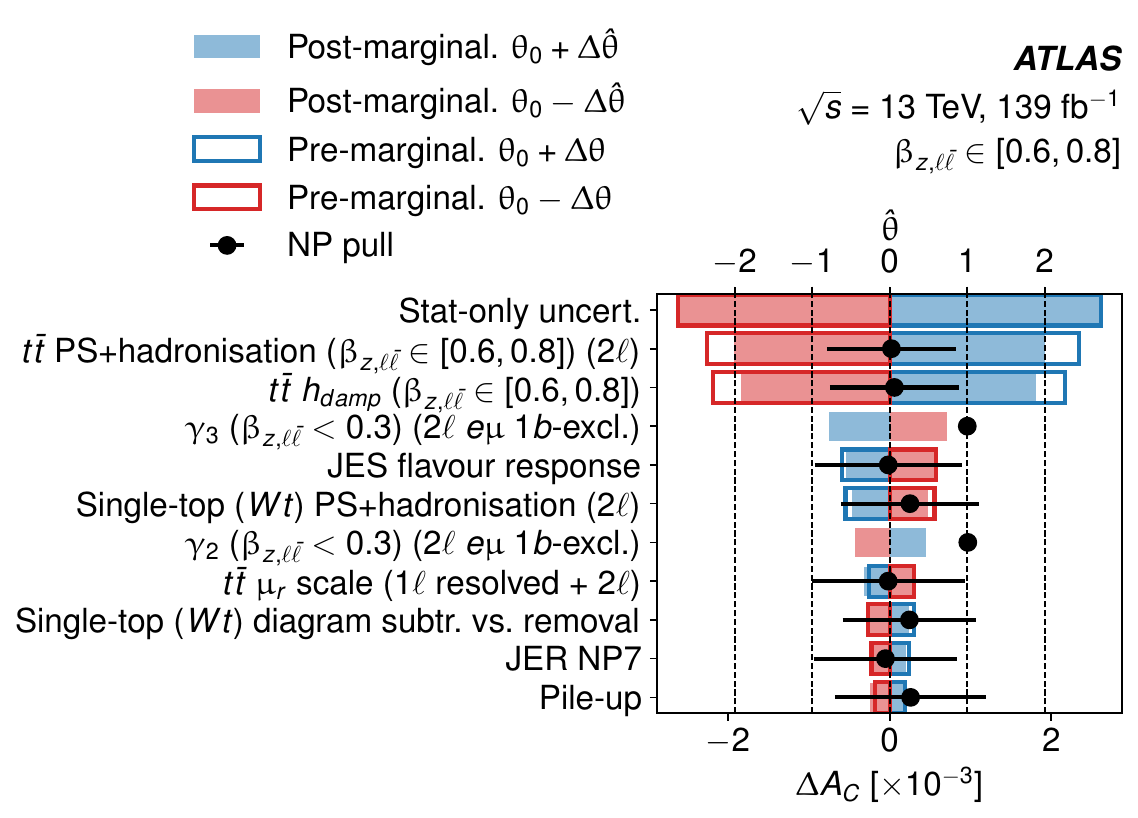}}
\caption{Ranking of the systematic uncertainties with marginalisation (a) for $\betall \in [0.3, 0.6]$ and (b) for $\betall \in [0.6, 0.8]$. Only the ten highest-ranking uncertainties are shown, ordered according to their post-marginalisation ranking, which considers the constraining power of the marginalisation for the data. The red and blue bars show the effect on unfolded \Ac of downward and upward variation of the systematic uncertainty, respectively. The points show the pulls and constraints of the ranked NPs obtained from data.}
\end{figure}
 
\begin{figure}[H]
\centering
\subfigure[][\betall > 0.8]{\includegraphics[width=0.68\textwidth]{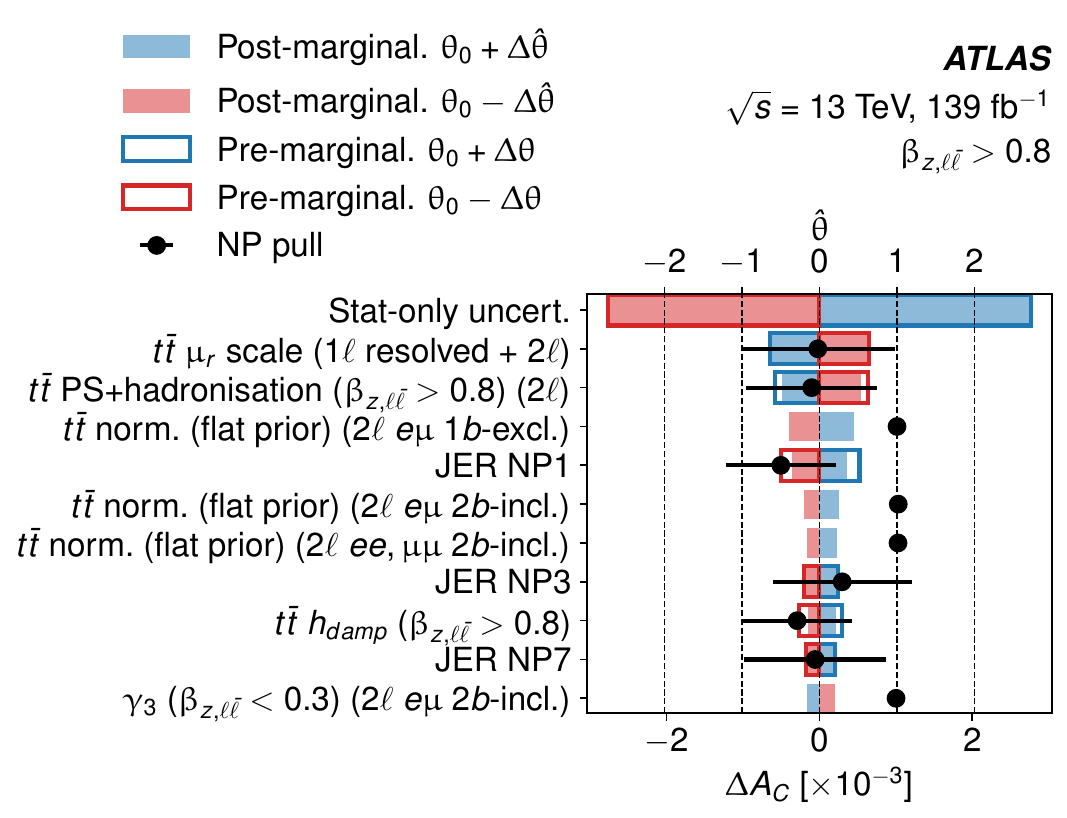}}\\
\subfigure[][\mll < 200 GeV]{\includegraphics[width=0.7\textwidth]{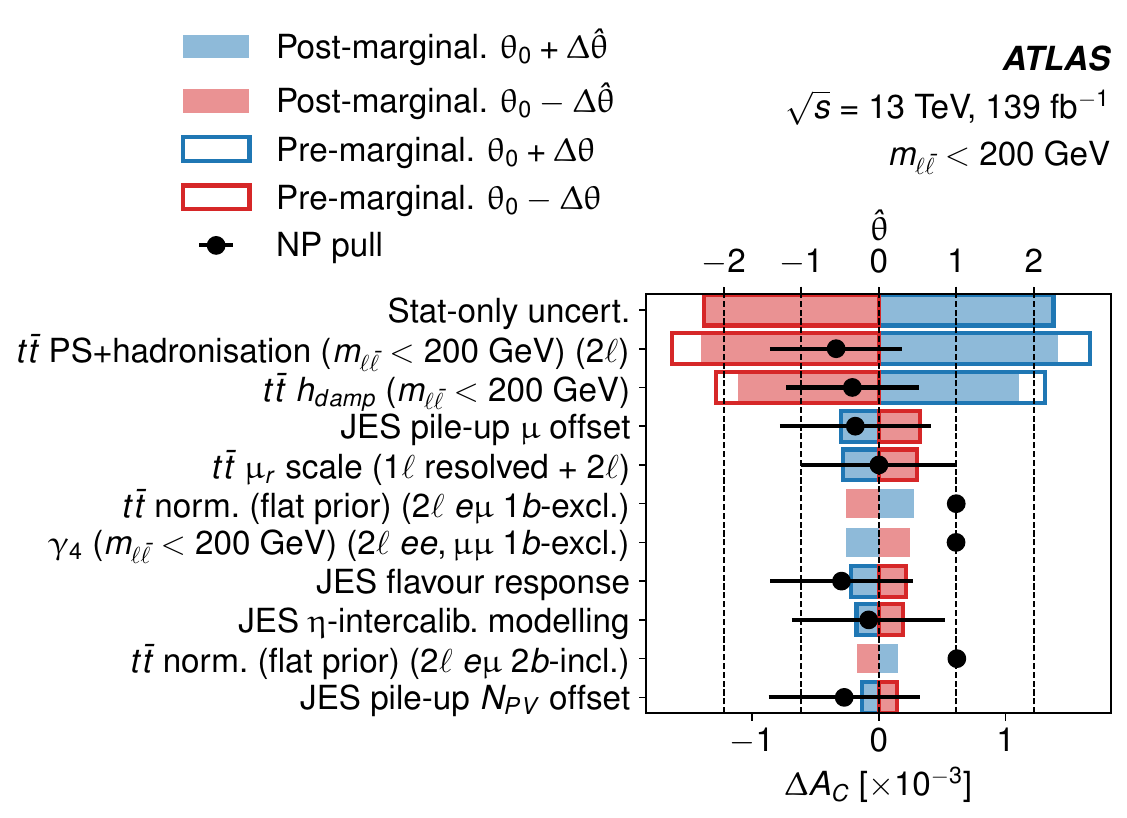}}
\caption{Ranking of the systematic uncertainties with marginalisation (a) for $\betall > 0.8$ and (b) for $\mll \in [0, 200]$~\GeV. Only the ten highest-ranking uncertainties are shown, ordered according to their post-marginalisation ranking, which considers the constraining power of the marginalisation for the data. The red and blue bars show the effect on unfolded \Ac of downward and upward variation of the systematic uncertainty, respectively. The points show the pulls and constraints of the ranked NPs obtained from data.}
\end{figure}
 
\begin{figure}[H]
\centering
\subfigure[][200 < \mll < 300 GeV]{\includegraphics[width=0.75\textwidth]{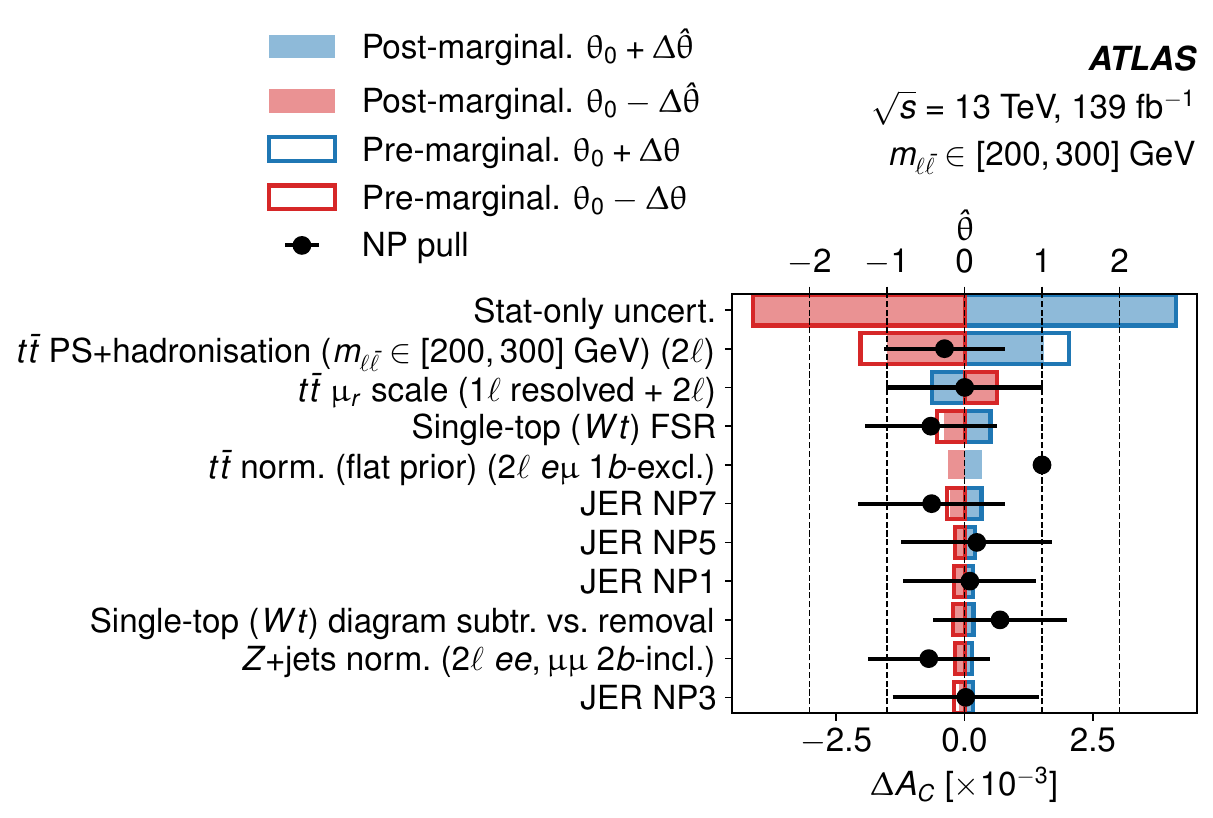}} \\
\subfigure[][300 < \mll < 400 GeV]{\includegraphics[width=0.75\textwidth]{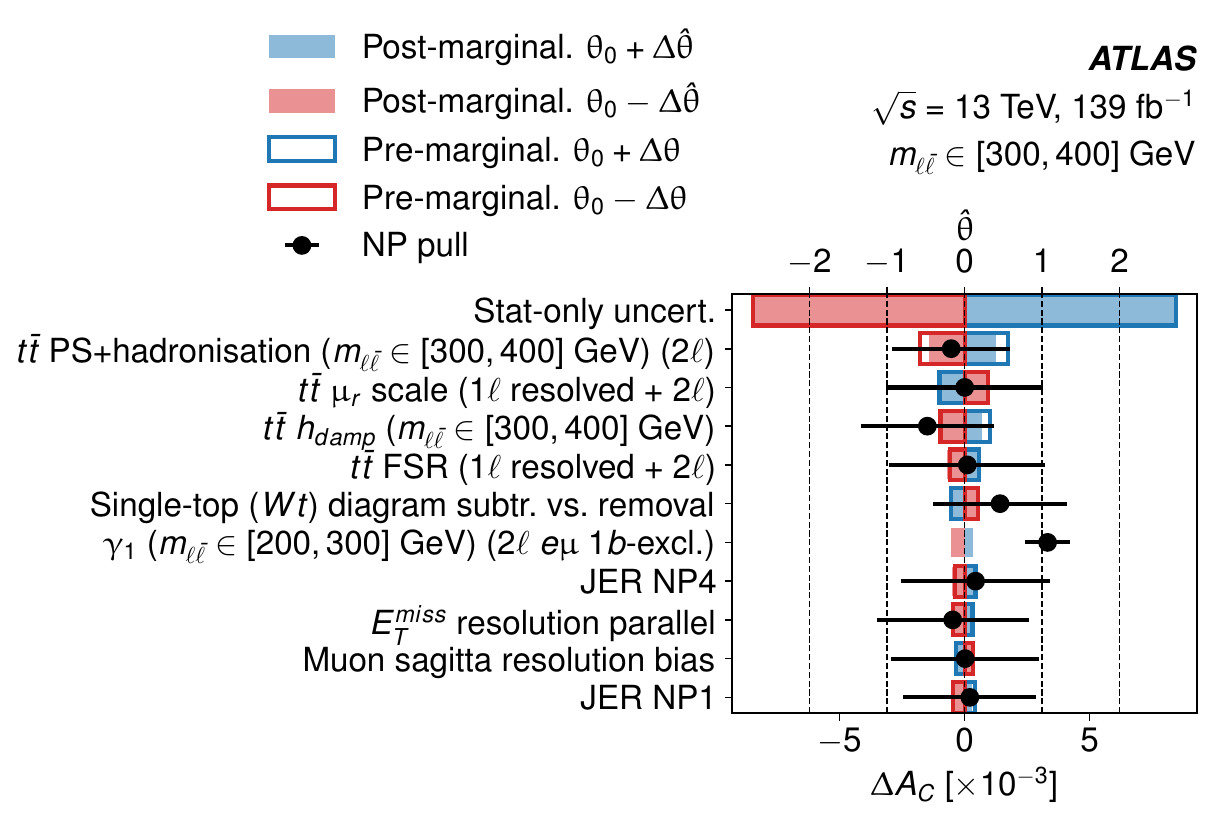}}
\caption{Ranking of the systematic uncertainties with marginalisation (a) for $\mll \in [200, 300]$~\GeV\ and (b) for $\mll \in [300, 400]$~\GeV. Only the ten highest-ranking uncertainties are shown, ordered according to their post-marginalisation ranking, which considers the constraining power of the marginalisation for the data. The red and blue bars show the effect on unfolded \Ac of downward and upward variation of the systematic uncertainty, respectively. The points show the pulls and constraints of the ranked NPs obtained from data.}
\end{figure}
 
\begin{figure}[H]
\centering
\subfigure[][\mll > 400 GeV]{\includegraphics[width=0.7\textwidth]{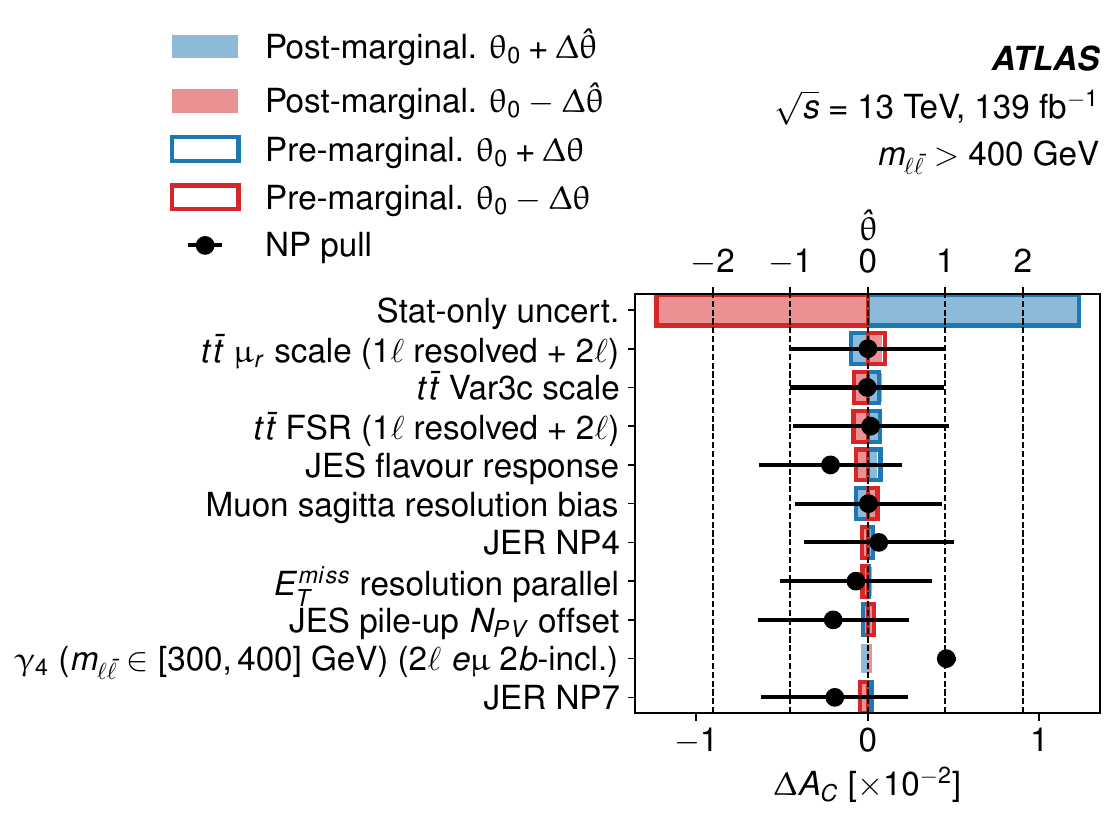}}\\
\subfigure[][\ptll < 20 GeV]{\includegraphics[width=0.7\textwidth]{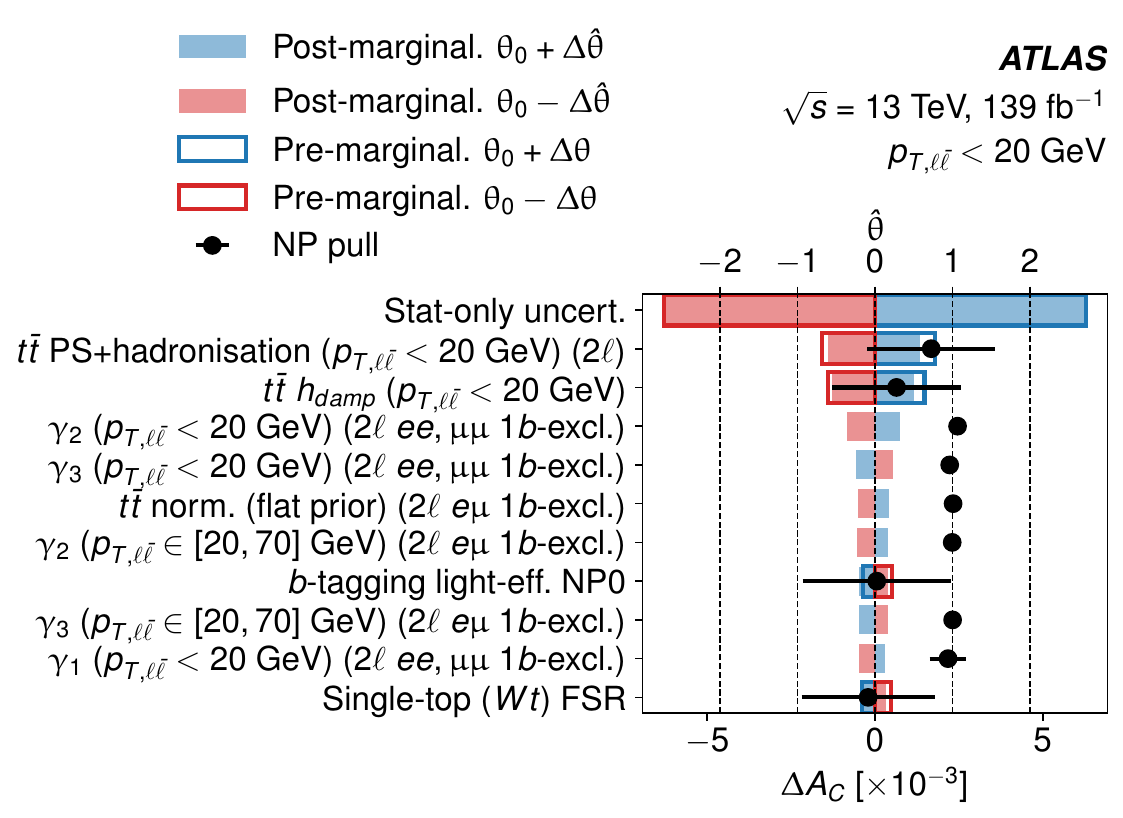}}
\caption{Ranking of the systematic uncertainties with marginalisation (a) for $\mll > 400$~\GeV\ and (b) for $\ptll < 20$~\GeV. Only the ten highest-ranking uncertainties are shown, ordered according to their post-marginalisation ranking, which considers the constraining power of the marginalisation for the data. The red and blue bars show the effect on unfolded \Ac of downward and upward variation of the systematic uncertainty, respectively. The points show the pulls and constraints of the ranked NPs obtained from data.}
\end{figure}
\begin{figure}[H]
\centering
\subfigure[][20 < \ptll < 70 GeV]{\includegraphics[width=0.71\textwidth]{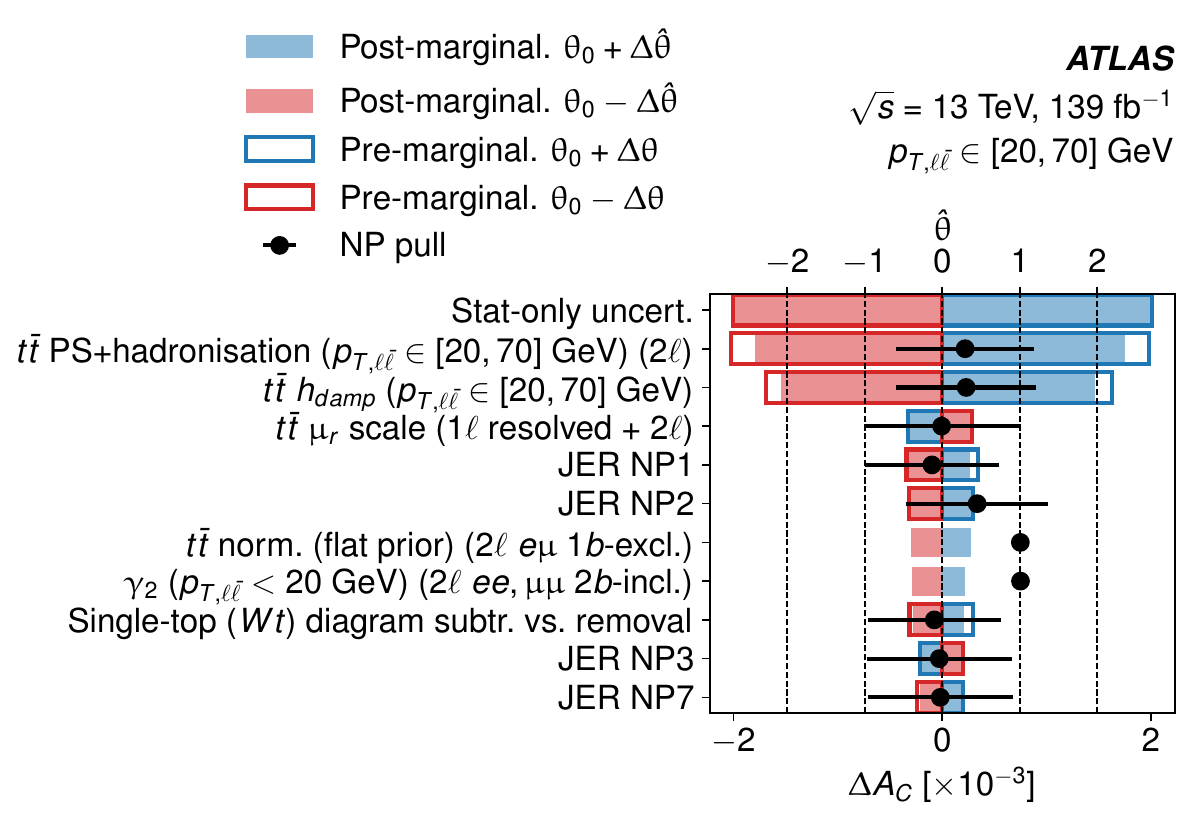}} \\
\subfigure[][\ptll > 70 GeV]{\includegraphics[width=0.7\textwidth]{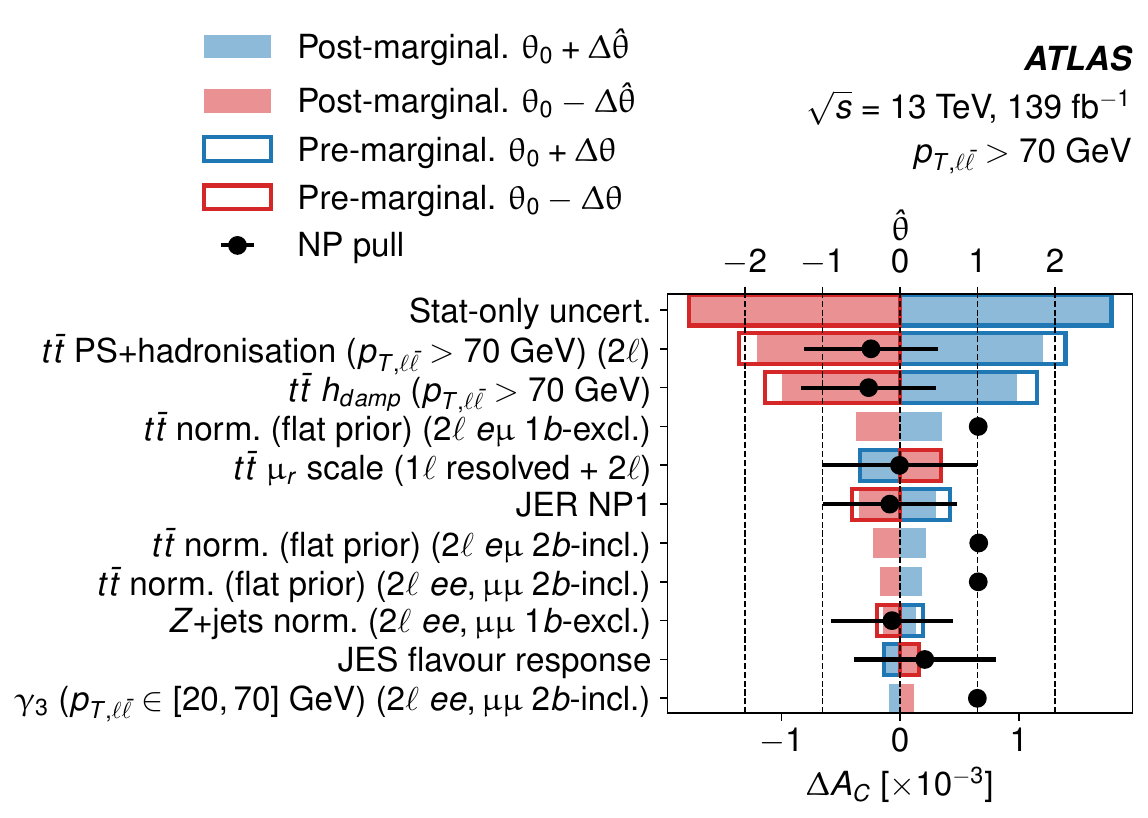}}
\caption{Ranking of the systematic uncertainties with marginalisation (a) for $\ptll \in [20, 70]$~\GeV\ and (b) for $\ptll > 70$~\GeV. Only the ten highest-ranking uncertainties are shown, ordered according to their post-marginalisation ranking, which considers the constraining power of the marginalisation for the data. The red and blue bars show the effect on unfolded \Ac of downward and upward variation of the systematic uncertainty, respectively. The points show the pulls and constraints of the ranked NPs obtained from data.}
\label{fig:NP_rankpull_dyptll_leptonic_bin2}
\end{figure}


\printbibliography

\clearpage
 
\begin{flushleft}
\hypersetup{urlcolor=black}
{\Large The ATLAS Collaboration}

\bigskip

\AtlasOrcid[0000-0002-6665-4934]{G.~Aad}$^\textrm{\scriptsize 101}$,
\AtlasOrcid[0000-0002-5888-2734]{B.~Abbott}$^\textrm{\scriptsize 119}$,
\AtlasOrcid[0000-0002-7248-3203]{D.C.~Abbott}$^\textrm{\scriptsize 102}$,
\AtlasOrcid[0000-0002-1002-1652]{K.~Abeling}$^\textrm{\scriptsize 55}$,
\AtlasOrcid[0000-0002-8496-9294]{S.H.~Abidi}$^\textrm{\scriptsize 29}$,
\AtlasOrcid[0000-0002-9987-2292]{A.~Aboulhorma}$^\textrm{\scriptsize 35e}$,
\AtlasOrcid[0000-0001-5329-6640]{H.~Abramowicz}$^\textrm{\scriptsize 150}$,
\AtlasOrcid[0000-0002-1599-2896]{H.~Abreu}$^\textrm{\scriptsize 149}$,
\AtlasOrcid[0000-0003-0403-3697]{Y.~Abulaiti}$^\textrm{\scriptsize 116}$,
\AtlasOrcid[0000-0003-0762-7204]{A.C.~Abusleme~Hoffman}$^\textrm{\scriptsize 136a}$
% ...
\bigskip
\\

% ...
$^{29}$Physics Department, Brookhaven National Laboratory, Upton NY; United States of America.\\
% ...
$^{35}$$^{(a)}$Facult\'e des Sciences Ain Chock, R\'eseau Universitaire de Physique des Hautes Energies - Universit\'e Hassan II, Casablanca;$^{(b)}$Facult\'{e} des Sciences, Universit\'{e} Ibn-Tofail, K\'{e}nitra;$^{(c)}$Facult\'e des Sciences Semlalia, Universit\'e Cadi Ayyad, LPHEA-Marrakech;$^{(d)}$LPMR, Facult\'e des Sciences, Universit\'e Mohamed Premier, Oujda;$^{(e)}$Facult\'e des sciences, Universit\'e Mohammed V, Rabat;$^{(f)}$Institute of Applied Physics, Mohammed VI Polytechnic University, Ben Guerir; Morocco.\\
% ...
$^{55}$II. Physikalisches Institut, Georg-August-Universit\"{a}t G\"ottingen, G\"ottingen; Germany.\\
% ...
$^{101}$CPPM, Aix-Marseille Universit\'e, CNRS/IN2P3, Marseille; France.\\
$^{102}$Department of Physics, University of Massachusetts, Amherst MA; United States of America.\\
% ...
$^{116}$Department of Physics, New York University, New York NY; United States of America.\\
% ...
$^{119}$Homer L. Dodge Department of Physics and Astronomy, University of Oklahoma, Norman OK; United States of America.\\
% ...
$^{136}$$^{(a)}$Departamento de F\'isica, Pontificia Universidad Cat\'olica de Chile, Santiago;$^{(b)}$Millennium Institute for Subatomic physics at high energy frontier (SAPHIR), Santiago;$^{(c)}$Instituto de Investigaci\'on Multidisciplinario en Ciencia y Tecnolog\'ia, y Departamento de F\'isica, Universidad de La Serena;$^{(d)}$Universidad Andres Bello, Department of Physics, Santiago;$^{(e)}$Instituto de Alta Investigaci\'on, Universidad de Tarapac\'a, Arica;$^{(f)}$Departamento de F\'isica, Universidad T\'ecnica Federico Santa Mar\'ia, Valpara\'iso; Chile.\\
% ...
$^{149}$Department of Physics, Technion, Israel Institute of Technology, Haifa; Israel.\\
$^{150}$Raymond and Beverly Sackler School of Physics and Astronomy, Tel Aviv University, Tel Aviv; Israel.\\
% ...
$^{a}$ Also Affiliated with an institute covered by a cooperation agreement with CERN.\\
$^{b}$ Also at An-Najah National University, Nablus; Palestine.\\
$^{c}$ Also at Borough of Manhattan Community College, City University of New York, New York NY; United States of America.\\
$^{d}$ Also at Bruno Kessler Foundation, Trento; Italy.\\
$^{e}$ Also at Center for High Energy Physics, Peking University; China.\\
$^{f}$ Also at Centro Studi e Ricerche Enrico Fermi; Italy.\\
% ...

\end{flushleft}


\end{document}